\documentclass[fleqn,usenatbib]{mnras}

\usepackage{newtxtext,newtxmath}

\usepackage[T1]{fontenc}

\DeclareRobustCommand{\VAN}[3]{#2}
\let\VANthebibliography\thebibliography
\def\thebibliography{\DeclareRobustCommand{\VAN}[3]{##3}\VANthebibliography}

\usepackage{graphicx}	
\usepackage{amsmath}	

\defcitealias{Constantin23}{C23}
\defcitealias{Roper22}{R22}

\title[The sizes of bright LBGs at $z\simeq3-5$ with PRIMER]{The sizes of bright Lyman-break galaxies at $z\simeq3-5$ with JWST PRIMER}

\author[R. G. Varadaraj et al.]{R. G. Varadaraj,$^{1}$\thanks{rohan.varadaraj@physics.ox.ac.uk}
R. A. A. Bowler,$^{2}$
M. J. Jarvis,$^{1,3}$
N. J. Adams,$^{2}$
N. Choustikov,$^{1}$
A. M. Koekemoer,$^{4}$
\newauthor
A. C. Carnall,$^{5}$
D. J. McLeod,$^{5}$
J. S. Dunlop,$^{5}$
C. T. Donnan,$^{5}$
N. A. Grogin$^{4}$
\\
$^{1}$Sub-department of Astrophysics, University of Oxford, Denys Wilkinson Building, Keble Road, Oxford, OX1 3RH, UK\\
$^{2}$Jodrell Bank Centre for Astrophysics, University of Manchester, Oxford Road, Manchester, M13 9PL, UK\\
$^{3}$Department of Physics, University of the Western Cape, Bellville 7535, South Africa\\
$^{4}$Space Telescope Science Institute, 3700 San Martin Drive, Baltimore, MD 21218, USA\\
$^{5}$Institute for Astronomy, University of Edinburgh, Royal Observatory, Edinburgh, EH9 3HJ\\
}

\date{Accepted XXX. Received YYY; in original form ZZZ}

\pubyear{2024}

\begin{document}
\label{firstpage}
\pagerange{\pageref{firstpage}--\pageref{lastpage}}
\maketitle

\begin{abstract}
We use data from the \textit{JWST} Public Release IMaging for Extragalactic Research (PRIMER) survey to measure the size scaling relations of 1668 rest-frame UV-bright Lyman-break galaxies (LBGs) at $z=3-5$ with stellar masses $\mathrm{log}_{10}(M_{\star}/M_{\sun}) > 9$.
The sample was selected from seeing-dominated ground-based data, presenting an unbiased sampling of the morphology and size distributions of luminous sources.
We fit S\'ersic profiles to eight NIRCam bands and also measure a non-parametric half-light radius. 
We find that the size distributions with both measurements are well-fit by a log-normal distribution at all redshifts, consistent with disk formation models where size is governed by host dark-matter halo angular momentum.
We find a size-redshift evolution of $R_{e} = 3.51(1+z)^{-0.60\pm0.22}$ kpc, in agreement with \textit{JWST} studies.
When considering the \textit{typical} (modal) size over $z=3-5$, we find little evolution with bright LBGs remaining compact at $R_{e}\simeq0.7-0.9$ kpc. Simultaneously, we find evidence for a build-up of large ($R_{e} > 2$ kpc) galaxies by $z=3$.
We find some evidence for a negatively sloped size-mass relation at $z=5$ when S\'ersic profiles are used to fit the data in F200W.
The intrinsic scatter in our size-mass relations increases at higher redshifts.
Additionally, measurements probing the rest-UV (F200W) show larger intrinsic scatter than those probing the rest-optical (F356W).
Finally, we leverage rest-UV and rest-optical photometry to show that disky galaxies are well established by $z=5$, but are beginning to undergo dissipative processes, such as mergers, by $z=3$. 
The agreement of our size-mass and size-luminosity relations with simulations provides tentative evidence for centrally concentrated star formation at high-redshift.
\end{abstract}




\section{Introduction}

The diversity of galaxy shapes and size have fascinated astronomers for decades. 
Early visual classification of structure within galaxies established the Hubble tuning fork, a classification system still used to this day \citep{Hubble1926}.
In the distant Universe, where galaxies are still evolving into the structures we see locally, their sizes are a highly useful probe into the fundamental physics of their growth \citep[e.g. see][]{Conselice14}.

Luminous, massive ($\mathrm{log}_{10}(M_{\star}/M_{\sun}) > 9$) galaxies at high-redshift ($z\gtrsim3$) are excellent laboratories for galaxy physics. 
These galaxies populate the most massive dark matter haloes \citep{wechsler18} and must be more evolved than their low-mass counterparts to accumulate such large stellar mass. 
This means they have had time to undergo major mergers, and there may have been enough time for star-forming clumps to coalesce and form bulges and disks \citep[e.g.][]{Dekel09, Rujopakarn23}. 
Massive galaxies are amongst the first to undergo a quenching of their star formation \citep[e.g.][]{Cowie96, Peng10, McLeod21}. 
Studying the resolved properties of these galaxies can thus enable us to learn about the earliest structural formation and feedback processes. 
The search for these rare massive galaxies was revolutionised by the advent of degree-scale ground-based imaging \citep[e.g.][]{miyazaki02, uds, UltraVISTA, VIDEO}, necessary due to the low surface densities of high-redshift Lyman break galaxies (LBGs). 
Since then, various number statistics of these rare high-redshift galaxies have been well-constrained at $z\simeq3-10$, such as the rest-UV luminosity function \citep[e.g.][]{Bowler14, ono18, stefanon19, Bowler20, bouwens21, harikane22, donnan22, adams23, Varadaraj23} and the two-point correlation function \citep[clustering, e.g.][]{adelberger05, bian13, Durkalec15, Hatfield18, harikane22}. 
However, since ground-based images are often seeing-dominated \citep[e.g.][]{Bowler17}, resolved properties have hitherto been difficult to measure.

Space-based telescopes such as the \textit{Hubble Space Telescope} (\textit{HST}) do not suffer from atmospheric seeing, providing a clear view of galaxy morphology \citep[e.g.][]{Ferguson04, Trujillo06, mosleh10, law12, Grazian12, vanderWel14, Shibuya15, whitney19}, but rest-frame optical-wavelength shapes were only observable out to $z\simeq3$. 
This limited the comparison of high-redshift morphologies with local optical observations, as well as the comparison between young and old stellar populations (as probed by the rest-UV and rest-optical respectively) in high-redshift galaxies. 
\textit{JWST}'s infrared coverage and unparalleled resolution has opened the window to rest-UV \textit{and} rest-optical for the first time at $z\gtrsim3$ \citep[e.g.][]{Ferreira23, kartaltepe23, Leethochawalit23, Ormerod23} revealing early mass assembly and a significant fraction of disk galaxies by $z\simeq6$.

Models describe galaxy disk formation by creating rotationally supported exponential disks from baryons. 
These disks have mass and angular momentum that are some fixed fraction of the host dark matter (DM) halo. 
The disk scale radius is proportional to a dimensionless `spin parameter' $\lambda$, which is related to the angular momentum, mass and energy of the system \citep{Fall80, Mo98, Bullock01}. 
These models can be tested with various size scaling relations. 
First, if the size of galaxies are determined from their host halo angular momentum, then we expect the size distribution of galaxies to follow a log-normal distribution \citep{deJong00}.
Second, the disk model of \citet{Fall80} predicts that the disk size of a galaxy scales with redshift as $r\propto(1+z)^{-1}$ in the case of a fixed circular velocity of the host DM halo, and as $r\propto(1+z)^{-1.5}$ in the case of fixed virial mass.
Prior to \textit{JWST}, various studies have supported each relation, as well as for values in-between \citep[e.g.][]{Hathi08, Oesch10, mosleh10}. 
Furthermore, \citet{curtis_lake16} argue that if one considers the \textit{typical} (modal) sizes of galaxies then no strong evolution is seen, with the effect driven by incompleteness in \textit{HST} samples at $z\gtrsim3$.
Third, the size-mass relation (and the related size-luminosity relation) gives insight into quenching mechanisms. 
In the local Universe, quiescent galaxies follow a steeper size-mass relation than star-forming galaxies \citep[SFGs,][]{vanderWel14}. 
The onset of quenching may thus be seen in an evolution of the high-redshift size-mass relation.
Additionally, within the framework of these disk formation models, \citet{vanderWel14} find no evolution in the size-mass relation of SFGs, suggesting that their sizes are still controlled by their host DM halos. 
Note however that large scatter and uncertainty exists in these measurements \citep[e.g.][]{huang13, curtis_lake16, Shibuya15, whitney19}.

\textit{JWST} has opened a new dimension to the study of size scaling relations via their wavelength dependence. 
Since rest-optical and rest-UV observations probe stellar populations of different ages, we may expect differences based on how and where stars form in high-redshift galaxies. 
Simulations of LBGs at a broad range of wavelengths have made enticing predictions for the \textit{JWST} era of astronomy. 
\citet{Garcia-Argumanez23} simulate observations from the Cosmic Evolution Early Release Science (CEERS) Survey \citep{CEERS} with Illustris-1 \citep{Vogelsberger14} and find that progenitors of local galaxies with $M>10^{11}M_{\sun}$ form $\sim25\%$ of their $z\sim1$ stellar masses by $z\simeq2.7$, suggesting significant mass build-up at high redshift.
\citet{Shen24} use the THESAN simulation \citep{Kannan22} to show that massive ($M_{\star} > 10^{9}M_{\sun}$) galaxies become increasingly compact over $z=10$ to $z=6$.
Additionally, their sizes roughly agree with analytical predictions from disk formation models, governed by the spin parameter $\lambda$ \citep{Fall80, Mo98, Bullock01}.

In the First Light And Reionization Epoch Simulations \citep[FLARES,][]{FLARES}, \citet{Roper22} (hereby \citetalias{Roper22}) find intrinsically negative size-luminosity relations.
At $z=5-10$, the inclusion of dust attenuation predicts observed size-luminosity relations with increasingly positive slopes at shorter wavelengths, building on similar predictions by \citet{Marshall22}. 
\citet{Constantin23} (hereby \citetalias{Constantin23}) also simulate CEERS observations with Illustris TNG50 \citep{TNG50} at $z=3-6$ and predict a size-redshift evolution in the rest-optical consistent with fixed circular velocity of DM haloes, as well as negative observed size-mass relations with wavelength at $z=5-6$. 
Both \citetalias{Roper22} and \citetalias{Constantin23} explain their findings by requiring that massive galaxies undergo episodes of centrally concentrated star formation, leading to small sizes at high mass and high dust attenuation in the cores of massive galaxies. 
Studying massive galaxies with \textit{JWST} allows us to test these predictions.

Having established the versatility of galaxy size measurements for understanding galaxy evolution, we now discuss how sizes are measured. 
Galaxy sizes can be complex to define depending on the morphology, so comparing like-for-like with other observations and simulations can often be complicated by the choice of definition.
The \cite{Sersic63} brightness profile is a common functional form fit to galaxies following $r^{1/n}$ for a given `S\'ersic index' $n$. 
This profile assigns galaxies an effective radius $R_{e}$ which contains half the total light. 
Various values of $n$ are effective at fitting both late- and early-type massive galaxies in the local Universe \citep[e.g.][]{Lange15}. 
Additionally, \textit{JWST} has continued to find that galaxies at high-redshift which appear disky are well-fit by a S\'ersic profile with $n\simeq1$ \citep{Adams23_prop, kartaltepe23, Ormerod23}. 
Alternatively, non-parametric measurements using some form of a circularised half-light radius are a popular choice in simulations at high redshift \citep{Roper22, katz23}. 
Such definitions make no assumptions on the shape of the galaxy, allowing a measurement to be made for disturbed and clumpy morphologies which are commonly seen at high redshift \citep{Shibuya16, Huertas_company23, kartaltepe23, Jacobs23}. 
It is thus vital that studies on galaxy size consider various definitions to test differences between parametric and non-parametric fitting, allow for a diverse range of galaxy morphologies, and provide like-for-like comparison with simulations and between redshifts.

In this work we measure the sizes of massive, rest-UV bright galaxies at $z=3-5$ selected from ground-based imaging. 
We use NIRCam imaging from Public Release IMaging for Extragalactic Research \citep[PRIMER,][]{PRIMER} that overlaps with the ground-based imaging from Visible and Infrared Survey Telescope for Astronomy (VISTA) Deep Extragalactic Observations \citep[VIDEO,][]{VIDEO}, UltraVISTA \citep{UltraVISTA} and the Hyper-Suprime Cam Subaru Strategic Program \citep[HSC-SSP,][]{AiharaDR3} to measure sizes across a broad wavelength range, $0.9-4.4$\micron. 
We use S\'ersic and non-parametric size fitting to measure the redshift evolution of sizes, as well as the size-mass and size-luminosity relations. 
We also test the wavelength dependence of the size-mass and size-luminosity relations, comparing with predictions from simulations.

The layout of this paper is as follows: in Section \ref{sec:data} we outline the ground-selected catalogues and \textit{JWST} imaging used in this work.
Section \ref{sec:methods} describes the creation of postage stamp images and the size fitting of the sample. 
Our results are presented in Section \ref{sec:results}, where we discuss the size-redshift evolution, the size-mass relations and the size-luminosity relations. 
In Section \ref{sec:discussion} we discuss our results in the context of disk formation models.
We also explore difficulties in measuring sizes and the size-redshift evolution. 
We then examine the comparison of our results to simulations, after which we discuss the implications of our results and simulation comparison for the scenario of compact central star formation at high-redshift. 
We finally conclude in Section \ref{sec:conclusions}.
We use a $\Lambda$ cold dark matter (CDM) cosmology with $\mathrm{H}_{0}=70 \ \mathrm{km \ s^{-1} \ Mpc^{-1}}, \Omega_{\mathrm{M}}=0.3, \Omega_{\Lambda} = 0.7$. Magnitudes are reported in the AB system \citep{oke83}.

\section{Data and sample}
\label{sec:data}

\begin{figure*}
\newcommand{\fitfigwidth}{1}
\centering
\includegraphics[width=\textwidth]{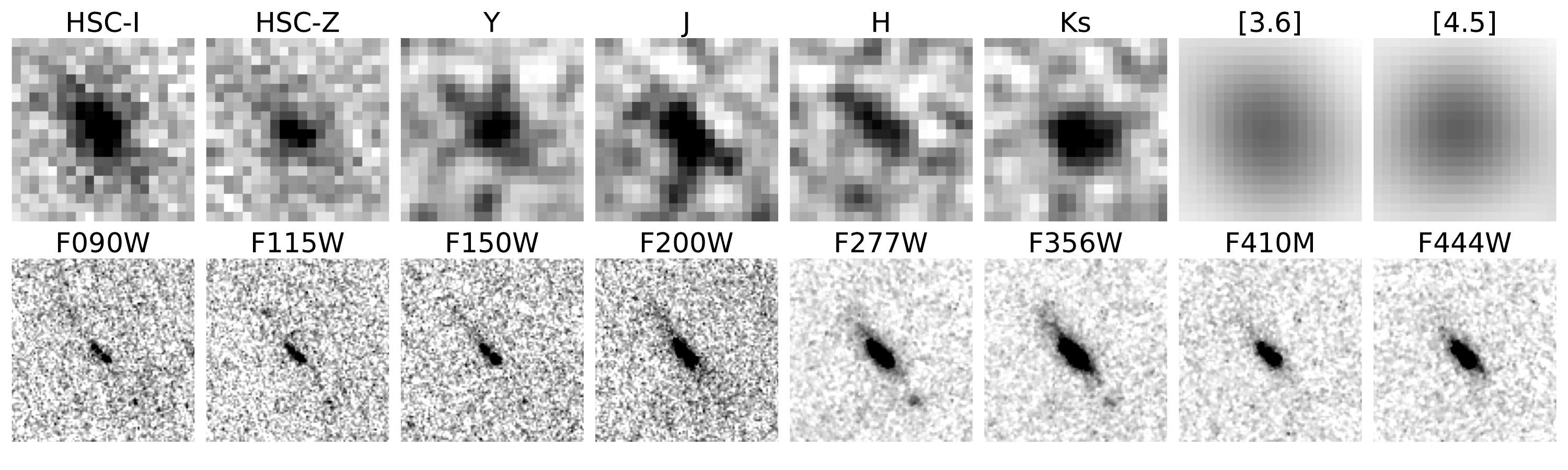}

\caption{Stamps of an example galaxy at $z=4.58$. \textbf{Top:} the ground-based imaging from HSC and VISTA used to select this object, as well as \textit{Spitzer}/IRAC imaging of this galaxy at $3.6\micron$ and $4.5\micron$. \textbf{Bottom:} the \textit{JWST} NIRCam imaging of the same object. The stamps are ordered in increasing wavelength. The resolution is $\sim$ five times better with NIRCam for the seeing-dominated $IZYJHK_{s}$ bands, and the large FWHM of the \textit{Spitzer} point spread function (PSF) in the [3.6] and [4.5] bands means no morphological information could previously be interpreted. With \textit{JWST}, we can qualitatively see a disk-like morphology in the reddest bands, and a decrease in size to shorter wavelengths. Note that the stamps above/below one another do not necessarily correspond to the same wavelength. The stamps are $4\times4$ arcsec. North is up and east is to the left. All stamps are scaled with a lower limit at $2\sigma$ below the noise level (white). The galaxy (black) saturates at $5\sigma$ above the noise level in the top row, $4\sigma$ in the short-wavelength NIRCam bands and $10\sigma$ in the long-wavelength NIRCam bands.}
\label{fig:resolution_example}
\end{figure*}

In this section we outline the deep degree-scale ground-based imaging and derived catalogues, as well as the \textit{JWST} imaging used in this work. 
Although the area covered by PRIMER is much smaller than ground-based surveys, the improvement in resolution allows us to measure resolved properties in what was previously a seeing-limited sample.
We do not conduct a new selection with the \textit{JWST} data to make use of this fact - a seeing-dominated sample is less biased by the sizes, with galaxies being purely selected by being bright in the rest-frame UV, providing a better representation of typical galaxy sizes and avoiding selection effects.

\subsection{Ground-based catalogues}
\label{sec: groundbased catalogues}

We make use of rest-UV selected catalogues in the 1.5 $\mathrm{deg}^2$ COSMOS field and 4.3 $\mathrm{deg}^2$ \textit{XMM}-Large Scale Structure (XMM-LSS) field created by \citet{adams23}. 
The full details of the selection can be found within their section 2, however we summarise it here. 
Ground-based near-infrared imaging from the UltraVISTA survey \citep{UltraVISTA} in COSMOS and VIDEO \citep{VIDEO} in XMM-LSS is combined with deep optical data from the Canada-France-Hawaii-Telescope Legacy Survey \citep[CFHTLS,][]{CFHTLS} and the Hyper Suprime-Cam Subaru Strategic Program Data Release 2 \citep[HSC-SSP DR2,][]{HSCSSP_DR2} to create multi-wavelength photometric catalogues spanning $0.3-2.4\micron$.
The XMM-LSS field contains the Ultra-Deep Survey \citep[UDS,][]{uds}.
We update the photometry with the most up-to-date survey data, including HSC-SSP Data Release 3 \citep{AiharaDR3} and UltraVISTA Data Release 5. 
We show an example galaxy at $z=4.58$ in these filters in Fig. \ref{fig:resolution_example}.
Fluxes are measured in 2 arcsec diameter circular apertures. 
The \textsc{LePhare} spectral energy distribution (SED) fitting code \citep{arnouts99, ilbert06} is used to calculate photometric redshifts and classify objects into galaxies, active galactic nuclei (AGN) and Milky Way stars. 
Objects are split into redshift bins at $z\simeq3$, 4 and 5 ($2.75 < z < 3.5$, $3.5 < z < 4.5$ and $4.5 < z < 5.5$ respectively). 
They are then required to be detected at a $5\sigma$ threshold in the deepest band containing the rest-frame ultraviolet continuum emission. 
These are HSC-$R$, HSC-$I$ and HSC-$Z$ for each redshift bin in the regions containing the \textit{JWST} imaging, outlined in Section \ref{sec:primer imaging}.
The $5\sigma$ depths of the $R, I$ and $Z$ bands are 27.1, 26.9 and 26.5 in COSMOS and 26.4, 26.3 and 25.7 in UDS based on a 2 arcsec diameter circular aperture.
For the $z\simeq5$ sample, a $<3\sigma$ limit is imposed on the CFHT-$u$ band to ensure a strong Lyman break and to remove further low-redshift interlopers. 
A cut of $\chi^{2}<100$ is applied to the best-fitting SED solution to remove artefacts and contaminants. 
\textit{Spitzer}/IRAC data are not used due to blending issues introduced by the large point spread function (PSF).
The catalogues are then cross-matched with a spectroscopic sample to test the reliability of the photometric redshifts.
The outlier rate (the fraction of photometric redshifts which differ from the spectroscopic redshift by more than $0.15 \times (1+z)$) is no more than 4.5\% in any of the redshift bins.
The recovery rate of objects in the spectroscopic sample is also greater than 80\% for the redshifts of interest.
The COSMOS and XMM-LSS catalogues contain 175,508, 47,978 and 14,875 objects in the $z=3,4,5$ bins respectively.

\subsection{Photometry and SED fitting}
\label{sec: photometry and sed fitting}

\begin{figure}
    \centering
    \includegraphics[width=\columnwidth]{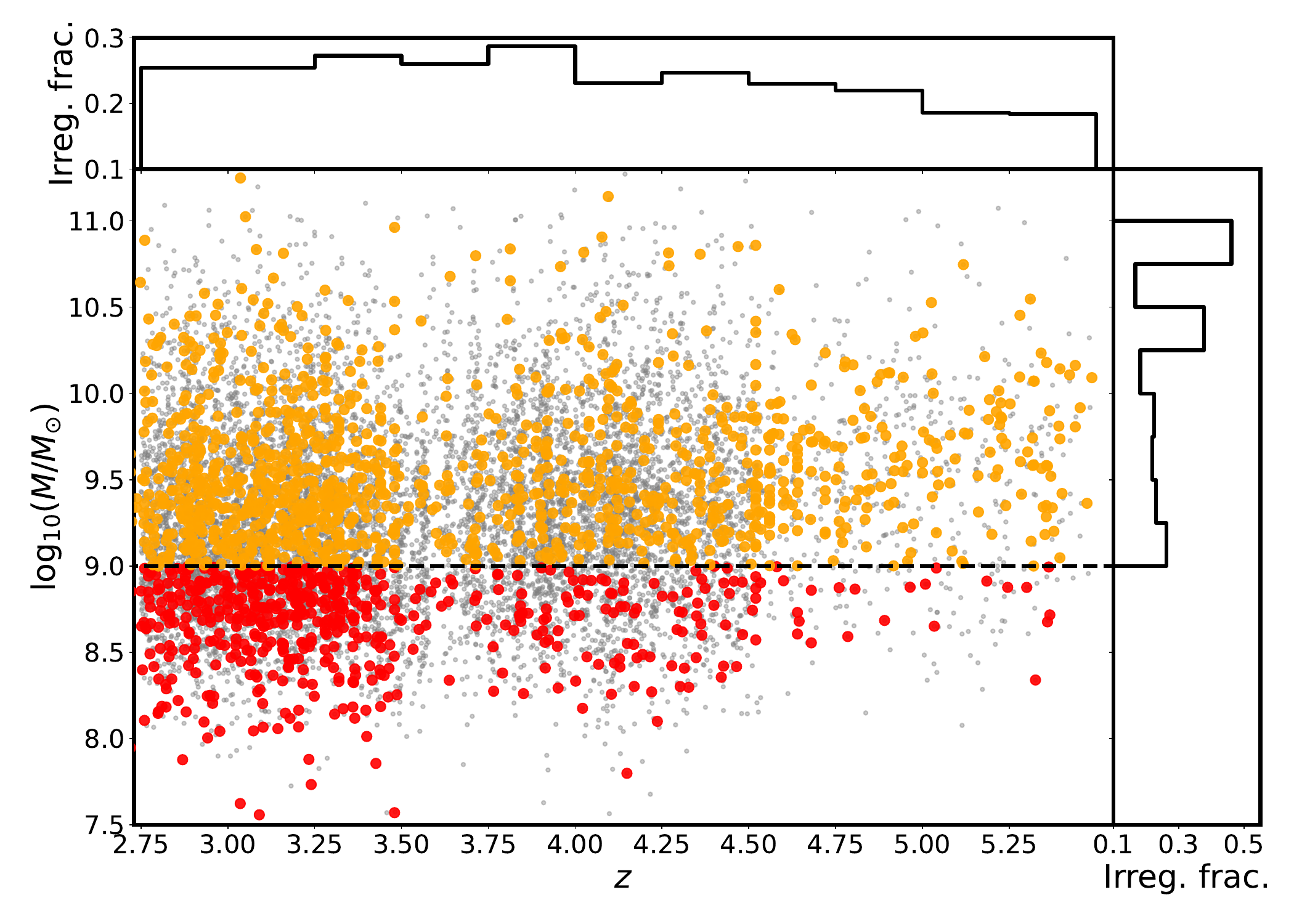}
    \caption{The stellar masses (measured using a CSFH with \textsc{BAGPIPES}) of the COSMOS and XMM-LSS catalogues from \citet{adams23} that overlap with the PRIMER imaging, in terms of their photometric redshift. 
    We apply a stellar mass cut at $\mathrm{log}_{10}(M_{\star}/M_{\sun}) > 9$. 
    The gray points are stellar masses for objects in a wider region around the PRIMER imaging.
    The histograms show the fraction of galaxies removed because the \textsc{PyAutoGalaxy} S\'ersic fit does not converge, in bins of redshift and mass. 
    These poor fits are used as a proxy of irregular morphology (see Section \ref{sec:irregular fraction}), and hence this fraction is a proxy for the irregular fraction.
    Note the different axis scales for each histogram.}
    \label{fig:sample_mass_z}
\end{figure}

Some objects, particularly at $z\simeq3$, have effective radii significantly larger than the 2 arcsec diameter aperture used to measure their photometry, resulting in underestimates of the total flux. 
We use \textsc{SExtractor} \citep{sextractor} to measure \textsc{FLUX\_AUTO} (which uses a flexible elliptical aperture to estimate total flux) in each ground-based filter. 
We then apply a correction to the aperture fluxes in each band based on the ratio of \textsc{FLUX\_AUTO} to the aperture flux in $K_{s}$ \citep{Adams23_epochs}. 
We find that the ratio is similar in the other bands.
We then use this corrected ground-based photometry to estimate properties of the sample using a SED fitting analysis with Bayesian Analysis of Galaxies for Physical Inference and Parameter EStimation \citep[\textsc{bagpipes, }][]{BAGPIPES}{}{}. 
We use a fiducial constant star formation history (CSFH). Although this choice may not capture complex star formation histories and episodes, it has been found that a CSFH finds similar stellar masses to non-parametric SFHs for very high-redshift galaxies \citep{whitler23}. 
We set the redshift to be the photometric redshift found by \citet{adams23} with \textsc{LePhare}, allowing it to vary by $\pm 0.1$. We adopt a uniform prior on the ionization parameter, $-4 \le \mathrm{log}U \le -2$. 
The metallicity is allowed to vary in the range $0.2 \le Z/Z_{\sun} \le 1$, and we assume $Z_{\sun} = 0.02$. 
Dust reddening was prescribed by the \citet{calzetti} dust law, with attenuation in the range $0.0 \le A_{V} \le 4.0$.
 The time since star formation switched on varies between the start of the Universe and the redshift being considered.
Fig. \ref{fig:sample_mass_z} shows the mass and photometric redshift distribution of the sample.
For our analysis we take objects with stellar masses $\mathrm{log_{10}}(M_{\star}/M_{\sun}) > 9$ to focus on the massive objects whilst balancing against incompleteness towards lower mass, particularly at higher redshift.
By comparing to the stellar mass distribution in a wider area around the PRIMER imaging, we estimate that the $z=5$ bin is 85\% complete.
Additionally, the $\mathrm{log_{10}}(M_{\star}/M_{\sun}) > 9$ is the same cut used by \citetalias{Constantin23}, allowing for a direct comparison to their simulation results.
This leaves 1429, 366 and 236 objects in the $z=3,4,5$ bins. 
We then estimate the \textit{rest-frame} magnitudes of objects in each NIRCam band by convolving the SED found from the ground-based photometry with the NIRCam filter transmission curves.
This method is used to remain consistent with the determination of $M_{\mathrm{UV}}$ by \citet{adams23} of placing a top-hat filter at 1500\AA \ with width 100\AA \ on the rest-frame SED, and provides robust errors.

\subsection{JWST PRIMER imaging}
\label{sec:primer imaging}

Public Release IMaging for Extragalactic Research (PRIMER) is a \textit{JWST} Cycle-1 program (GO PID: 1837, PI: J. Dunlop) focusing on two key \textit{HST} CANDELS \citep{Candels1, Candels2} fields, COSMOS and UDS. 
This work focuses on the overlap between the COSMOS and XMM-LSS catalogues produced by \citet{adams23} and the $340 \ \mathrm{arcmin}^{2}$ CANDELS-COSMOS and CANDELS-UDS fields imaged by \textit{JWST}, respectively. 
These fields contain 1572, 1056 and 275 objects from the ground-based catalogues in the $z=3,4,5$ bins.
Imaging is taken in ten NIRCam+MIRI bands. We make use of the eight deep NIRCam filters, namely F090W, F115W, F150W and F200W from the short-wavelength detector and F270W, F356W, F410M and F444W from the long-wavelength detector, allowing us to probe the rest-frame UV and optical emission of $z\simeq3-5$ galaxies.
We show an example galaxy at $z=4.58$ in these filters in Fig. \ref{fig:resolution_example}, and show a comparison to the ground-based filters outlined in Section \ref{sec: groundbased catalogues}.
In this work we primarily make use of F115W, F200W, F356W and F444W.
The rest-frame wavelengths covered by these filters at $z=3-5$ are $\lambda\simeq0.2-0.3\micron, \ 0.3-0.5\micron, \ 0.6-0.9\micron$ and $0.7-1.1\micron$ respectively.
We use version 0.5 of the PRIMER reduction, which contains observations between Dec. 2022 and Jan. 2023. 
The NIRCam data is reduced using the "jwst\_1039.pmap" calibration reference file, and the \textit{JWST} Pipeline version 1.9.4 \citep{Bushouse23}. 
Custom steps are taken to remove excess $1/\mathrm{f}$ striping, mask residual snowball artifacts and remove wisp artifacts in F150W and F200W. 
The astrometry is aligned to \textit{Gaia} Data Release 3 \citep[][]{GaiaDR3}{}{}. 
The pixel scale of the imaging is 0.03 arcsec per pixel.
The $5\sigma$ depths of the images used in this work are 27.4 in F115W, 27.9 in F200W, 28.5 in F356W and 28.5 in F444W, measured in 0.3 arcsec diameter circular apertures.

\section{Methods}
\label{sec:methods}

In this section we describe how the NIRCam imaging is used for the morphological analysis. The longest wavelength NIRCam filters (F356W, F410M and F444W) provide us with a new view into the rest-frame optical emission of $z\simeq3-5$ galaxies previously only covered by \textit{Spitzer}/IRAC, opening up a new resolved rest-optical view of high-redshift LBGs.

\subsection{NIRCam stamps and PSFs}
\label{sec:stamps_and_psf}

In order to make use of the superior resolution of \textit{JWST} compared to ground-based imaging, we crossmatch the ground-based COSMOS and UDS catalogue with the PRIMER imaging to extract $20 \times 20$ arcsec postage stamps of the objects.
This corresponds to $667 \times 667$ pixels.
We then use \textsc{Photutils} \citep{photutils} to perform a 2D background subtraction with a box size of 50 pixels, masking objects and bad pixels at a $2\sigma$ threshold.
We convolve the error images with a Gaussian filter (with standard deviation of two pixels) to remove erroneous very low values which cause the morphological fits to fail.

The \textit{JWST} PSF is required for measuring the sizes of galaxies in the sample.
A popular tool for modelling the \textit{JWST} PSF is WebbPSF \citep{WebbPSF}. 
However, it has been found that empirical PSFs constructed from NIRCam imaging often have broader FWHMs than constructed by WebbPSF \citep[e.g.][]{Ding22, ono23_psf, tacchella23}, likely because PSFs generated by WebbPSF are not drizzled. 
Additionally, \citet{zhuang23} find strong variation in the NIRCam PSF over the field of view, with variations increasing at shorter wavelengths. 
They recommend using \textsc{PSFEx} \citep{psfex} to construct a grid of empirical PSFs across the image.
We use a global PSF model due to the limited number of stars in the field, identified from the FWHM-magnitude diagram with FWHMs taken from \textsc{SExtractor}.
\citet{zhuang23} find that a global PSF still leads to good results, with an average $\lesssim 0.02$ mag systematic offset and $\lesssim 0.05$ mag random scatter when later attempting AGN-host decomposition (see Section \ref{sec:parametric sizes}). 
The PSF FWHMs as measured by PSFEx are $0.07\arcsec$, $0.09\arcsec$, $0.16\arcsec$ and $0.17\arcsec$ in F115W, F200W, F356W and F444W respectively (the filters used in this work). 
These are shown for F200W and F356W on Fig. \ref{fig:mass_relations}.

\subsection{Size fitting}
\label{sec:size fitting}

In order to compare directly with other studies and with simulations, we use both a parametric and non-parametric method to measure the sizes of galaxies in our sample.

\begin{figure*}
\newcommand{\fitfigwidth}{1}
\centering
\includegraphics[width=\fitfigwidth\columnwidth]{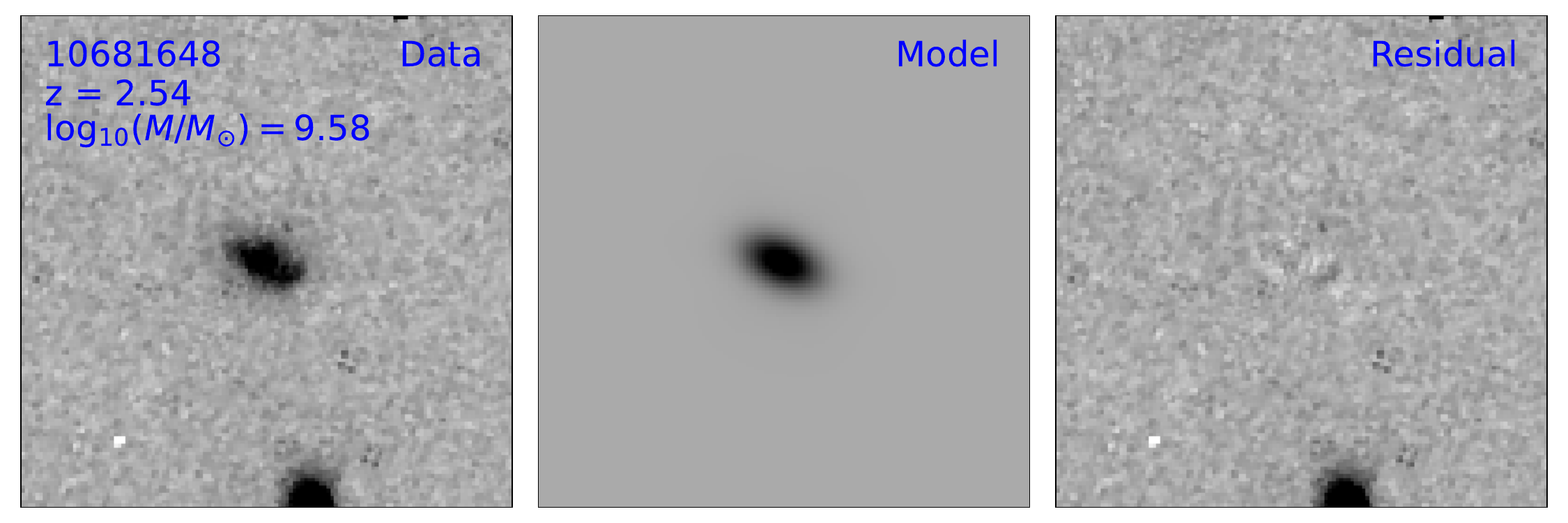}
\includegraphics[width=\fitfigwidth\columnwidth]{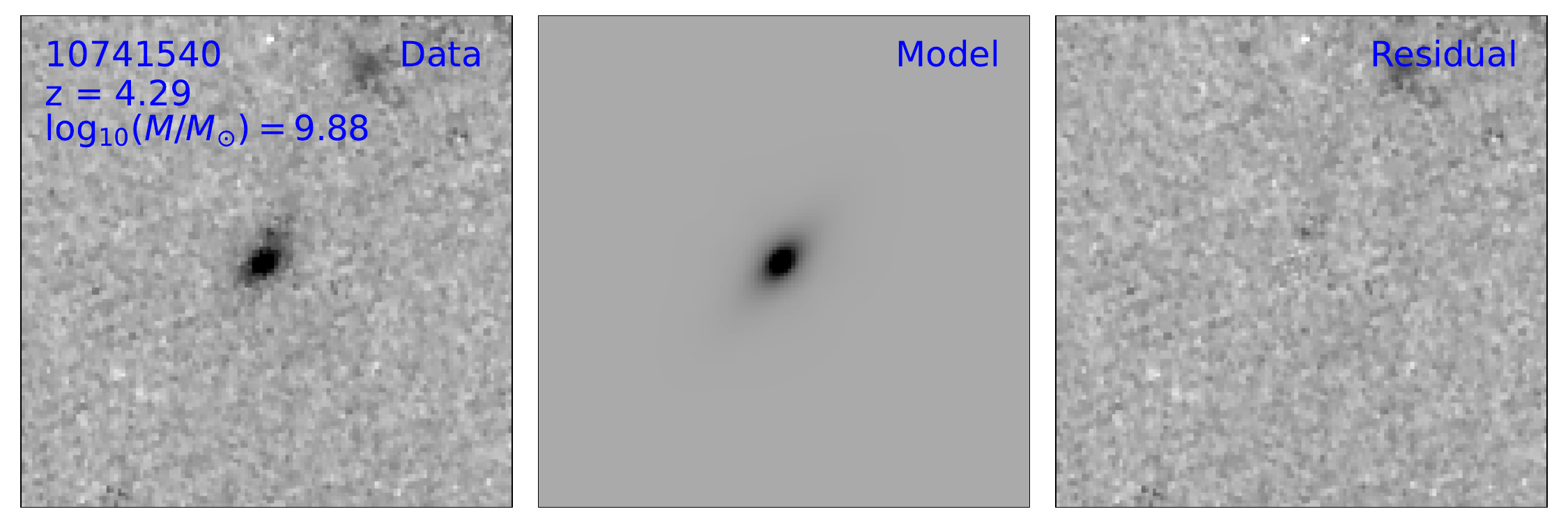}

\includegraphics[width=\fitfigwidth\columnwidth]{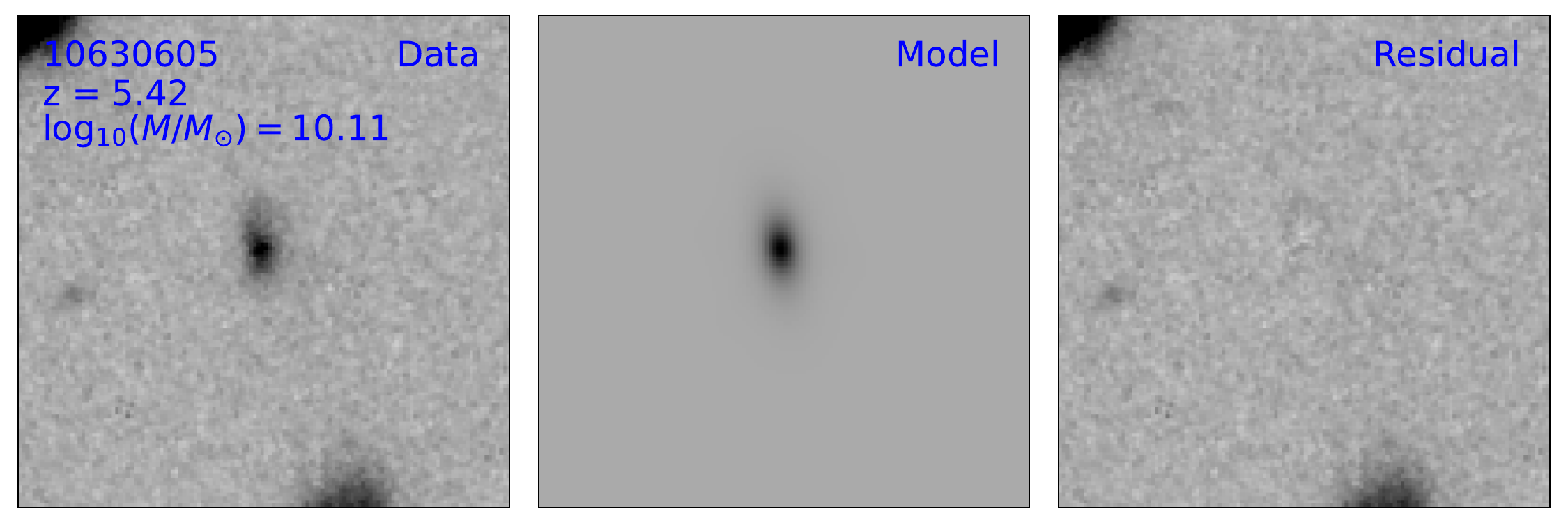}
\includegraphics[width=\fitfigwidth\columnwidth]{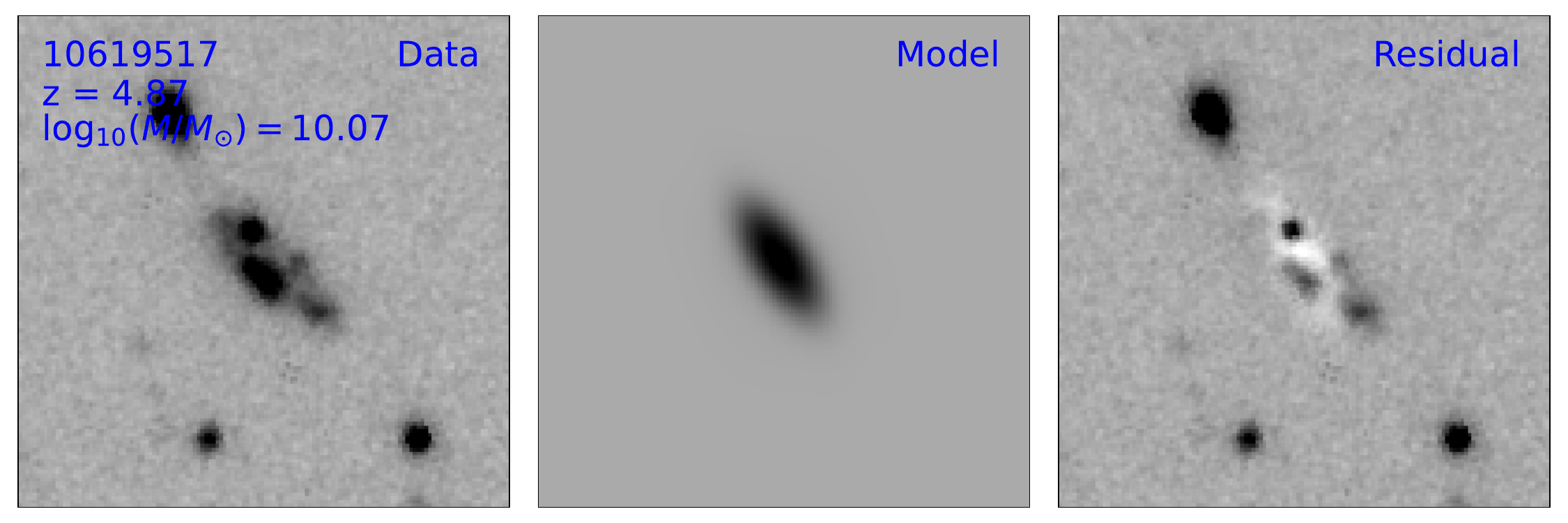}

\caption{Four examples of S\'ersic fitting to different galaxies in the sample. In each plot we show the data, model and the residual (data minus model). The ID, photometric redshift and stellar mass of the object are labelled on the data stamp of each galaxy. The top row and bottom left plots show examples from the $z=3, 4, 5$ bins respectively. The bottom right plot shows an example of a clumpy galaxy where the S\'ersic fitting does not perform well, leading to large residuals. The stamps are $4\times4$ arcsec. North is up and east is to the left. The scaling is set to a lower limit of $-0.1$ MJy/sr and saturates at 0.3 MJy/sr, which highlights the residuals in the case of poor fits.}
\label{fig:fitting_examples}
\end{figure*}

\subsubsection{Parametric sizes}
\label{sec:parametric sizes}

We make use of \textsc{PyAutoGalaxy}\footnote{https://github.com/Jammy2211/pyautogalaxy}
\citep[version 2022.07.11.1,][]{pyautogalaxy}{}{}  to measure the sizes of galaxies in the NIRCam filters. 
This takes in the image, noise map and PSF and uses fully automated Bayesian model fitting of galaxy two-dimensional surface brightness profiles to return the best-fit parameters, allowing us to extract the Bayesian Information Criterion (BIC) and assess model fits.
We first mask the postage stamp of each object to a circular region of radius 0.7 arcsec, expanding to $1.8$ arcsec for the largest objects at $z\simeq3$. 
We find that these sizes are sufficient for including the entire object whilst balancing against the number of pixels that are fitted for.
These choices also do not affect the results as the best-fit sizes are much smaller than these circular regions.
We use a simple S\'ersic profile to measure the effective radius $R_{e}$.
We employ a Gaussian prior on the S\'ersic index $n$, centred at $n=1$ with a standard deviation of $\sigma_{n}=2$ and truncated at an upper limit at $n=4$ \citep[][]{kartaltepe23}{}{}. 
We place a uniform prior on the centre of the object, allowing it to vary within the central 0.2 arcsec to account for any differences in the centroid between the ground-based and \textit{JWST} imaging. 
The effective radius is allowed to vary between $0-3$ arcsec, and we allow the ellipticity to vary freely. 
We make use of the linear light profiles available in \textsc{PyAutoGalaxy} where the intensity of the profile is solved for implicitly, reducing the dimensionality of the fitting and removing degeneracies that can occur between intensity and other profile parameters. 
However, we note that degeneracies can still occur between S\'ersic index and size.
We input the empirical PSF described in Section \ref{sec:stamps_and_psf} to convolve with the model.
We discard bad fits by inspection of residuals in F115W and F444W to check the most extreme wavelengths probed in this work. These bad fits almost always arise in the case of clumpy or irregular morphologies, leading to very large residuals. This leaves 1203, 278 and 187 objects in the $z=3,4,5$ bins.
This means we remove 16\%, 24\% and 21\% of the objects that remain from the mass cut in each redshift bin respectively.
We show four examples of the S\'ersic fitting in Fig. \ref{fig:fitting_examples}, including one case of a poor fit to a clumpy galaxy.

Since the sample selected in \citet{adams23} included both AGN and SFGs, we need to account for objects that may contain accretion-dominated emission and contaminate the total flux, leading to overestimates of stellar mass and inaccurate size measurements. 
We run a second round of fitting with a PSF+S\'ersic profile to account for this, using the initial round of fitting to further inform the priors. 
We place a Gaussian prior on the centre of the PSF and S\'ersic components with a standard deviation of 0.05 arcsec, centred on the position found by the S\'ersic-only fitting. 
Similarly, we employ a Gaussian prior on the ellipticity and S\'ersic index, with standard deviations of 0.2 and 0.5 respectively, centred on the values found in the previous round of fitting.
These are generally wide enough to account for large differences in the best-fit parameters.
We no longer use linear light profiles for the fitting, instead allowing the intensity of both profiles to vary freely so that the contribution from each component can be measured. 
We apply this fitting in F115W and F444W in order to check for a PSF contribution in both the rest-frame UV \textit{and} the rest-frame optical, motivated by recent discoveries of highly reddened AGN found by \textit{JWST} \citep[e.g.][]{Kocevski23, matthee23, Labbe23, Juodzbalis23, Scholtz23}. 
We compute the BIC for both rounds of fitting for each object in order to compare the models. 
We find that when the PSF+S\'ersic fit is preferred according to the BIC, it is preferred in \textit{both} NIRCam filters, revealing no hidden, highly reddened AGN candidates in this UV-selected sample. 
Objects with a preferred PSF+S\'ersic fit (2\% at $z=5$, 6\% at $z=4$, 1\% at $z=3$) are discarded from the sample. 
We find that the fraction of sources with a PSF component increases towards brighter $M_{\mathrm{UV}}$.
They are interesting in their own right as Type I AGN candidates, but in this work the sample size of such candidates is small. 

We also fit a S\'ersic model to the PSFs from the filters used in this work to test the smallest size measurable by \textsc{PyAutoGalaxy}. The S\'ersic indexes are typically $\sim0.2$. The modal effective radius is $0.15\arcsec$ and they range from $0.013\arcsec$ to $0.021\arcsec$, comparable to the pixel scale of the imaging ($0.03\arcsec$). These sizes are significantly smaller than the PSF FWHM (see Section \ref{sec:stamps_and_psf}). At $z=3,4,5$ the maximum size of $0.021\arcsec$ corresponds to $\mathrm{log}_{10}(R_{e}/\mathrm{kpc}) = -0.79, -0.84, -0.88$ respectively (see Fig. \ref{fig:mass_relations}).

\subsubsection{Non-parametric sizes}

We measure a non-parametric size following the pixel-based method of \citetalias{Roper22} to compare like-for-like size-luminosity relations. 
This method is better suited to deal with clumpy and irregular morphologies. We select pixels in the image at a $5\sigma$ threshold and then place a circular aperture with a diameter of 2 arcsec on each object, which is large enough to capture the entire object. 
We then order the pixels in this aperture from most to least luminous. We take the number of pixels containing half the luminosity, and use the pixel area $A$ to compute the radius using $R = \sqrt{A/\pi}$.
\citetalias{Roper22} use an aperture size enclosing 30 pkpc which is too large for our objects ($\sim4-5$ arcsec diameter) as neighbouring objects occasionally enter the aperture. 
The number of pixels containing half the luminosity is generally much smaller than the total number of pixels in the aperture, so our choice of aperture radius does not impact the non-parametric size measurements. 
The size measured is also broadened by the PSF. 
To allow for comparison with simulations, we need to account for this effect. 
We first scale the PSFs constructed in Section \ref{sec:stamps_and_psf} to match the luminosity of the object in each filter. 
We then measure the size of the PSF in the same manner as before, and then deconvolve assuming the PSF is a Gaussian (by subtracting the PSF size from the initial size in quadrature).
In Appendix \ref{sec:size measurement comparison} we present a brief comparison of the S\'ersic and non-parametric sizes.

\section{Results}
\label{sec:results}

In this section we present the results of the size fitting and the dependence of size on redshift, stellar mass and luminosity. The sizes reported in the log-normal distributions and the size-redshift relations are measured in F356W to compare directly with the predictions of \citetalias{Constantin23}, probing $0.6-0.9 \ \micron$ in the rest frame. Similarly, the S\'ersic size-mass relations are reported in F200W (probing rest-frame $0.3-0.5 \ \micron$) and F356W to compare directly with \citetalias{Constantin23}. For the size-luminosity relations (with both size measurement methods) we use F115W and F444W, which probe $0.2-0.3 \ \micron$ and $0.7-1.1 \ \micron$ respectively, to maximise the difference in wavelength to provide a clearer comparison to the predictions of \citetalias{Roper22}.


\begin{figure}
    \centering
    \includegraphics[width=\columnwidth]{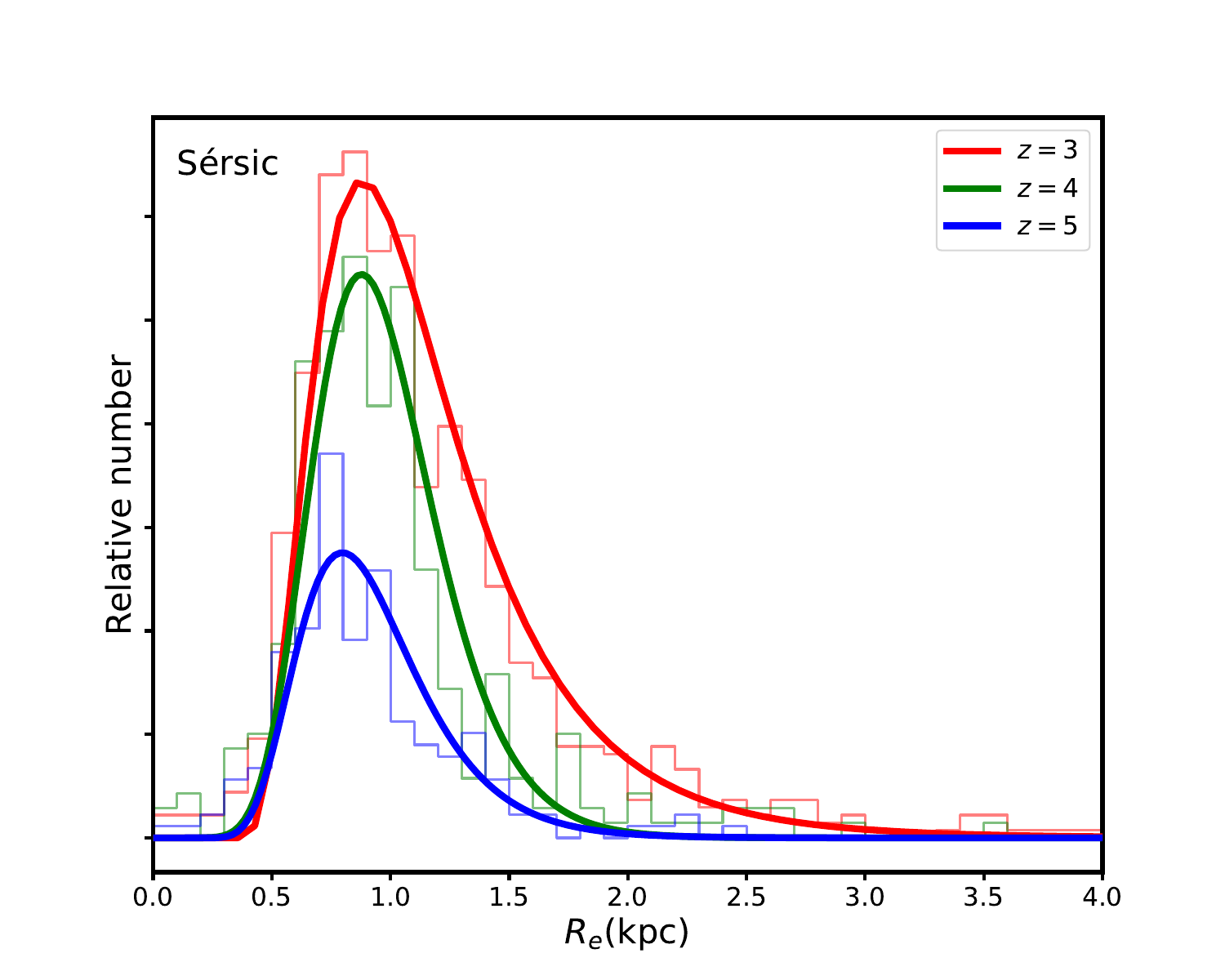}
    \includegraphics[width=\columnwidth]{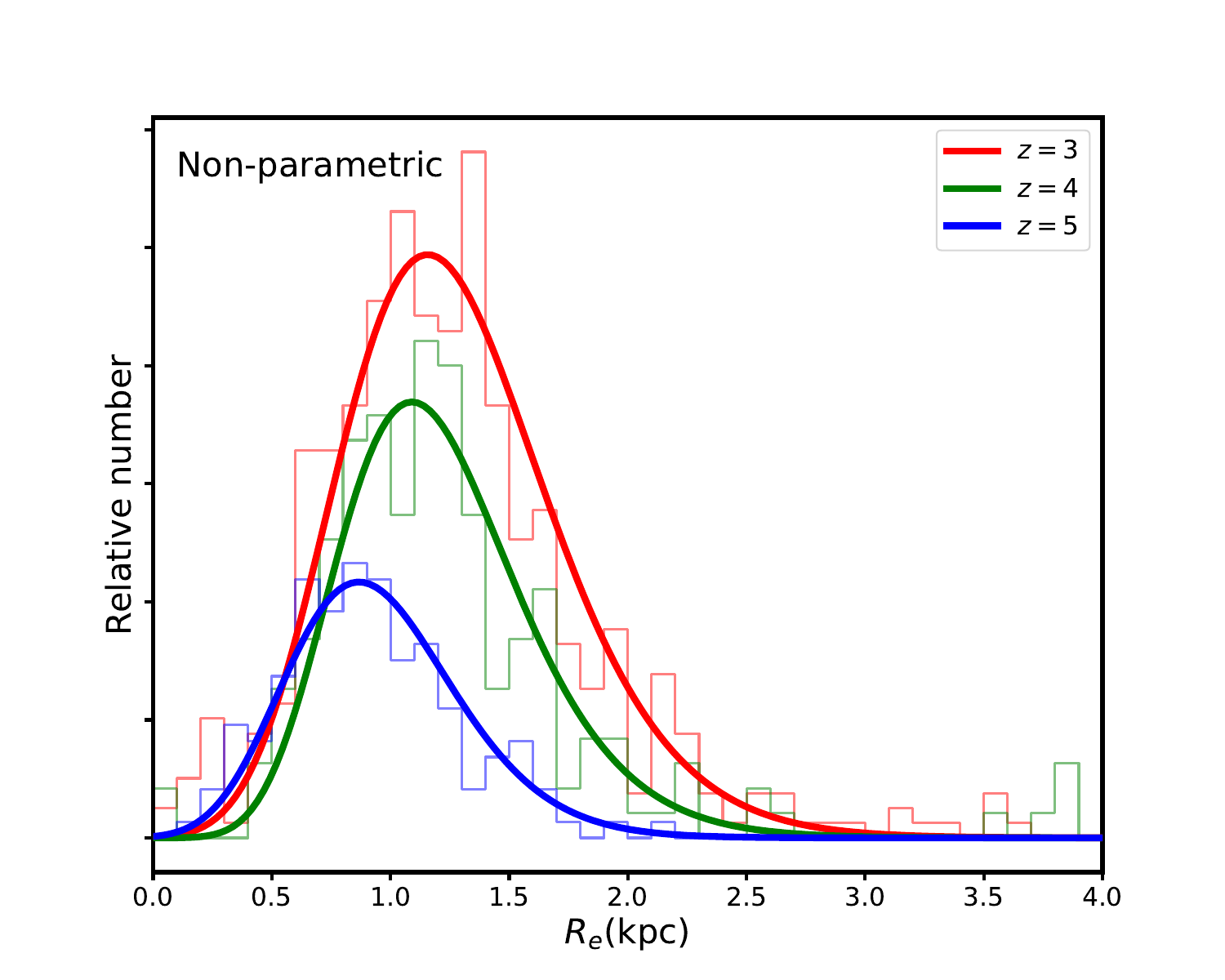}
    \caption{The size distribution of galaxies in each of our redshift bins measured in F356W. \textbf{Top:} S\'ersic size fitting. \textbf{Bottom:} non-parametric size fitting. We show the distributions as histograms, and then plot the log-normal fit to the distributions. The $z=4$ and 5 distributions are scaled by a factor of two and three respectively to emphasise the relative positions of the peaks. This peak is taken as the size of a typical galaxy in each redshift bin.}
    \label{fig:lognormal_dist}
\end{figure}

\begin{figure*}
    \centering
    \includegraphics[width=\columnwidth]{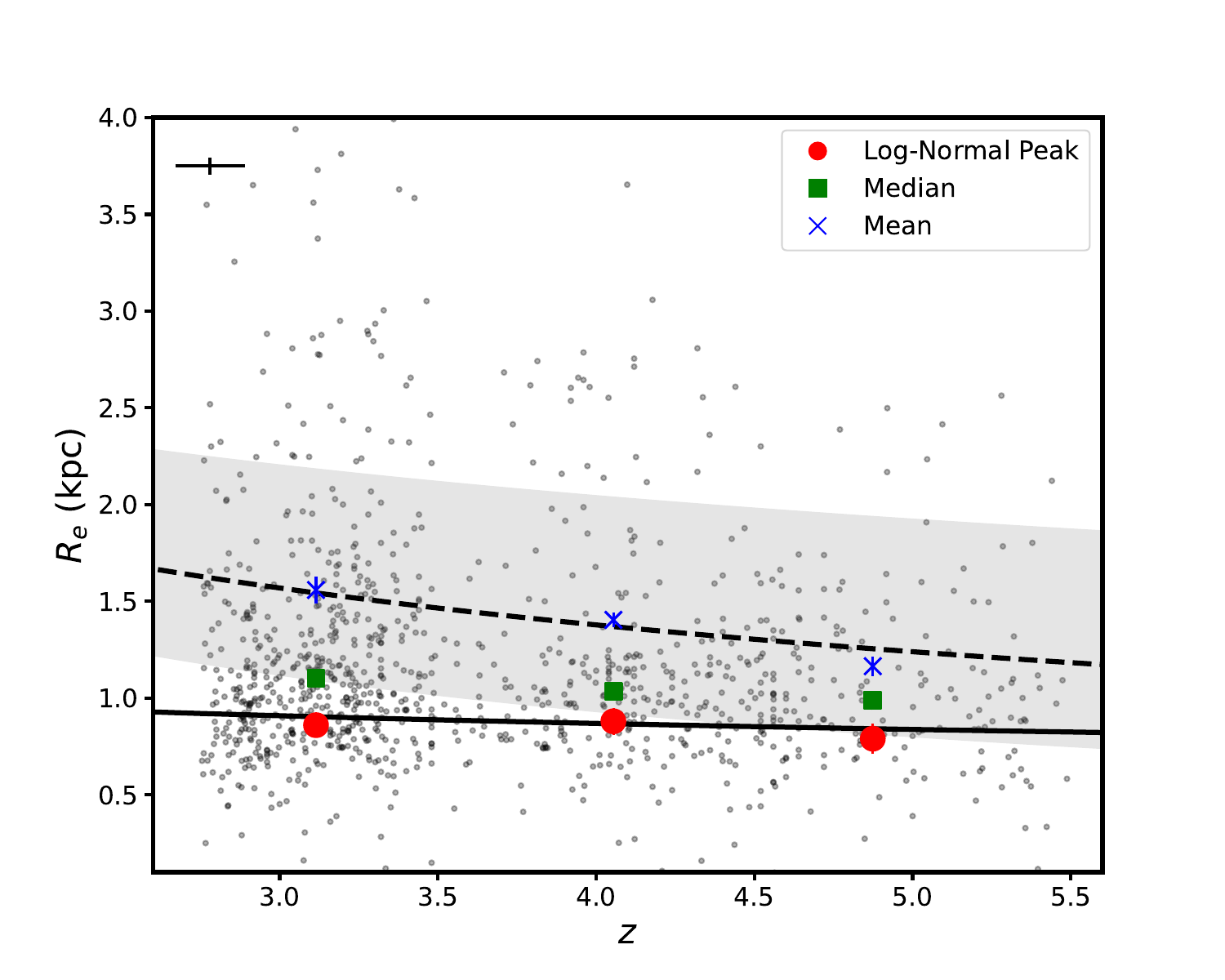}
    \includegraphics[width=\columnwidth]{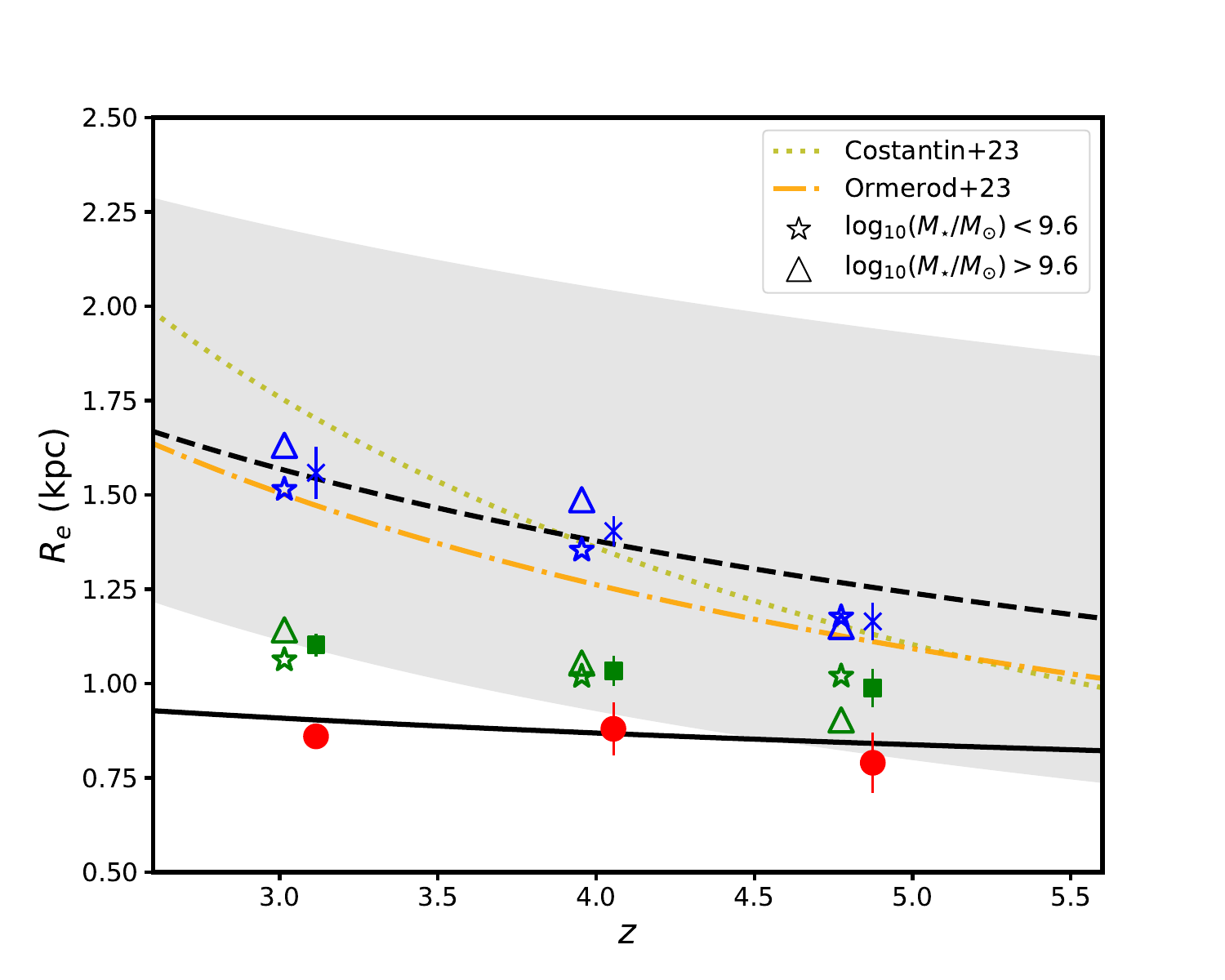}
    \caption{The size-redshift distribution of our galaxies as measured in F356W. \textbf{Left:} Individual galaxy size and photometric redshift measurements are shown by the grey circles. The typical error is shown in the top left. At the mean redshift of each redshift bin ($\Bar{z} = 2.96, 4.04, 4.91$) we plot different measures of the typical size and their errors. Circles represent the fit of the log-normal fit to the S\'ersic size distribution, squares represent medians and crosses represent means (values are reported in Table \ref{tab:average sizes}). The black dashed line represents the best-fit redshift evolution to all the data without binning, $R_{e} = 3.51(1+z)^{-0.60\pm0.22}$ kpc, and the shaded region shows the error. The solid black line shows the best-fit line if we fix the power $\alpha$ in this relation to $\alpha=-0.20$, following the evolution found by \citet{curtis_lake16} at $z=4-8$. \textbf{Right}: we compare our best-fit evolution, median and log-normal peak measurements as in the left panel to results from \citet{Ormerod23} and predictions from \citetalias{Constantin23}. We also plot the median and mean values when the sample is split into two mass bins at $\mathrm{log}_{10}(M_{\star}/M_{\sun}) = 9.6$. Note the narrower $R_{e}$ axis for clarity.}
    \label{fig:redshift_evolution}
\end{figure*}

\begin{table}
\centering
\caption{The mean, median and modal (log-normal peak) S\'ersic sizes of galaxies in this work in redshift bins at $z=3,4,5$. }
\begin{tabular}{cccc}
\hline
 & Mean (kpc) & Median (kpc) & Log-normal peak (kpc) \\
\hline
$z = 3$ & $1.56\pm0.07$ & $1.10\pm0.03$  & $0.86\pm0.03$\\
$z = 4$ & $1.40\pm0.04$ &  $1.03\pm0.04$& $0.88\pm0.07$\\
$z = 5$ & $1.16\pm0.05$&  $0.98\pm0.05$ & $0.79\pm0.08$\\
\hline
\end{tabular}

\label{tab:average sizes}
\end{table}

\subsection{Log-normal distribution fitting}
\label{sec:lognormal results}

In the top panel of Fig. \ref{fig:lognormal_dist} we show histograms of S\'ersic sizes in each redshift bin. 
We also show the log-normal fit to each distribution.
We use \textsc{SciPy} \citep{SciPy} to fit a log-normal function to the size distributions in each redshift bin. 
Errors on the peak of the fit are estimated with bootstrapping.
We find that the size distributions are well-described by the log-normal function, with reduced $\chi^{2}$ values of $\chi^{2}_{\mathrm{red}}= 1.16, 1.11, 1.08$ in the $z=3,4,5$ bins respectively. 
Using the peak as a measure of typical (modal) galaxy size in each bin, we find sizes of $\Bar{R}_{e}=0.79\pm0.08$ kpc at $z\simeq5$, $\Bar{R}_{e}=0.88\pm0.07$ kpc at $z\simeq4$ and $\Bar{R}_{e}=0.86\pm0.03$ kpc at $z\simeq3$. 
These values are consistent with no evolution in the typical size of galaxies over $z=3-5$. 
This weak evolution extends the findings of \citet{curtis_lake16} at $z=4-8$ to lower redshifts (although we note the different size definition of circularised half-light radii in that work). 
\citet{Shibuya15} also find compact S\'ersic sizes ($R_{e} < 1$kpc) at $z>3$ using the peak of the log-normal distribution.
They find this for a bright ($L_{\mathrm{UV}}=0.3-1L^{*}_{z=3}$) \textit{HST}-selected sample, where $L^{*}_{z=3}$ is the characteristic luminosity of LBGs at $z=3$, corresponding to $M_{\mathrm{UV}}=-21$ \citep{Steidel99}.

There appears to be some evolution in the tail of the distribution via emergence of objects with $R_{e} \gtrsim 2$ kpc between $z\simeq3-4$. 
To test whether this is a genuine build-up in the tail or due to the limited volume probed and the evolving rest-UV luminosity function over $z=3-5$ \citep{adams23}, we scale the $z=3$ distribution by the total number of galaxies at $z=4$ and 5 and calculate the expected number of large galaxies by integrating under the curve at $R_{e} > 2$ kpc. 
If the intrinsic distributions are the same across redshift and only changes due to the limited volume, then the expected numbers from the scaled distribution should match the observed distributions. 
At $z=4$ and $z=5$ we expect $5^{+12}_{-4}$ and $6^{+21}_{-5}$ galaxies  with $R_{e}>2$ kpc respectively by integrating the tail of the curve. 
The scaled $z=3$ distribution predicts $133^{+17}_{-16}$ and $77^{+10}_{-9}$ galaxies with $R_{e}>2$ kpc at $z=4$ and $z=5$. 
This points to a genuine build-up in the tail of the distribution in the high-mass end, suggesting a transition between $z=3$ and $z=4$, where much larger, extended structures begin to develop within galaxies.

A source of bias in measuring this tail of large galaxies comes from cosmological surface brightness dimming.
This makes low-surface brightness regions of galaxies more difficult to detect at higher redshift.
To test this, a sub-sample of the $z=3$ sources (including $R_{e}>2$ kpc galaxies) were redshifted out to $z=5$. 
The scatter in the redshifted sizes is 20\%, consistent with the error on the size measure.
There is no dependence on the initial size at $z=3$, i.e. the scatter is not larger for galaxies with extended morphologies.
We conclude that cosmological dimming does not significantly bias our results.

\subsection{Irregular galaxies}
\label{sec:irregular fraction}
We note that the S\'ersic fitting employed here fails on disturbed/irregular and clumpy galaxies, and such fits are discarded. 
\citet{Huertas_company23} conduct a morphological analysis of $\mathrm{log}_{10}(M_{\star}/M_{\sun}) > 9$ galaxies in CEERS at $z=0-6$. 
They find that the disturbed fraction of galaxies in F150W at $z\simeq3-6$ remains high at {$\sim70\%$} for $\mathrm{log}_{10}(M_{\star}/M_{\sun}) \sim 9-10.5$. 
Although we measure sizes in F356W where we find a lower fraction of disturbance, we may still underestimate a tail of large objects due to the failure of the S\'ersic fitting for irregular morphologies. 
We therefore also carry out the log-normal fitting on our non-parametric sizes, shown in the bottom panel of Fig. \ref{fig:lognormal_dist}.
The distributions are still well-fit by the log-normal distribution, with reduced $\chi^{2}$ values of $\chi^{2}_{\mathrm{red}}= 1.11, 1.13, 1.07$ in the $z=3,4,5$ bins respectively. 
There appears to be a mild evolution in the typical size when clumpy/irregular galaxies are included, with $\Bar{R}_{e}=0.86\pm0.11$ kpc at $z\simeq5$, $\Bar{R}_{e}=1.09\pm0.07$ kpc at $z\simeq4$ and $\Bar{R}_{e}=1.15\pm0.03$ kpc at $z\simeq3$.
The distributions are also broader than their S\'ersic counterparts due to the contribution by large ($R_{e} > 1$ kpc) clumpy/irregular systems.

We can use the fraction of bad S\'ersic fits as a proxy for the irregular fraction. Here, irregular includes disturbed and clumpy morphologies.
In Fig. \ref{fig:sample_mass_z} we show the irregular fraction binned in both redshift and stellar mass.
The irregular fractions drops from 0.25 at $z=5$ to 0.20 at $z=3$.
This agrees with visual classification results from \citet{Ferreira23}, who find that galaxies with $\mathrm{log}_{10}(M_{\star}/M_{\sun}) > 9$ at $z>3$ are not dominated by irregular structures.
Our results are also in agreement with \citet{kartaltepe23}, who find that the irregular fraction increases from 0.2 at $z=5$ to 0.3 at $z=3$.
Our results differ to the visual classification by \citet{Jacobs23}, who find a very high peculiar fraction of $\sim0.9$ at $z\approx4.9$, dropping to nearly zero at $z\approx2.5$ for $\mathrm{log}_{10}(M_{\star}/M_{\sun}) > 9.5$, although the number of galaxies in their high-mass bins is small.
They see a constant peculiar fraction of $\sim0.6$ over $z=3-5$ in their lower-mass bin, $\mathrm{log}_{10}(M_{\star}/M_{\sun}) < 9.5$. This higher value compared to our fraction may be caused by the inclusion of low-mass galaxies ($\mathrm{log}_{10}(M_{\star}/M_{\sun}) < 9$) where we apply a cut.
In terms of stellar mass, our irregular fraction decreases slightly between $10^{9}M_{\sun}$ and $10^{10}M_{\sun}$.
This trend broadly agrees with \citet{Jacobs23}, who find that the fraction of peculiars is higher at $\mathrm{log}_{10}(M_{\star}/M_{\sun}) < 9.5$.
However, our irregular fraction spikes up to larger values at $M>10^{10.5}M_{\sun}$.
In the high-mass regime, a visual inspection reveals that the irregular galaxies in our sample appear to be merging systems, such as the bottom-right galaxy in Fig. \ref{fig:fitting_examples}.

\subsection{Size-redshift evolution}
\label{sec:size redshift evolution}

In Fig. \ref{fig:redshift_evolution} we show the size evolution of our sample with redshift. 
The left panel shows the individual measurements for each LBG, as well as various measures of the average size (mean, median and log-normal peak) in each redshift bin.
These values are reported in Table \ref{tab:average sizes}.
The mean and median values in each bin are significantly larger than the log-normal peaks (based on S\'ersic sizes), due to being skewed by large objects. 
The black dashed line shows the best fit derived from fitting to the full sample (all galaxies with $\mathrm{log}_{10}(M_{\star}/M_{\sun}) >9$) using a power law parameterisation $R_{e} = R(1+z)^{\alpha}$. 
Our best fit finds $R_{e} = 3.51(1+z)^{-0.60\pm0.22}$ kpc. 
Comparing this fit to the results binned by redshift suggests the evolution is more rapid than implied by the log-normal peaks by $z\simeq3$.
However, due to large scatter in the sizes there is large uncertainty in the power-law fitting.

In the right-hand panel we compare our results to other studies using \textit{JWST}. 
Our power-law fit and mean sizes in each redshift bin are remarkably consistent with the power-law fitting of \citet{Ormerod23}, who measure the size evolution of galaxies across $z=0-8$.
Their relation is based on the mean size of galaxies with redshift. 
We note that they use a slightly smaller area than this study ($64 \ \mathrm{arcmin}^{2}$).
Their power-law index is in agreement with ours with $\alpha=0.71\pm0.19$.
Our results also agree with the evolution determined by \citet{Ward23}, who find a size evolution with power-law index of $\alpha=0.63\pm0.07$ using $97\ \mathrm{arcmin}^{2}$ of \textit{JWST} imaging.
We note that their determination differs slightly - they find the evolution at a fixed stellar mass of $5\times10^{10}M_{\sun}$.

The build-up in the tail of the log-normal distribution (see Section \ref{sec:lognormal results}) is complemented by a size-redshift relation that finds larger sizes at lower redshift. 
Together, they suggest an emergence of large ($R_{e} > 2$) galaxies at $z=3$.
We can complement this result with a measure of the evolution of \textit{typical} galaxy sizes by focusing on our log-normal peaks. 
Our use of log-normal distribution fitting is based on the results of \citet{curtis_lake16}, which is physically motivated (see Section \ref{sec:disk formation}) and better accounts for the typical size of galaxies by incorporating a tail for the largest objects. 
Additionally, using the log-normal peak of the distribution allows for a determination of the typical size of the galaxy in samples that are down to 50\% complete.
\citet{curtis_lake16} mention that selection effects need to be accounted for carefully with certain flux cuts to ensure the typical sizes of galaxies can be recovered. 
They find that selection effects in previous works \citep[e.g.][]{Shibuya15} preferentially biases towards smaller sizes due to their choice in selection methodology. They require a signal-to-noise of 15 in a 0.35 arcsec diameter aperture, tending to select smaller galaxies. 
We do not expect to be preferentially skewed to smaller galaxies because the $5\sigma$ depths of the PRIMER imaging is $\sim0.3-3$ mag deeper than the ground-based filters used to select the sample.
Additionally, the ground-based selection is seeing-dominated and therefore less affected by size.
\citet{curtis_lake16} find a redshift evolution $R_{e} \propto (1+z)^{-0.20}$ at $z=4-8$ when fitting to their log-normal peaks, a much weaker evolution than \citet{Shibuya15} and \citetalias{Constantin23}. 
If we fix the exponent $\alpha=-0.20$, also shown in Fig. \ref{fig:redshift_evolution} and fit for the power law coefficient to our log-normal peaks, we find a relation that agrees with our typical galaxy sizes, extending the results of \citet{curtis_lake16} to $z=3$.

\subsubsection{Comparison with other studies}

The illustrisTNG simulation \citep{illustrisTNG} is used by \citetalias{Constantin23} to measure the sizes of $\mathrm{log}_{10}(M_{\star}/M_{\sun}) >9$ galaxies with artificial noise added to match the depth of the Cosmic Evolution Early Release Science survey \citep[CEERS, ][]{CEERS}. 
Our mean sizes are 15\% smaller than the predictions of \citetalias{Constantin23} at $z=3$.
They conclude that their upturn in the size evolution is attributed to more massive galaxies becoming larger at $z\simeq3$, and our best-fit derived from the full sample recovers a similar trend, but not as steep (although agreeing within the errors). 
Similarly, \citet{Ormerod23} do not observe a redshift evolution as steep in their \textit{HST}+\textit{JWST} sample.
The CEERS fields have an area of $\sim100 \ \mathrm{arcmin}^{2}$, a factor of $\sim3.8$ less than PRIMER. 
Although scaling the log-normal distributions (see Section \ref{sec:lognormal results}) suggests a genuine build-up of large galaxies, we caution that conclusions regarding the growth of high-mass galaxies require a similar analysis on larger volumes. 

\citetalias{Constantin23} measure median values in three mass bins centred at $\mathrm{log}_{10}(M_{\star}/M_{\sun}) = 9.1, 9.4$ and $10.0$. 
In Figure \ref{fig:redshift_evolution} we show our mean and median sizes when split into two mass bins at $\mathrm{log}_{10}(M_{\star}/M_{\sun}) = 9.6$. 
We use two bins because there are too few objects in the low- and high-mass bins when using the same binning as \citetalias{Constantin23}. 
Additionally, this value approximately splits the sample in half.
We find that at $z=5$, both the mean and median in the low-mass bin are larger than in the high-mass bin. 
At $z=3-4$ the mean and median are larger in the high-mass bin. 
A similar trend is seen in \citetalias{Constantin23}, and this reflects the flat/negative size-mass relations (see Section \ref{sec:size-mass}). 
\citetalias{Constantin23} also see a broadening in their median sizes towards lower redshift and a convergence to similar values at higher redshift when binning in mass (see their fig. 8).
Although we see this to some extent, the broadening of sizes between the bins by $z=3$ is not as pronounced - they differ by no more than 0.15 kpc at any redshift (compared to $>0.5$ kpc in \citetalias{Constantin23}).
This suggests a weak mass dependence of the size-redshift evolution in our sample.

To summarise, we find that whilst the \emph{typical} size of galaxies does not evolve significantly over $z=3-5$ (as measured by the log-normal peaks of S\'ersic sizes in each redshift bin), there is evidence for an emergence of large ($R_{e} > 2$ kpc) galaxies when we fit for the entire sample (i.e. all galaxies with $\mathrm{log}_{10}(M_{\star}/M_{\sun}) > 9$).

\subsection{Size-mass relations}
\label{sec:size-mass}

\begin{figure*}
\newcommand{\fitfigwidth}{1}
\centering

\includegraphics[width=\fitfigwidth\columnwidth]{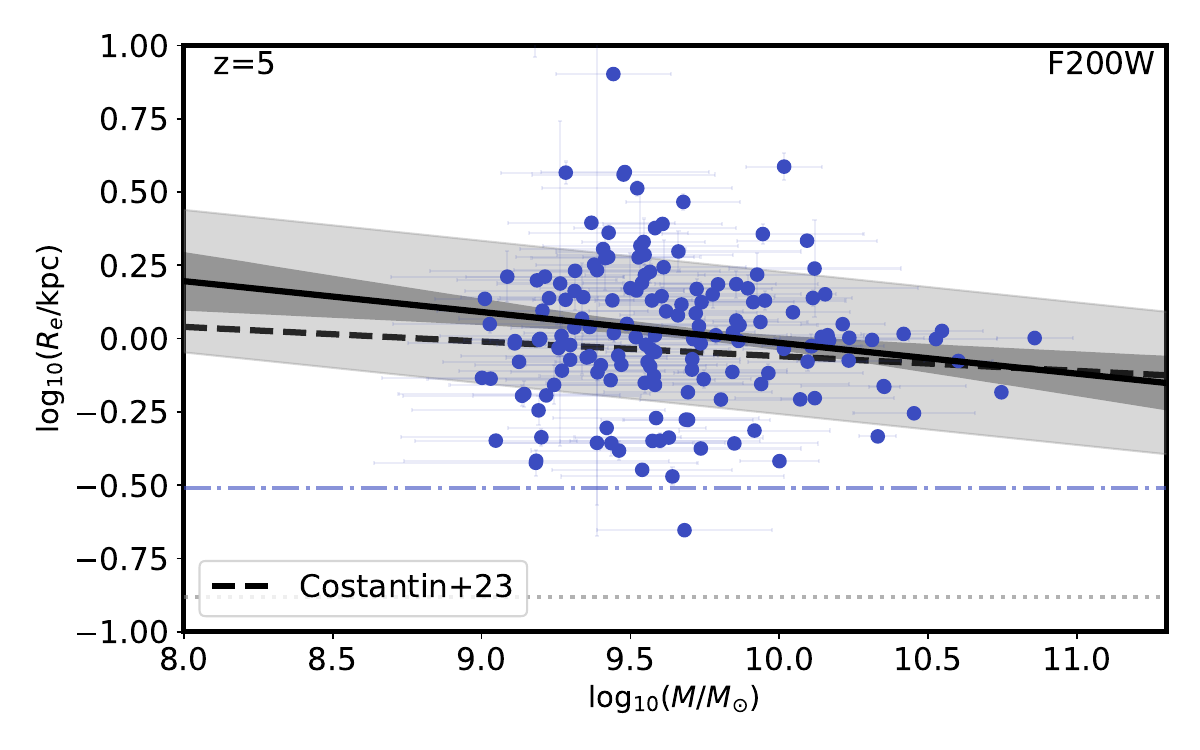}
\includegraphics[width=\fitfigwidth\columnwidth]{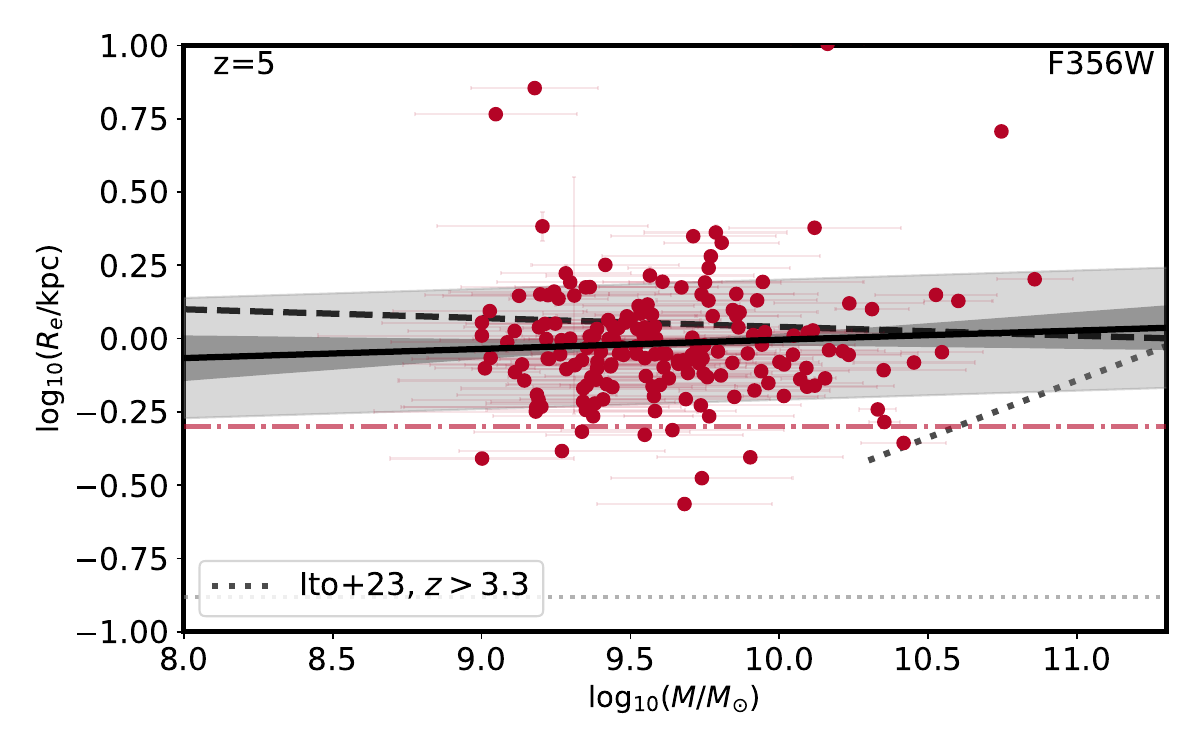}

\includegraphics[width=\fitfigwidth\columnwidth]{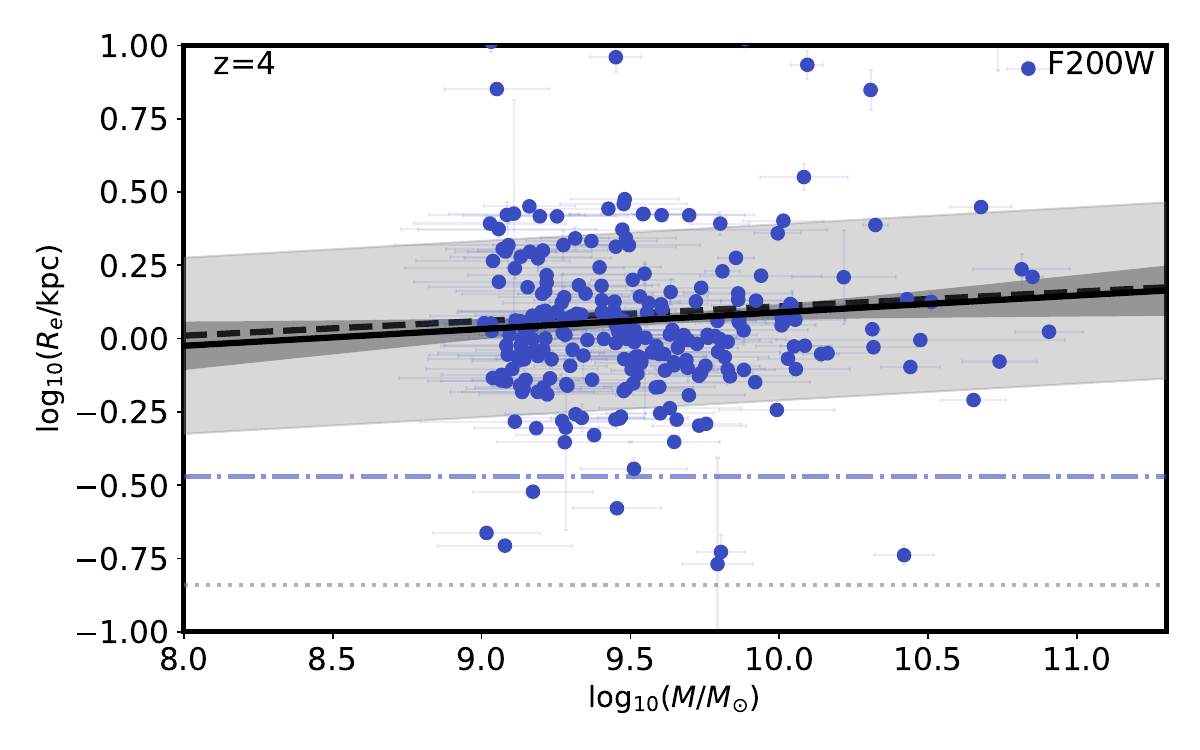}
\includegraphics[width=\fitfigwidth\columnwidth]{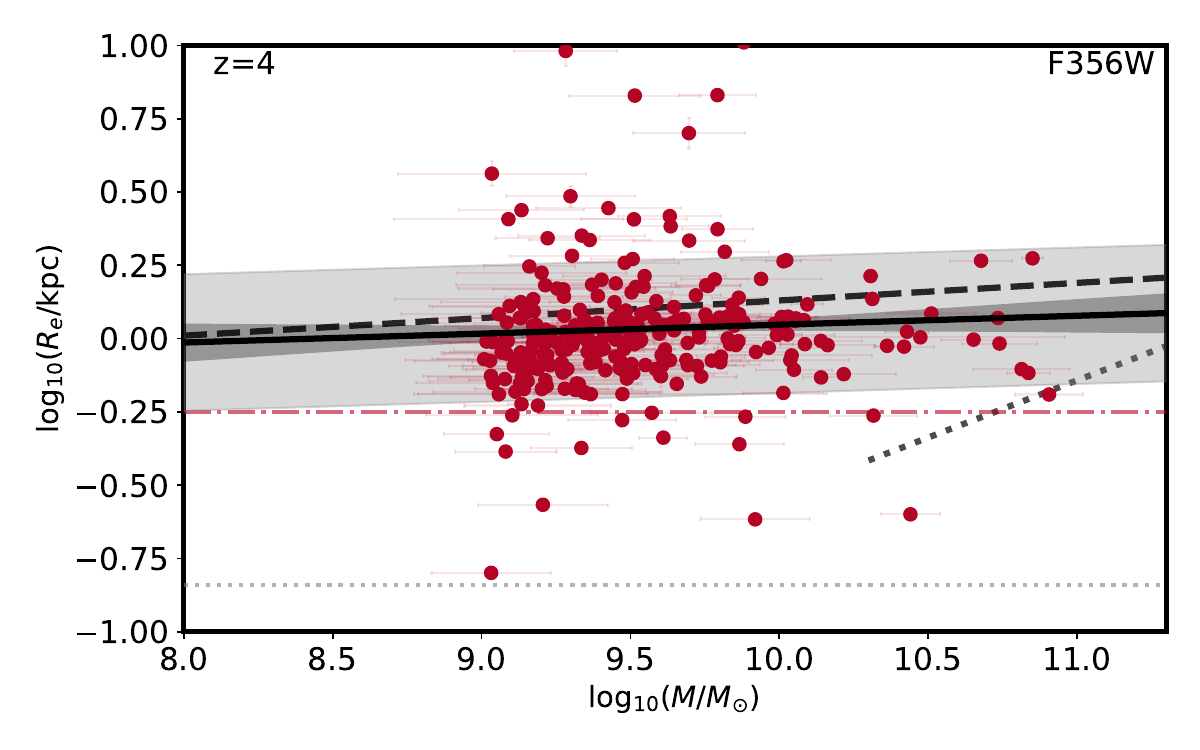}

\includegraphics[width=\fitfigwidth\columnwidth]{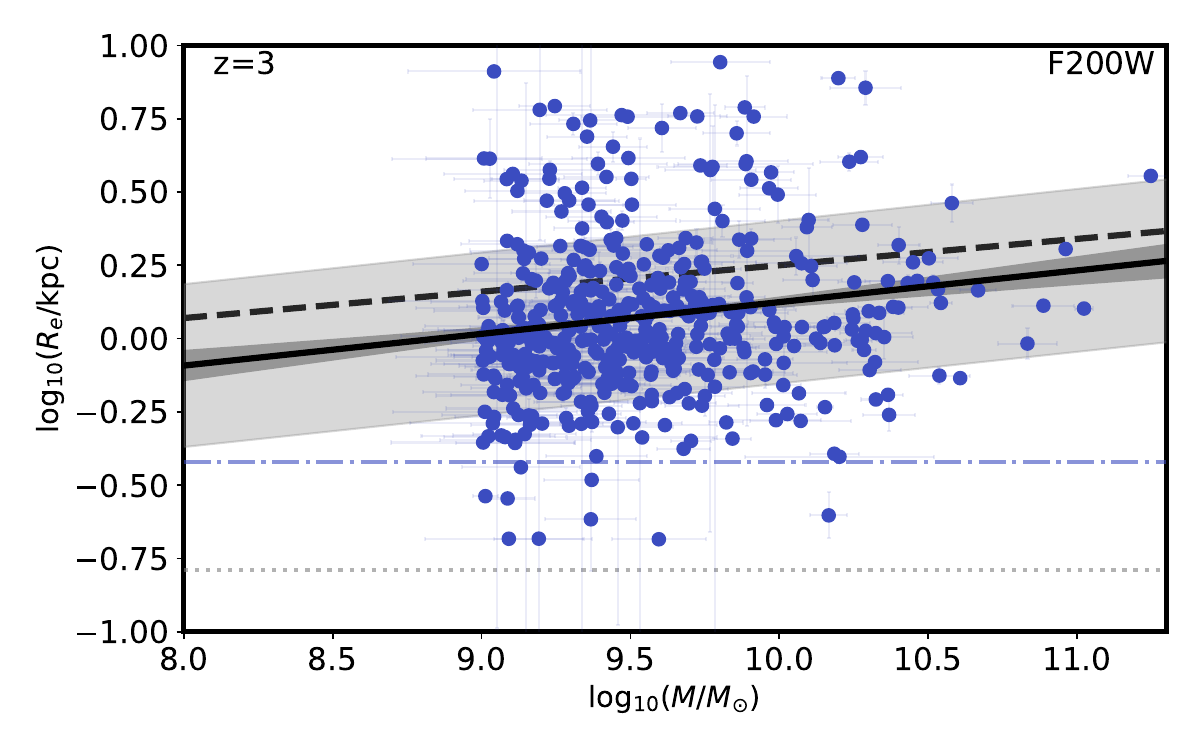}
\includegraphics[width=\fitfigwidth\columnwidth]{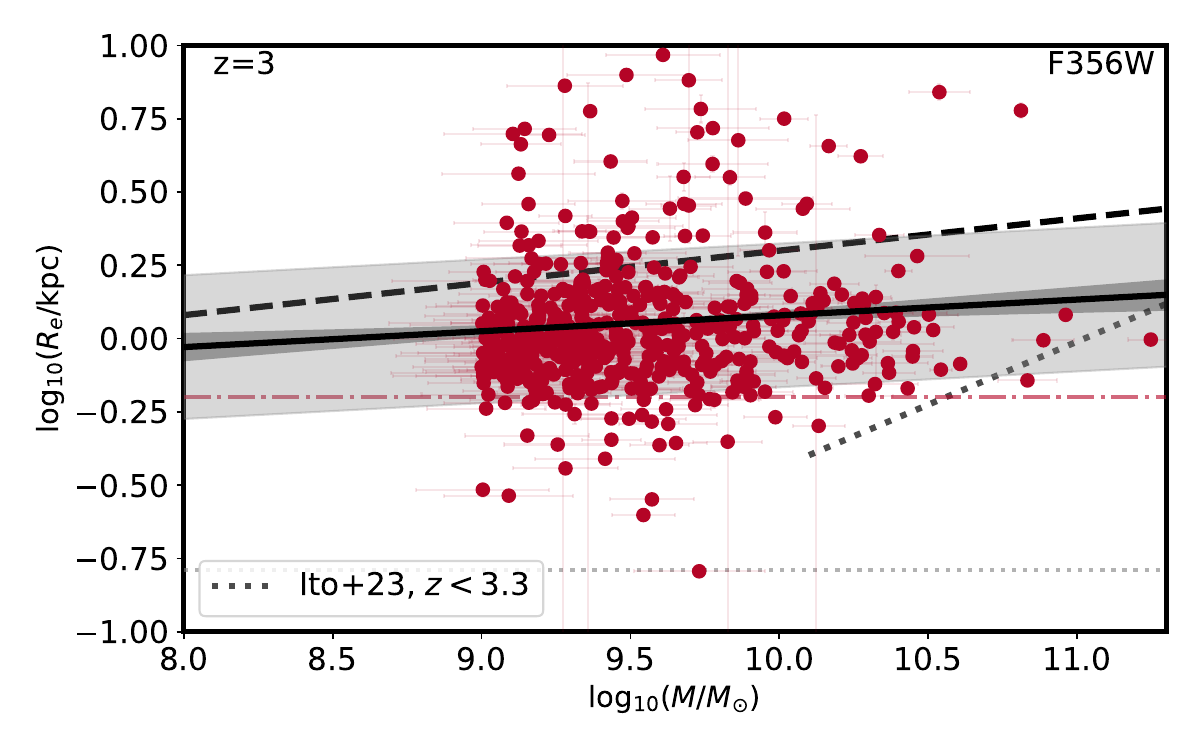}

\caption{The S\'ersic size-mass relations of our sample. Each row shows a redshift bin (labelled in the top-left of each panel). The left column shows size measurements made in F200W, and the right column shows F356W. The points and errors show individual measurements, and the black solid line shows the best-fit line. The light-gray shaded region represents the intrinsic scatter, and the dark-gray shaded region represents the statistical error. The dashed black line shows predictions for the size-mass relations in each filter and redshift bin from \citetalias{Constantin23}. The horizontal dash-dotted line represents the PSF FWHM of each filter, converted to an equivalent radius at each redshift. The horizontal dotted gray line at the bottom of each panel represents the size inferred for a point source by \textsc{PyAutoGalaxy} (see Section \ref{sec:parametric sizes}). On the F356W results we also show the rest-optical size-mass relations at $z=3-5$ for massive quiescent galaxies found by \citet{Ito23}.}
	\label{fig:mass_relations}
\end{figure*}

\begin{table}
\centering
\caption{The best-fit values for the size-mass relation given by $\mathrm{log}_{10}R_{e} = a\times\mathrm{log}_{10}(M_{\star}/M_{\sun}) + b$ and the intrinsic scatter $\sigma_{{\mathrm{log}R_{e}}}$ in each redshift bin and for each filter used.}
\begin{tabular}{cccc}
\hline
F200W & \textbf{a} & \textbf{b} & $\sigma_{{\mathrm{log}R_{e}}}$ (dex) \\
\hline
$z = 3$ & $0.11\pm0.05$ & $-0.96\pm0.32$  & $0.28\pm0.01$\\
$z = 4$ & $0.06\pm0.05$ &  $-0.48\pm0.48$& $0.30\pm0.01$\\
$z = 5$ & $-0.11\pm0.06$&  $1.04\pm0.56$ & $0.24\pm0.01$\\
\hline
F356W & \textbf{a} & \textbf{b} & $\sigma_{{\mathrm{log}R_{e}}}$ (dex) \\
\hline
$z = 3$ & $0.05\pm0.03$ & $-0.46\pm0.29$  & $0.25\pm0.01$\\
$z = 4$ & $0.03\pm0.04$& $-0.26\pm0.38$ & $0.23\pm0.01$\\
$z = 5$ & $0.03\pm0.05$&  $-0.32\pm0.44$ & $0.21\pm0.01$\\
\hline

\end{tabular}

\label{tab:size-mass}
\end{table}

In Fig. \ref{fig:mass_relations} we present the S\'ersic size-mass relations in each redshift bin for both F200W and F356W in order to compare directly with the predictions of \citetalias{Constantin23}.
They use the illustrisTNG simulation \citep{illustrisTNG} to measure the S\'ersic sizes of $\mathrm{log}_{10}(M_{\star}/M_{\sun}) >9$ synthetic galaxies matching those found in the Cosmic Evolution Early Release Science survey \citep[CEERS,][]{CEERS}. 
We fit straight lines of the form $\mathrm{log}_{10}R_{e} = a\times\mathrm{log}_{10}(M_{\star}/M_{\sun}) + b$.
The independent variables in our size-mass and size-luminosity relations themselves possess errors, complicating the task of fitting a straight line.
\citet{roxy} find that common approaches to dealing with both $x$ and $y$ errors can ignore the underlying distribution of the true independent variable values, leading to biased results (see references therein). 
They find that using a single Gaussian prior (with mean and variance to be determined as part of the inference) on `latent variables' describing the true values of the independent variables performs best, and they dub this `Marginalised Normal Regression'. 
It is key that this works in the presence of intrinsic scatter, an additional property we want to measure. 
We therefore use their \textit{Regression and Optimisation with X and Y errors}  (\textsc{ROXY}) package\footnote{https://github.com/DeaglanBartlett/roxy} to fit the size relations.
We measure the intrinsic scatter and compare it to the statistical error of the fitting. We present the best-fit values in Table \ref{tab:size-mass}.
In order to assess the robustness of measuring very small sizes, we compare the PSF sizes (see Section \ref{sec:stamps_and_psf} and Fig. \ref{fig:mass_relations}) to the sizes of our sample.
The PSF FWHMs correspond to sizes of 0.38 kpc (0.63 kpc) at $z=3$ in F200W (F356W), 0.34 kpc (0.56 kpc) at $z=4$ in F200W (F356W) band, and 0.31 kpc (0.50 kpc) at $z = 5$ in F200W (F356W).
In these bands, 95\% of the sample is larger than the PSF FWHM, meaning the majority of sources are resolved.
We verified that removing these sources from the sample does not significantly change the size-mass relation.
We note that any observational biases towards larger sizes in this work due to the PSF minimum size are also present in \citetalias{Constantin23}, still allowing for a direct comparison. 
IllustrisTNG indeed predicts sizes smaller than the PSF FWHM (see their fig. 7), which are not measurable in the mock observations.

Interestingly, at $z=5$ we see some evidence for a negative size-mass relation in F200W, consistent with the prediction of negative slopes by \citetalias{Constantin23}. 
A negative slope is allowed in F356W within errors.
At $z=4$, the slopes in F200W and F356W are both positive and consistent with one another. 
We also find that the slopes are consistent with \citetalias{Constantin23}. 
At $z=3$, the slope in F356W appears to be flatter than that in F200W. 
We test whether the fit is being skewed by the relatively compact objects at $\mathrm{log}_{10}(M_{\star}/M_{\sun}) \gtrsim 10.5$ in F356W by re-fitting without them, and find that the results do not change significantly. 
The relation in F200W is consistent with \citetalias{Constantin23}, and the F356W relation appears flatter but is consistent within $2\sigma$.
Our slopes are flatter than those found by \citet{Ward23}, who measure the rest-$5000$\AA \ sizes of star-forming galaxies in $97 \mathrm{arcmin}^{2}$ of CEERS imaging. 
They find a gradient of $a=0.25$ in a redshift bin of $z=3.0-5.5$, and a gradient of $a=0.15$ in a redshift bin of $z=2.0-3.0$.
We note some differences between our methodologies - our redshift bins span almost the whole range of their highest redshift bin, and their $z=2.0-3.0$ bin will contain objects at lower redshifts than we can select for, complicating the comparison.
Broadly speaking, \citet{Ward23} find a flattening of the size-mass relation at lower redshifts, and we see hints of this from $z=5$ to $z=4$.
Similarly, \citet{Pandya24} find some evidence for a flattening size-mass relation over $z=3-8$.

\subsubsection{Intrinsic scatter}
One notable result is that the intrinsic scatter of objects about the best-fit relation increases with decreasing redshift, and is lower in F356W than in F200W by 0.03-0.07 dex.
The values are reported in Table \ref{tab:size-mass}.
By plotting both the intrinsic scatter and statistical error in Fig. \ref{fig:mass_relations}, we can see that the intrinsic scatter dominates over measurement errors in the fitting. 
\citet{morishita23} measure the size-mass relation in star-forming galaxies at $5<z<14$ and find a redshift-corrected intrinsic scatter of $\sigma_{\mathrm{log}R_{e}} = 0.30\pm0.01$ dex, consistent with our findings at $z=3-4$ in F200W but higher than our findings in F356W (see Table \ref{tab:size-mass}). 
Our intrinsic scatters increase by 0.04 dex from $z=5$ to $z=3$ in both F200W and F356W. 
The values in F200W are larger than low-redshift ($z<1$) results for both star-forming and quiescent galaxies, being more consistent with our resuts in F356W \citep[$\sigma_{\mathrm{log}R_{e}}\simeq0.20$ dex,][]{Kawinwanichakij21}.
Whilst our values in F356W broadly agree with \citet{Ward23}, the increase we find is in disagreement with their constant intrinsic scatter ($\sim0.2$ dex) out to $z=5.5$. This is likely due to the fact that their highest redshift bin nearly encapsulates all three of our redshift bins.

\subsubsection{Comparison with quiescent galaxies}
The infrared coverage provided by \textit{JWST} has allowed for the selection of massive quiescent galaxies at $z>3$, implying extremely early quenching episodes at $z\gtrsim5$ \citep[e.g. ][]{Carnall23, Carnall23Nature, Long23, Ito23, Ji24}. 
We can thus compare our results for star-forming LBGs to the quiescent population in the rest-optical at $z=3-5$. 
In Fig. \ref{fig:mass_relations} we show the findings of \citet{Ito23} to compare to our F356W results. 
Although our number counts are low at $\mathrm{log}_{10}(M_{\star}/M_{\sun}) > 10.5$, extrapolating our results would suggest that quiescent galaxies tend to be more compact than star-forming galaxies, as seen at lower redshifts \citep{vanderWel14}.
Additionally, \citet{Ji24} also find compact sizes for massive ($\mathrm{log}_{10}(M_{\star}/M_{\sun}) > 10$) quiescent galaxies at $z=1.6-5$.

\citet{Ito23} note that AGN may result in apparent compact sizes for some objects, but do not remove any objects since none have X-ray detections.
They also find that quiescent galaxies are more compact with increasing wavelength, which could be caused by inside-out quenching that produces compact morphologies for evolved stellar populations. 
Although wider areas are needed to better constrain the positions of the most massive objects in the size-mass plane, perhaps this explains the flatter relation we find using F356W compared to F200W at $z=3$ where massive galaxies are still undergoing intense star formation, but have begun to quench in their centres. 
\citet{Ormerod23} find a similar trend for both star-forming and passive galaxies at $z<3$, also interpreting this as evolved stars occupying the central regions of galaxies with star formation occuring at larger radii.
Similarly, at $z=3-4$ Allen et al. (in prep.) find a flattening of the size-mass relation at longer wavelengths.
Correspondingly, at $z=1-2.5$ \citet{Suess19a} find that most star-forming and quiescent galaxies have negative colour gradients (i.e. redder in their centres and bluer at larger radii). They also find that the colour gradient evolves rapidly over this redshift range \citep{Suess19b}, broadly fitting into the picture of inside-out growth and quenching.

Overall, we find good agreement of our results with predictions from simulations. Although the slope of the relations appear to transition from positive to negative at $z=4-5$, a lack of dynamic range in mass limits this conclusion. However, we do find an increase in the intrinsic scatter with increasing redshift.
Better dynamic range in massive objects is required to better constrain and confirm the negative slopes at $z=5$, the flattening of the F356W slope, and the evolution of the gradients. 
We note that in order to compare directly with \citetalias{Constantin23} we use the same filters across the full redshift range, meaning there is a shift in the rest-frame wavelengths probed ($\lambda \simeq 0.3-0.5\micron, \ 0.6-0.9\micron$ for F200W and F356W respectively). 
A good alternate probe of the evolution would be with consistent rest-frame wavelengths (Allen et al., in prep.). 

\subsection{Size-luminosity relations}
\label{sec:size-luminosity}

\begin{figure*}
    \centering
    \includegraphics[width=\textwidth]{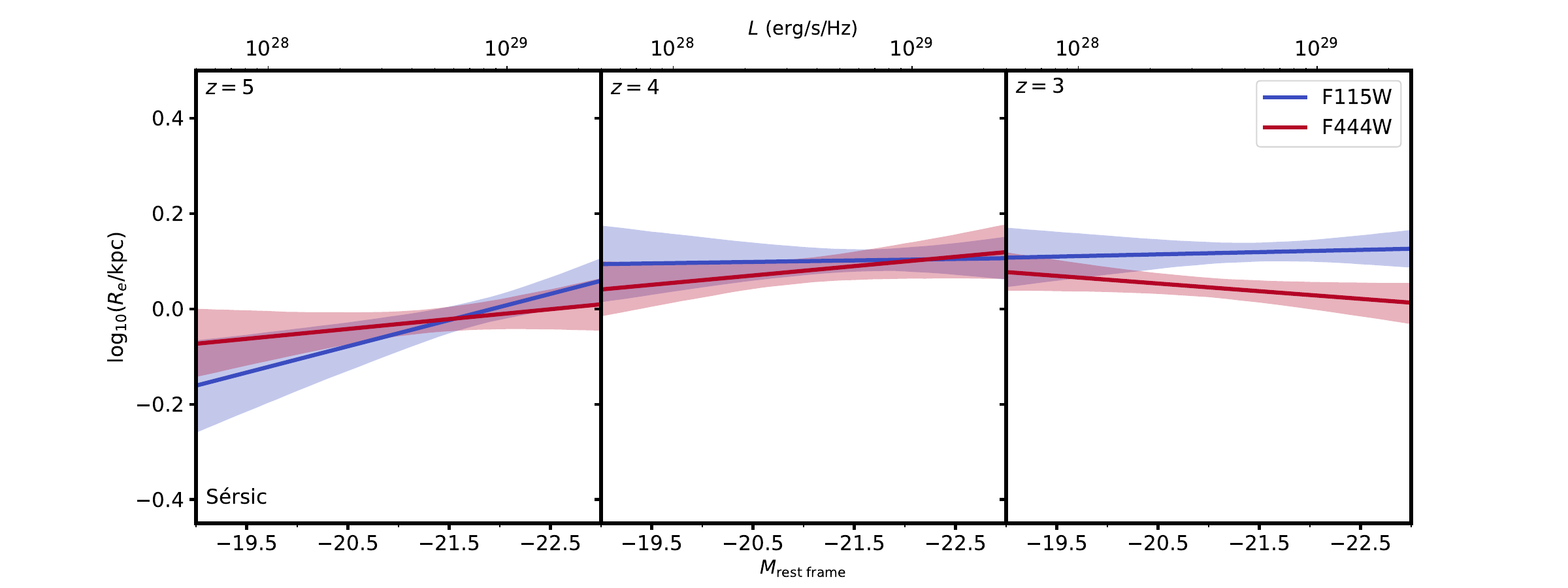}

    \includegraphics[width=\textwidth]{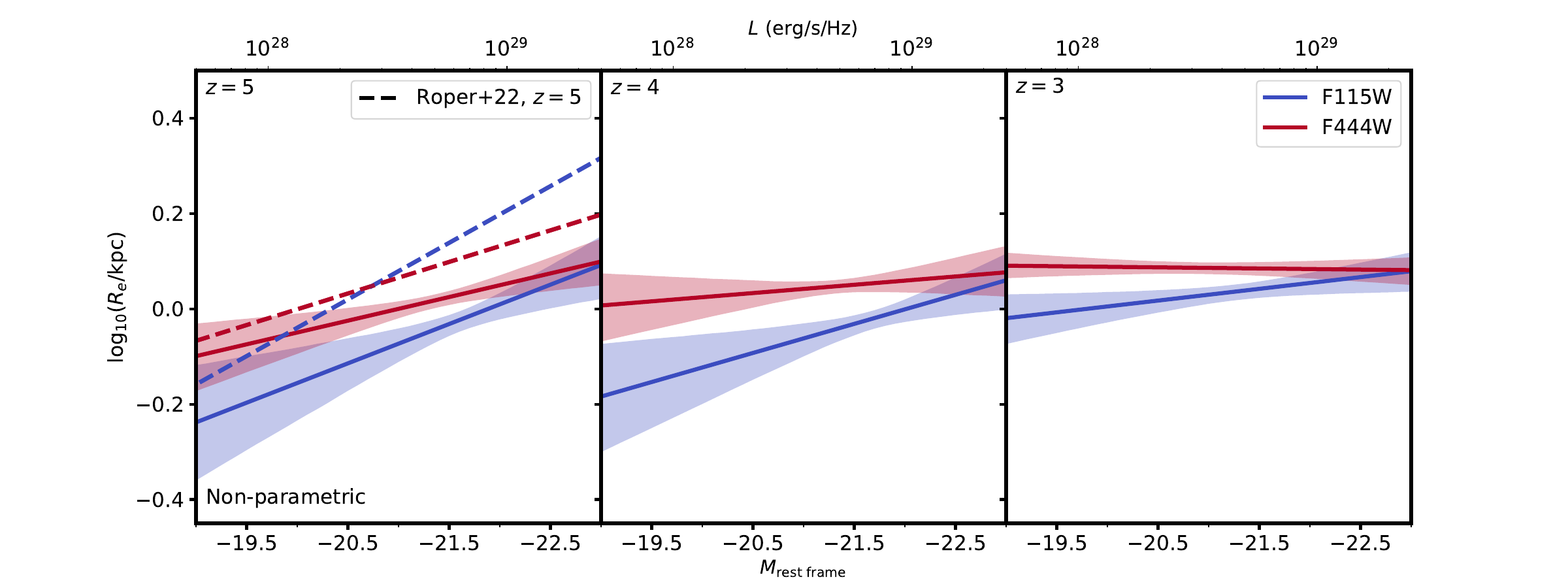}
    
    \caption{The size-luminosity relations in each redshift bin (labelled in the top-left of each panel) in F115W and F444W. \textbf{Top:} results from our S\'ersic fitting. The rest-frame absolute magnitudes calculated for each of F115W and F444W (see Section \ref{sec: photometry and sed fitting}) are shown on the x-axis, with the equivalent luminosity on the top axis. We also plot the $1\sigma$ error contours. \textbf{Bottom:} results from our non-parametric size fitting. The plots are the same as the top panel, but we also show predictions from the FLARES model of \citetalias{Roper22} at $z=5$ with the dashed lines.}
    \label{fig:luminosity_relation}
\end{figure*}

\begin{table}
\centering
\caption{The best-fit values for the size-luminosity relation givens by $\mathrm{log}_{10}R_{e} = a \times M_{\mathrm{rest \ frame}} + b$ in each redshift bin and for each filter used. The top table shows the results from S\'ersic size fitting, and the bottom table shows results from the non-parametric sizes.}
\begin{tabular}{ccc}
\textbf{S\'ersic} & & \\
\hline
F115W & a & b \\
\hline
$z=3$ & $-0.00\pm0.02$ & $0.02\pm0.50$ \\
$z=4$ & $-0.00\pm0.03$ & $0.03\pm0.63$ \\
$z=5$ & $-0.06\pm0.03$ & $-1.23\pm0.72$ \\
\hline
F444W & a & b \\
\hline
$z=3$ & $0.02\pm0.02$ & $0.40\pm0.38$ \\
$z=4$ & $-0.02\pm0.03$ & $-0.33\pm0.55$ \\
$z=5$ & $-0.02\pm0.03$ & $-0.46\pm0.62$ \\
\hline
\\
\textbf{Non-parametric} & & \\
\hline
F115W & a & b \\
\hline
$z=3$ & $-0.02\pm0.02$ & $-0.49\pm0.47$ \\
$z=4$ & $-0.06\pm0.04$ & $-1.35\pm0.93$ \\
$z=5$ & $-0.08\pm0.05$ & $-1.80\pm1.00$ \\
\hline
F444W & a & b \\
\hline
$z=3$ & $0.00\pm0.01$ & $0.15\pm0.26$ \\
$z=4$ & $-0.02\pm0.03$ & $-0.35\pm0.65$ \\
$z=5$ & $-0.05\pm0.03$ & $-1.05\pm0.62$ \\
\hline

\end{tabular}

\label{tab:size-luminosity}
\end{table}

In Fig. \ref{fig:luminosity_relation} we present the results of our size-luminosity fitting in the F115W and F444W filters to the functional form $\mathrm{log}_{10}R_{e} = a \times M_{\mathrm{rest \ frame}} + b$, where $M_{\mathrm{rest \ frame}}$ is the rest-frame absolute magnitude in each filter. 
We choose these filters to maximise the difference in wavelength between the relations, providing a better test of the wavelength dependence predicted by \citetalias{Roper22}.
These filters probe $0.2-0.3\micron$ and $0.7-1.1\micron$ in the rest frame, respectively.
The top panel shows the results for the S\'ersic fitting, and the bottom panel shows the non-parametric results.
The best-fit values are presented in Table \ref{tab:size-luminosity}. 
As in Section \ref{sec:size-mass}, we briefly compare our non-parametric sizes with the PSF sizes. 
At $z=3-4$, 97\% of the sizes are larger than the PSF FWHMs in F115W and F444W. At $z=5$, nearly 93\% of the sizes are larger. The majority of the sample is therefore resolved.
We verified that removing sources consistent with the PSF FWHM from the sample does not significantly change the size-luminosity relations.
First looking at the S\'ersic results, at $z=4-5$ we find that the relations are consistent with one another.
It also appears that for a given magnitude, galaxies become larger between $z=4-5$.
At $z=3$ the F444W relation becomes negative and the sizes are more compact than in F115W.

The results from non-parametric size fitting described in Section \ref{sec:size fitting} are slightly different.
The non-parametric size-luminosity relations generally show a flattening of the relation with decreasing redshift, and the sizes in F115W are more compact than in F444W for any given luminosity.
We do not find significant differences between the gradients in these bands at a given redshift. 

Comparing our two size-fitting methods, we find that at $z=4-5$ the non-parametric sizes have slightly steeper relations than the S\'ersic sizes. 
This is likely due to the inclusivity of the non-parametric method to massive, clumpy/irregular galaxies. 
Whilst the S\'ersic fitting may fit to an individual clump or fail altogether (see Fig. \ref{fig:fitting_examples}), the non-parametric method better accounts for this, meaning we do not have to discard any fits.
Indeed, if we remove clumpy galaxies the non-parametric size-luminosity relations flatten slightly, but the results are not changed significantly.

Massive galaxies can thus contribute to the population with larger sizes. 
By $z=3$ the relations for both methods are relatively flat. 
The offsets between the bands differs across the two methods. In the S\'ersic fitting, the offset between the red and blue bands disappears by $z=4$. For the non-parametric relations a significant offset remains between the two bands at $z=3-4$.
Since the non-parametric method is agnostic to where the light in the galaxy lies, it may perform better on disturbed profiles in F115W where the S\'ersic fitting converges, but may not be the most appropriate size measure. 
A visual inspection of the galaxies in each NIRCam filter reveals that objects which appear disk-like at long wavelengths can often have slightly irregular morphologies at short wavelengths. 
For example, in Fig. \ref{fig:resolution_example} the object appears to be disk-like in the long-wavelength filters, whereas in the short-wavelength filters it appears to be comprised of two components.
In these cases we find that the S\'ersic fitting ignores the clumps and still fits a disk, similar to the bottom right panel in Fig. \ref{fig:fitting_examples} but to a less extreme extent. 
This may slightly overestimate these sizes, leading to the agreement in the S\'ersic relations across wavelength.
We thus conclude that galaxies tend to be more compact in their rest-UV light compared to their rest-optical light.
There also appears to be a flattening of the relations towards lower redshift across both measurement methods.
In Appendix \ref{sec:size measurement comparison} we compare the S\'ersic and non-parametric sizes for galaxies that have a good S\'ersic fit.

\subsubsection{Comparison with the literature}
We can also compare our results from both size-fitting methods to the literature. Assuming $R\propto L^{-\beta}$, \citet{huang13} find power laws of $\beta=0.22^{+0.058}_{-0.056} \ \mathrm{and} \ 0.25^{+0.15}_{-0.14}$ at $z=4$ and $z=5$ respectively for LBGs using S\'ersic fitting. 
Converting our relations from magnitude to luminosity, the non-parametric relations in F115W have power laws with exponent $\beta=0.15\pm0.10$ and $0.20\pm0.08$ at $z=4$ and $z=5$ respectively. 
The S\'ersic relations in F115W have exponents $\beta=0.00\pm0.08$ and $0.15\pm0.08$ at $z=4$ and $z=5$ respectively. 
Our S\'ersic results are thus slightly shallower in slope than \citet{huang13} (although the errors are large), and our non-parametric results are consistent within the errors.
The non-parametric results are also consistent with \citet{Shibuya15} and \citet{curtis_lake16} who both find values spanning $\beta \simeq 0.1-0.27$ over $z=3-5$. 
We also note that these results are consistent with size-luminosity relations of bright galaxies at $z>5$ found by, e.g., \citet{Holwerda15, Shibuya15}. 
However, results from lensed galaxies at $z>5$ (thus probing intrinsically fainter galaxies) tend to show much steeper slopes around $\beta\sim0.5$ \citep[e.g. ][]{Grazian12, kawamata18, Bouwens22, Yang22}.
\citet{Yang22} find that different lensing models all produce the same steeper relation, and \citet{bouwens21} still find steep slopes after accounting for surface-brightness selection effects.
\citet{kawamata18} recover this steeper relation in a simple analytic model where smaller galaxies have lower specific angular momenta.

Our results are complementary to those of \citet{Suess22}.
For a sample of $1179$ galaxies at $z=1-2.5$ with $\mathrm{log}_{10}(M_{\star}/M_{\sun})\ge9$ selected from CEERS, they compare the observed $1.5\micron$ ($\lambda_{\mathrm{rest}}\sim0.55\micron$) and $4.4\micron$ ($\lambda_{\mathrm{rest}}\sim1.6\micron$) sizes using the F150W and F444W NIRCam filters, respectively.
This work is similar to their results in that the rest-frame wavelengths vary over the redshift range due to the fixed choice of filters ($\lambda \simeq0.2-0.3\micron, 0.7-1.1\micron$ in F115W and F444W), thus the focus is on the relative differences between the relations.
They find that the observed $4.4\micron$ sizes are more compact than at $1.5\micron$, and that the size difference is more pronounced at higher stellar mass.
This is in agreement with our S\'ersic size-luminosity relation at $z=3$ - sizes in F444W are more compact than in F115W, and this size difference increases towards brighter magnitudes.

\subsubsection{Comparison with \citet{Roper22}}
We also compare our results to predictions from \citetalias{Roper22}, who use the FLARES simulation \citep{FLARES} to predict the size-luminosity relation of massive galaxies at $z=5-10$ as a function of wavelength. 
The non-parametric size measurement we use follows their method, and hence they can be directly compared (although see the discussion in Section \ref{sec:comparison with sims}). 
Note that we can only compare our results against their $z=5$ predictions.
The closest matching synthetic filter to F115W in the rest-frame is their MUV-band, and the closest to F444W is their R-band.
Our gradients are consistent with \citetalias{Roper22}, with the relation in F115W being steeper than in F444W.
However, there is a significant offset between our relations and their predictions.
Their relations also cross, leading to objects appearing larger in F115W at the bright end (see Fig. \ref{fig:luminosity_relation}). 
This behaviour is not seen in our non-parametric results, but is seen in the S\'ersic results.
We note that the S\'ersic results use a sub-sample of all galaxies in the PRIMER imaging, as poor fits are discarded.
We test the impact of this by rerunning the non-parametric size-luminosity relations on this sub-sample. The relations flatten slightly but it does not change the results significantly.

\section{Discussion}
\label{sec:discussion}

Our results point to an emergence of large ($R_{e} > 2$ kpc) galaxies by $z=3$, whilst simultaneously pointing to little evolution in the \emph{typical} size of galaxies over $z=3-5$. 
Our S\'ersic size-mass relations show hints of a negative slope at $z=5$, intrinsic scatter that increases towards lower redshift and is larger in bluer bands, and some consistency with the predictions by \citetalias{Constantin23}.
Our size-luminosity relations exhibit a flattening towards lower redshift in both the S\'ersic and non-parametric sizes. 
The slopes of our non-parametric relations at $z=5$ are consistent with the predictions by \citetalias{Roper22} from FLARES, with an offset in the intercept.
In this section, we discuss and interpret these results in the context of disk formation models and explore the implications of the agreement of our results with simulation works.

\subsection{Formation of disks at high-redshift}
\label{sec:disk formation}

The use of log-normal fitting to the size distribution of galaxies is motivated by the disk formation model of \citet{Fall80}. 
In this theoretical framework the size of a galaxy is governed by its angular momentum, given to the system by tidal torques with neighbouring objects.
\citet{Peebles69} express the total angular momentum in terms of a dimensionless spin parameter $\lambda = JE^{1/2}M^{-5/2}G^{-1}$ where $J$ is the angular momentum of the system, $E$ is the total energy, $M$ is the total mass (all dominated by the DM halo before collapse) and $G$ is the gravitational constant. 
In the picture of hierarchical structure formation, the distribution of spin parameters follows a log-normal distribution. 
Assuming the disk is relaxed, we then expect $R_{e} \propto \lambda R_{\mathrm{vir}}$ where $R_{\mathrm{vir}}$ is the virial radius of the DM halo \citep{Bullock01, curtis_lake16}.
As described in Section \ref{sec:size redshift evolution} and shown in Fig. \ref{fig:lognormal_dist}, our size distributions are well-described by a log-normal distribution, consistent with these theoretical models out to $z=5$. 
Previous studies with \textit{HST} \citep{Shibuya15, curtis_lake16} also found strong agreement with a log-normal distribution out to $z\simeq6$ (beyond which sample sizes are small), but this was limited to rest-UV emission at $z\gtrsim3$. 
These results are thus complementary to \textit{HST} results, revealing consistency between rest-UV and rest-optical size distributions at these redshifts. 

The disk model of \citet{Fall80} and \citet{Mo98} also predicts that the size of a galaxy scales with redshift as $r\propto(1+z)^{-1}$ in the case of the host halo having a fixed circular velocity, and as $r\propto(1+z)^{-1.5}$ for a fixed halo mass.
However, as discussed by \citet{curtis_lake16}, rest-UV selected samples do not necessarily also select for DM halos with constant mass or circular velocity.
Therefore fitting this relation only reveals whether the rest-optical light of the UV-selected sample in this work best traces one of these scenarios, if certain assumptions hold.
For example, the disk mass and angular momentum are assumed to be fixed fractions $m_{\mathrm{d}}$ and $j_{\mathrm{d}}$ of the total mass $M$ and total angular momentum $J$, which are dominated by the dark matter halo. 
\citet{Mo98} assume $m_{\mathrm{d}}$ is the same for all disks, and they also largely assume $m_{\mathrm{d}} = j_{\mathrm{d}}$.
\citet{curtis_lake16} discuss results from \citet{Danovich15}, where high-redshift massive galaxies accrete angular momentum via cold gas streams, resulting in a high value of $j_{\mathrm{d}}$ and breaking the assumption that the disk is relaxed. 
This would drive up the value of the spin parameter $\lambda$ for our most massive sources, and this may partially explain the build-up of objects at $R_{e} > 2$ kpc at $z=3$ relative to $z=4-5$.
Prior to this build-up, within the THESAN simulation at $z\gtrsim6$ \citet{Shen23} find that the sizes of massive galaxies agree better with model predictions where $j_{\mathrm{d}} / m_{\mathrm{d}}$ is small, keeping sizes compact at earlier times.
However, this does not explain the shallower redshift evolution we find compared to the predictions. Our value of $R_{e} = 3.51(1+z)^{-0.60\pm0.22}$ is in tension with the fixed circular velocity scenario at the $2\sigma$ level.
\citet{Legrand19} find that for the least massive haloes ($\mathrm{log}_{10}(M_{h}/M_{\sun}) \lesssim 12.5$), the stellar-mass to halo-mass ratio decreases towards higher redshift.
Since we cut at a constant mass over all redshifts, by $z=3$ we may be sampling lower-mass haloes compared to $z=5$. 
Thus in our sample for a given low-mass halo, at $z=3$ the halo is more likely to be occupied by a more massive galaxy than at $z=5$. 
We therefore may be selecting lower-mass haloes towards lower redshift, which could explain the shallower evolution that we find.

We can gain further insight into disk formation via the evolution of the intrinsic scatter in the size-mass relations.
In disk formation models, the intrinsic scatter is related to the distribution of spin parameters \citep{Bullock01}. 
The larger scatter we see at lower redshift may thus reflect that galaxies have had more time to undergo processes that drive them away from the standard disk evolution-based relation, such as mergers. 
This is also higher in the rest-frame UV than the rest-frame optical, suggesting that by $z=5$ the rest-optical stellar populations are well established, but rest-UV emission exhibits a greater diversity in size perhaps caused by gas accretion driving star formation.
This fits well with the build-up of large ($R_{e} > 2$ kpc) galaxies seen in the tail of the log-normal distributions by $z=3$ - mechanisms forming much larger galaxies begin to dominate between $z=3-4$, driving up the intrinsic scatter.
We note that the Universe increases in age by $\sim40\%$ between $z=3-4$, a time step of $\sim600$ Myr.

Our findings of the fraction of irregular galaxies are consistent with an emerging picture of rest-optical morphology via visual classification with \textit{JWST} \citep[see Section \ref{sec:size redshift evolution},][]{kartaltepe23, Ferreira23, Jacobs23}.
Whilst our S\'ersic fitting fails on these objects, the evolution of the irregular fraction provides insight into processes that drive galaxies away from relaxed disks. 
A large increase in the irregular fraction towards higher mass (see Fig. \ref{fig:sample_mass_z}) could reflect a high fraction of merging systems. 
Mergers may be one cause of driving galaxies away from the size-mass relation at $z=3$, breaking the relaxed disk assumption in the formalism of \citet{Mo98} and causing the increase in intrinsic scatter.
Essentially, whilst disky galaxies are well-established by $z=5$ as shown by the well-fit log-normal distributions, by $z=3$ they begin to undergo processes such as mergers which disturb the disks \citep[e.g.][]{Ventou17}, and also push them to larger sizes as seen by the build-up in the tail of the $z=3$ log-normal distribution.

\subsection{Difficulties in size measurements}
\label{sec:size difficulties}

\begin{figure}
    \newcommand{\fitfigwidth}{0.49}
    \centering
    \includegraphics[width=\fitfigwidth\columnwidth]{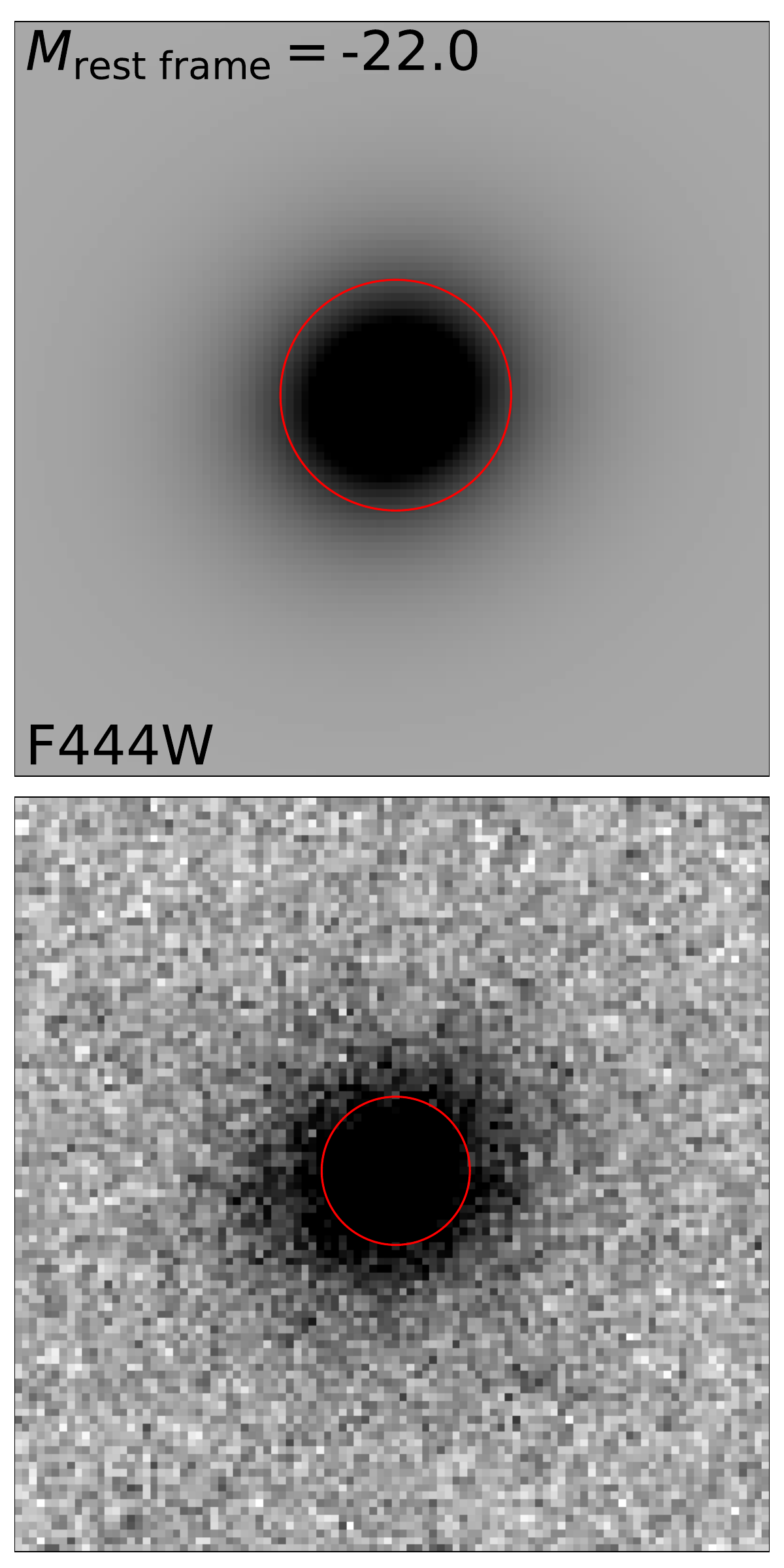}
    \includegraphics[width=\fitfigwidth\columnwidth]{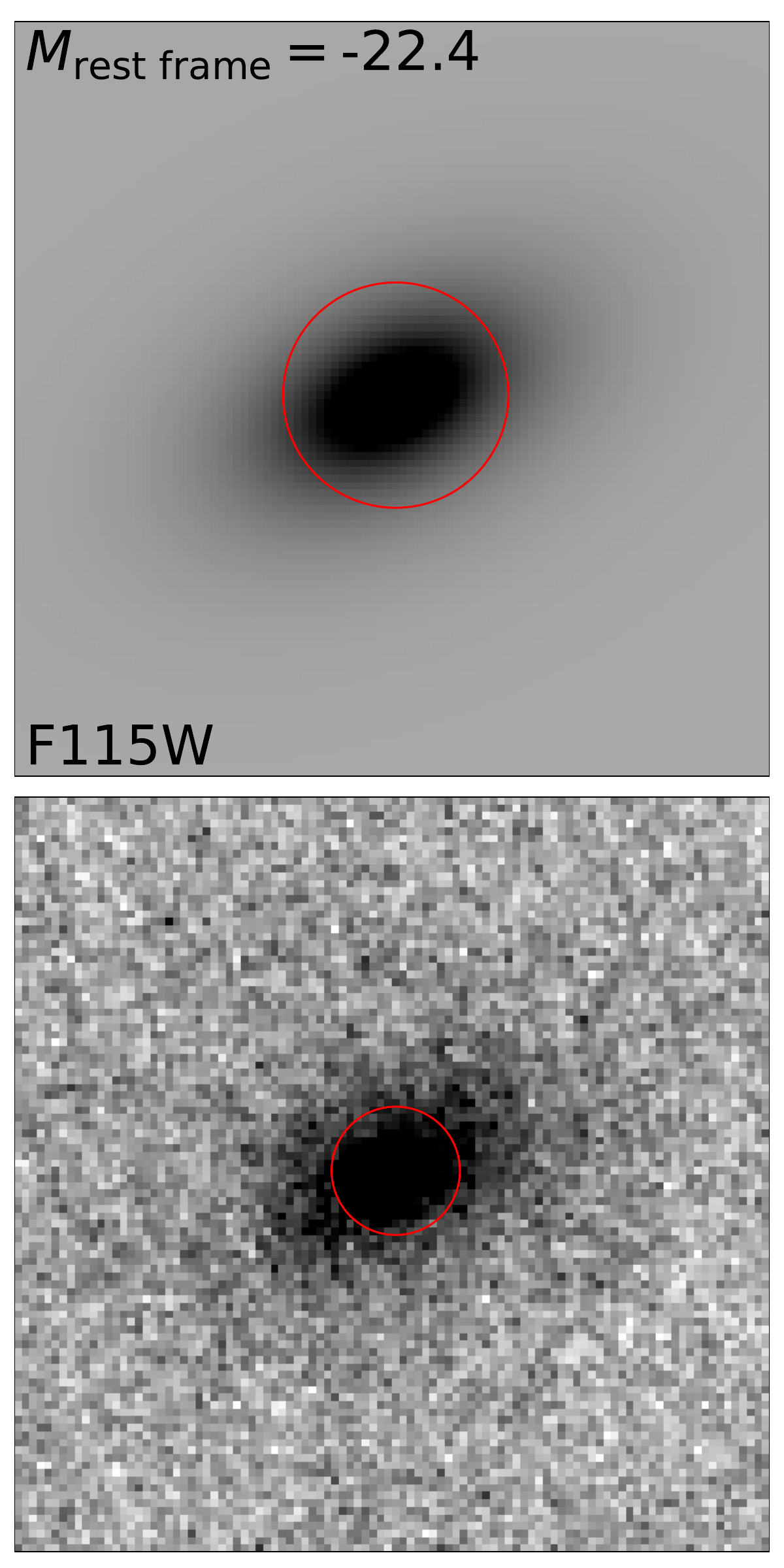}
    \caption{Results of non-parametric fitting on synthetic galaxies in FLARES \citep{Vijayan21}. The top row shows the synthetic galaxy image, and a circular aperture corresponding to the size measured in this noiseless image. We also note the rest-frame absolute magnitude and the filter the object is injected to in the bottom row, where we re-measure the size and plot the corresponding aperture. The modal $5\sigma$ depth of the PRIMER imaging measured in a 0.3" diameter aperture is 27.4 in F115W and 28.5 in F444W.}
    \label{fig:mock galaxy injection}
\end{figure}

Many studies attempt to fit a $r\sim(1+z)^\alpha$ relation to the average or peak size to test different disk formation model scenarios, and agreement has been found with both $r\propto(1+z)^{-1}$ in the case of the galaxy sample following a fixed circular velocity of DM halos, and $r\propto(1+z)^{-1.5}$ for a fixed halo mass. \citep[e.g][]{Hathi08, Oesch10, mosleh10}. 
The most recent determinations with \textit{JWST} are by \citet{Ormerod23}, who use $64 \ \mathrm{arcmin}^{2}$ to measure the size evolution of galaxies across $z=0-8$, and \citet{Ward23} who use $97 \ \mathrm{arcmin}^{2}$ to measure sizes from $z=0.5-5.5$.
Both studies find a weaker evolution in size than predicted by the disk formation models, $\alpha=0.71\pm0.19$ and $\alpha=0.63\pm0.07$.
They agree with our value of $\alpha=0.60\pm0.22$, which would appear to rule out the fixed virial mass scenario.
However, as we have shown in Section \ref{sec:size redshift evolution} and discussed in the previous section, there are several competing effects governing the strength of this evolution.
First, we do see evidence for a larger proportion of galaxies at $R_{e} > 2$ kpc at $z=3$ relative to $z=4-5$. 
Second, from the mass-size and mass-luminosity relations, we generally expect the largest galaxies to be the brightest and most massive, which are the rarest \citep{adams23}. 
Thus due to our limited area of $340 \ \mathrm{arcmin}^{2}$, coupled with an evolving luminosity function, it is difficult to draw conclusions on our value of the exponent (derived from fitting to the full sample) which is strongly affected by the largest galaxies.
Third, the constant stellar mass cut with redshift may result in a non-constant cut in DM halo masses.
Fourth, \citet{Ward23} find $\sim1000$ more galaxies than \citet{Ormerod23} in a narrower redshift range (with fairly similar areas), suggesting large differences in the way these galaxies are selected in \textit{HST} and \textit{JWST} data.
Arguably, our sample avoids issues surrounding the selection function with \textit{JWST} by using ground-selected galaxies. 
For example, \citet{curtis_lake16} discuss size-dependent incompleteness and under-estimation of the sizes of the largest \textit{HST}-selected galaxies in \citet{Shibuya15} caused by the use of high signal-to-noise required in small aperture sizes.
Due to the ground-based imaging being seeing-dominated, the selection of our sample is largely agnostic to size and morphology.
We thus can avoid these biases that may be introduced in a \textit{JWST}-selected sample.
Finally, large scatter in size leads to large uncertainty on our exponent of the size-redshift relation derived using the full sample.
However, based on the build-up of large ($R_{e} > 2$ kpc) galaxies at $z=3$, we may conclude that whilst the typical size of galaxies does not change, the same processes that increase the intrinsic scatter towards decreasing redshift also result in the establishment of a population with large sizes. 
Further constraining their evolution with redshift requires a wide-area study such as \textit{Euclid}.

An additional complication in size measurements arises from the various definitions used in the literature \citep[e.g.][]{Sersic63, Petrosian76, Kron80} as well as multiple non-parametric measurement methods \citep[e.g.][]{Conselice2000, Oesch10, Roper22}.
Comparing like-for-like with other studies and with simulations is thus rendered more difficult.
This motivated our choice of using a parametric \textit{and} non-parametric size fitting method in this work, allowing us to directly compare to current \textit{JWST} studies and simulations.
The S\'ersic profile choice is natural following from disk formation models. 
However, many fits must be discarded due to irregular morphology. 
The non-parametric method deals with this in a better manner, but arguably gives us less insight into disk formation at high-redshift.
These two choices allow us to compare directly to predictions from \citetalias{Roper22} and \citetalias{Constantin23}, who each use different size definitions to make predictions on wavelength dependence of size scaling relations.

\subsection{Comparison with simulations}
\label{sec:comparison with sims}

\begin{figure}
    \centering
    \includegraphics[width=\columnwidth]{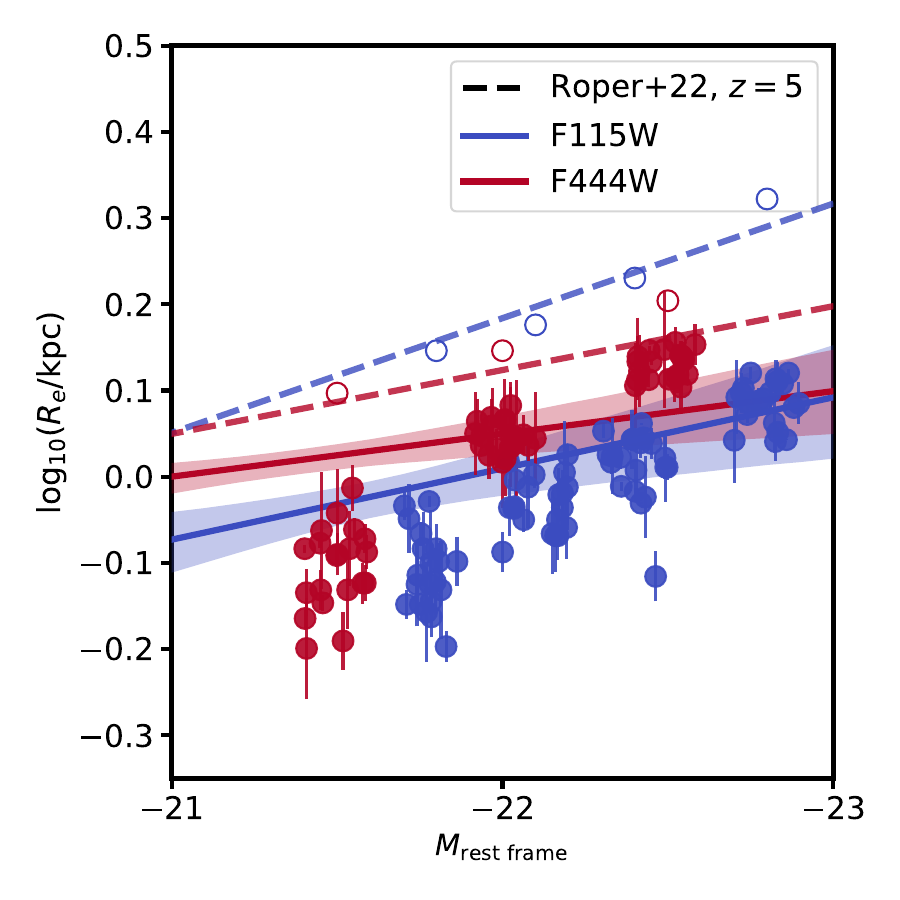}
    \caption{The size-luminosity relations at $z=5$, as shown in Fig. \ref{fig:luminosity_relation} but with a comparison to simulated mock galaxy images. The open circles represent mock galaxies created as in Fig. \ref{fig:mock galaxy injection}, and the filled circles show their size after injection into an empty region of our imaging. We offset the injected sizes in magnitude within 0.2 mag for clarity. The colours of the circles match the filters used.}
    \label{fig:matching sim to obs}
\end{figure}

The two key simulation studies we compare to in this work are that of \citetalias{Constantin23} and \citetalias{Roper22}.
\citetalias{Constantin23} use the illustrisTNG simulation \citep{illustrisTNG} to measure the sizes of $\mathrm{log}_{10}(M_{\star}/M_{\sun}) >9$ galaxies with noise added to match the depth of CEERS \citep{CEERS}. 
\citetalias{Roper22} use the FLARES simulation \citep{FLARES} to predict the size-luminosity relation of massive galaxies at $z=5-10$ as a function of wavelength.
We note that our measurements provide a like-for-like comparison with these two studies, avoiding possible systematics that arise from, for example, comparing half-mass and half-light radii \citep[e.g.][]{Wu20}.
Our size-redshift evolution is slightly weaker than predicted by \citetalias{Constantin23}, and reasons for this have been discussed in Sections \ref{sec:disk formation} and \ref{sec:size difficulties}. 
Looking to the size-mass relations, our results are in good agreement with the predictions, including a possibly negative relation at $z=5$. 
One reason for this is that the depths of CEERS are not too dissimilar to PRIMER. 
In a similar vein, our derived gradients of the size-luminosity match predictions from \citetalias{Roper22}, but our relations are offset to smaller sizes at the $2\sigma$ level. 
In this case, they do not simulate mock images at typical \textit{JWST} depths. 
When using the non-parametric size-fitting method, the addition of noise dominates over the low-surface brightness outer regions of galaxies, reducing the total light measured for the galaxy. 
This leads to a smaller number of pixels used to calculate the effective half-light radius.  
We generate mock galaxies with S\'ersic profiles using \textsc{Synthesizer} \citep{Vijayan21}, which uses the same FLARES simulation as in \citetalias{Roper22}. 
In Fig. \ref{fig:mock galaxy injection} we show the impact of this effect.
We use the same simple stellar population model as \citetalias{Roper22} - v2.2.1 of the Binary Population and Spectral Synthesis (\textsc{BPASS}) stellar population synthesis models \citep{BPASS} and assume a \citet{Chabrier03} initial mass function.
We measure the non-parametric size of the mock galaxy, inject it into empty regions of our F115W and F444W imaging, and then re-measure the size. 
We overplot an aperture corresponding to the measured size. 
The sizes are clearly smaller after injection into our noisy images, and this may explain the bulk of the offset. 
In Fig. \ref{fig:matching sim to obs} we show how seven objects on the \citetalias{Roper22} size-luminosity relations behave after they are each injected 20 times into different empty regions of our images.
We generate mock galaxies at rest-frame absolute magnitudes of $M_{\mathrm{F115W}}=-21.8, -22.1, -22.4, -22.8$ and $M_{\mathrm{F444W}}=-21.5, -22, -22.5$.
The $5\sigma$ depths of the PRIMER imaging measured in a 0.3 arcsec diameter aperture are 27.4 in F115W and 28.5 in F444W.
For brighter objects, the sizes are reduced onto our relations. 
Fainter objects fall below our relation, but noise in the images causes large scatter in the derived size since the depth of the imaging varies over the COSMOS and UDS fields.
Additionally, there is a large difference in depth between F115W and F444W, so the impact noise has on measured size is stronger in F115W. 
This explains slightly larger offsets seen in F115W compared to F444W between the clean mock galaxy sizes and the sizes after injection into noisy images, and suggests an ideal depth of $\sim28$ for such analyses, where our faintest sources are not washed out completely by the noise.
This simple simulation has highlighted that direct comparisons between observational data, influenced by background noise, and clean simulated galaxies are difficult, limiting what can be learned about the physics governing size scaling relations.
Comparisons to simulations where \textit{JWST}-like noise, PSFs and pixel scales have been applied may enable more reliable comparisons to be undertaken.

\subsection{Compact central star formation?}

Considering our tentative agreement with the simulation works of \citetalias{Roper22} (once instrumental noise is accounted for) and \citetalias{Constantin23} (a possible negative size-mass relation at $z=5$), it is natural to ask: what physical mechanisms drive the evolution of these simulated galaxies? 
To explain the negative size-mass relations at $z=5$ and steepening size-luminosity slopes with decreasing wavelength, both studies appeal to compact central star formation in the most massive galaxies. 
Specifically, \citet{Roper22, Roper23} find that the most massive galaxies at $z\gtrsim5$ develop dense cores leading to the negative slopes and to higher star formation. 
This enriches the cores of the galaxies, allows for metal-line cooling and inhibiting stellar/AGN feedback, driving further star formation, akin to the beginnings of inside-out galaxy formation and quenching \citep[e.g.][]{Pichon11, Nelson16, Baker23, Shen23}.
It is difficult to ascertain from our limited sample at $z=5$ whether there is a true negative relation. 
However, we note that the non-parametric size-luminosity relations show more compact sizes in the rest-UV compared to the rest optical.
This is not seen in the S\'ersic sizes, but as discussed in Section \ref{sec:size-luminosity} the blue S\'ersic sizes could be slightly over-estimated due to a larger profile being fit to more clumpy morphologies.
Further studies constraining the size-luminosity relations at higher redshift and allow for a wider dynamic range in mass and luminosity will provide stronger tests for these scenarios.
We note that other physical mechanisms may affect the sizes seen in \citetalias{Roper22}. 
For example, the H$\alpha$ emission line crosses into F444W at $z=5$ \citep{Davis23} and approximately traces star formation.
If not accounted for, this will act to decrease $R_{e}$.
Additionally, the manner in which dust is modelled in these simulations is important. 
\citetalias{Roper22} use a dust-to-metal ratio to calculate line-of-sight attenuation. 
They do not directly simulate dust production and destruction, instead using proxies from observed UV luminosity functions.
Direct simulation of dust production/creation, such as with \textsc{THESAN} \citep{THESAN22}, may be required if dust is the primary driver behind these size scaling relations.
The agreement of slopes with \citetalias{Roper22} is encouraging, but stronger differences between size-luminosity gradients with wavelength are predicted at $z>5$ which remain to be tested with observations.
Despite this, we have provided strong initial tests of predictions of the size-mass and size-luminosity relations by using size measurements that enable direct comparisons, providing tentative evidence for centrally compact star formation at high-redshift.

\section{Conclusions}
\label{sec:conclusions}

We present size measurements of 1668 rest-frame UV selected galaxies with stellar masses $\mathrm{log}_{10}(M_{\star}/M_{\sun}) > 9$ using data from PRIMER.
These sources were selected from the ground-based study of \citet{adams23}.
The sample was selected from seeing-dominated data, and hence presents an unbiased sampling of the morphology and size distributions of luminous sources.
Additionally, the wavelength coverage of \textit{JWST} NIRCam allows us to measure rest-optical sizes at $z>3$, previously not possible with \textit{HST}. 
We measure S\'ersic and non-parametric sizes to test models of disk formation, measure the size-mass and size-luminosity relations, and compare directly to predictions of these relations from simulations.

\begin{itemize}
    \item Galaxy sizes in F356W follow a log-normal distribution in both S\'ersic and non-parametric sizes, following predictions from \citet{Fall80} and \citet{Bullock01}, implying disk formation as early as $z=5$. 
    Whilst this has been seen in rest-UV morphologies out to $z=6$ with \textit{HST}, we confirm this result for rest-optical morphologies at $z=4-5$.
    \item We find evidence for the build-up of large ($R_{e} > 2$ kpc) galaxies in the tail of the log-normal distribution at $z=3$, implying a transition from compact to extended sizes from $z\simeq3-4$ for rarer objects. This is supported by a size-redshift evolution of $R_{e} = 3.51(1+z)^{-0.60\pm0.22}$ kpc derived from fitting to the full sample, consistent with other \textit{JWST} findings \citep{Ormerod23, Ward23}.  
    \item Simultaneously, following \citet{curtis_lake16}, if we fix the exponent of the evolution to $-0.20$, we find a size-redshift relation consistent with the peaks of our log-normal distributions. 
    This suggests that the size of a \emph{typical} galaxy does not evolve significantly over $z=3-5$, remaining compact with $R_{e}\simeq0.8-0.9$ kpc.
    \item We measure S\'ersic size-mass relations in F200W and F356W and find agreement with predictions from \citetalias{Constantin23} who use Illustris TNG50 to predict the size-mass relation in CEERS. 
    We find the intrinsic scatter is larger in F356W than in F200W in each redshift bin, and the intrinsic scatter is larger at lower redshift. 
    We speculate that this is because galaxies at $z=3$ have had more time to undergo processes such as mergers and gas accretion that drive them away from the relation.
    \item The gradients of our non-parametric size-luminosity relations agree with predictions from the FLARES simulation of \citetalias{Roper22}, but with an offset that we believe is driven by larger sizes measured for the simulations when instrumental noise is unaccounted for.
    \item Simulations suggest the scaling relations are driven by compact central star formation at high-redshift, and our results provide tentative evidence for this mechanism. Further observations that improve the dynamic range of this sample are needed to place stronger constraints on interesting cases such as negative size-mass relations and more compact size-luminosity relations in the rest-UV compared to the rest-optical. 


Additional results from PRIMER (Allen et al. in prep.) will produce a multi-wavelength view of galaxy sizes over a wide mass range, allowing us to understand size evolution in a diverse population.
Furthermore, a similar analysis in a larger volume will allow for stronger constraints on the emergence of large ($R_{e} > 2$kpc) galaxies at $z\simeq3-4$. 
Just as degree-scale ground-based imaging revolutionized constraints on high-redshift number statistics such as the bright-end of luminosity functions \citep[e.g.][]{Bowler14, stefanon19, harikane22}, missions such as \textit{Euclid}, which will achieve \textit{HST} resolution over degree-scale fields \citep{Euclid22}, will revolutionize our understanding of resolved properties of the very same galaxies.
\end{itemize}

\section*{Acknowledgements}
RGV would like to thank William Roper and Christopher Lovell for helpful discussion regarding the generation of mock galaxies with FLARES, and James Nightingale for helpful discussion regarding \textsc{PyAutoGalaxy}.

RGV acknowledges funding from the Science and Technology Facilities Council (STFC) [grant code ST/W507726/1]. 
RAAB acknowledges support from an STFC Ernest Rutherford Fellowship [grant number ST/T003596/1]. 
MJJ acknowledges support of the STFC consolidated grant [ST/S000488/1] and
[ST/W000903/1] and from a UKRI Frontiers Research Grant [EP/X026639/1]. MJJ also acknowledges support from
the Oxford Hintze Centre for Astrophysical Surveys which is funded
through generous support from the Hintze Family Charitable Foundation. 
NJA acknowledges support from the  European Research Council (ERC) Advanced Investigator Grant EPOCHS (788113).
ACC thanks the Leverhulme Trust for their support via the Leverhulme Early Career Fellowship scheme.
NC, DJM, JSD and CTD acknowledge the support of the Science and Technology Facilities Council.
JSD acknowledges the support of the Royal Society through a Royal Society Research Professorship

This work is based on observations made with the NASA/ESA/CSA James Webb Space Telescope. The data were obtained from the Mikulski Archive for Space Telescopes at the Space Telescope Science Institute, which is operated by the Association of Universities for Research in Astronomy, Inc., under NASA contract NAS 5-03127 for JWST. These observations are associated with program \#1837. The authors acknowledge the PRIMER team for developing their observing program with a zero-exclusive-access period.

This work is based on data products from observations made with ESO Telescopes at the La Silla Paranal Observatory under ESO programme ID 179.A-2005 and ID 179.A-2006 and on data products produced by CALET and the Cambridge Astronomy Survey Unit on behalf of the UltraVISTA and VIDEO consortia.

The Hyper Suprime-Cam (HSC) collaboration includes the astronomical communities of Japan and Taiwan, and Princeton University. The HSC instrumentation and software were developed by the National Astronomical Observatory of Japan (NAOJ), the Kavli Institute for the Physics and Mathematics of the Universe (Kavli IPMU), the University of Tokyo, the High Energy Accelerator Research Organization (KEK), the Academia Sinica Institute for Astronomy and Astrophysics in Taiwan (ASIAA), and Princeton University. Funding was contributed by the FIRST program from the Japanese Cabinet Office, the Ministry of Education, Culture, Sports, Science and Technology (MEXT), the Japan Society for the Promotion of Science (JSPS), Japan Science and Technology Agency (JST), the Toray Science Foundation, NAOJ, Kavli IPMU, KEK, ASIAA, and Princeton University.

This paper is based on data collected at the Subaru Telescope and retrieved from the HSC data archive system, which is operated by the Subaru Telescope and Astronomy Data Center (ADC) at NAOJ. Data analysis was in part carried out with the cooperation of Center for Computational Astrophysics (CfCA), NAOJ.

This work made use of \textsc{dynesty} \citep{Speagle20, koposov23}, \textsc{numpyro} \citep{numpyro1, numpyro2}, \textsc{PyAutoFit} \citep{pyautofit}, and \textsc{Astropy}:\footnote{http://www.astropy.org}, a community-developed core Python package and an ecosystem of tools and resources for astronomy \citep{astropy:2013, astropy:2018, astropy:2022}.

\section*{Data Availability}

All imaging data was obtained from original sources in the public
domain. The catalogues used in this study are published through \citet{adams23}.



\bibliographystyle{mnras}
\bibliography{mnras_template} 

\begin{thebibliography}{}
\makeatletter
\relax
\def\mn@urlcharsother{\let\do\@makeother \do\$\do\&\do\#\do\^\do\_\do\%\do\~}
\def\mn@doi{\begingroup\mn@urlcharsother \@ifnextchar [ {\mn@doi@} {\mn@doi@[]}}
\def\mn@doi@[#1]#2{\def\@tempa{#1}\ifx\@tempa\@empty \href {http://dx.doi.org/#2} {doi:#2}\else \href {http://dx.doi.org/#2} {#1}\fi \endgroup}
\def\mn@eprint#1#2{\mn@eprint@#1:#2::\@nil}
\def\mn@eprint@arXiv#1{\href {http://arxiv.org/abs/#1} {{\tt arXiv:#1}}}
\def\mn@eprint@dblp#1{\href {http://dblp.uni-trier.de/rec/bibtex/#1.xml} {dblp:#1}}
\def\mn@eprint@#1:#2:#3:#4\@nil{\def\@tempa {#1}\def\@tempb {#2}\def\@tempc {#3}\ifx \@tempc \@empty \let \@tempc \@tempb \let \@tempb \@tempa \fi \ifx \@tempb \@empty \def\@tempb {arXiv}\fi \@ifundefined {mn@eprint@\@tempb}{\@tempb:\@tempc}{\expandafter \expandafter \csname mn@eprint@\@tempb\endcsname \expandafter{\@tempc}}}

\bibitem[\protect\citeauthoryear{{Adams} et~al.,}{{Adams} et~al.}{2023a}]{Adams23_epochs}
{Adams} N.~J.,  et~al., 2023a, \mn@doi [arXiv e-prints] {10.48550/arXiv.2304.13721}, \href {https://ui.adsabs.harvard.edu/abs/2023arXiv230413721A} {p. arXiv:2304.13721}

\bibitem[\protect\citeauthoryear{{Adams} et~al.,}{{Adams} et~al.}{2023b}]{Adams23_prop}
{Adams} N.~J.,  et~al., 2023b, \mn@doi [\mnras] {10.1093/mnras/stac3347}, \href {https://ui.adsabs.harvard.edu/abs/2023MNRAS.518.4755A} {518, 4755}

\bibitem[\protect\citeauthoryear{{Adams}, {Bowler}, {Jarvis}, {Varadaraj}  \& {H{\"a}u{\ss}ler}}{{Adams} et~al.}{2023c}]{adams23}
{Adams} N.~J.,  {Bowler} R.~A.~A.,  {Jarvis} M.~J.,  {Varadaraj} R.~G.,   {H{\"a}u{\ss}ler} B.,  2023c, \mn@doi [\mnras] {10.1093/mnras/stad1333}, \href {https://ui.adsabs.harvard.edu/abs/2023MNRAS.523..327A} {523, 327}

\bibitem[\protect\citeauthoryear{{Adelberger}, {Steidel}, {Pettini}, {Shapley}, {Reddy}  \& {Erb}}{{Adelberger} et~al.}{2005}]{adelberger05}
{Adelberger} K.~L.,  {Steidel} C.~C.,  {Pettini} M.,  {Shapley} A.~E.,  {Reddy} N.~A.,   {Erb} D.~K.,  2005, \mn@doi [\apj] {10.1086/426580}, \href {https://ui.adsabs.harvard.edu/abs/2005ApJ...619..697A} {619, 697}

\bibitem[\protect\citeauthoryear{{Aihara} et~al.,}{{Aihara} et~al.}{2019}]{HSCSSP_DR2}
{Aihara} H.,  et~al., 2019, \mn@doi [\pasj] {10.1093/pasj/psz103}, \href {https://ui.adsabs.harvard.edu/abs/2019PASJ...71..114A} {71, 114}

\bibitem[\protect\citeauthoryear{{Aihara} et~al.,}{{Aihara} et~al.}{2022}]{AiharaDR3}
{Aihara} H.,  et~al., 2022, \mn@doi [\pasj] {10.1093/pasj/psab122}, \href {https://ui.adsabs.harvard.edu/abs/2022PASJ...74..247A} {74, 247}

\bibitem[\protect\citeauthoryear{{Arnouts}, {Cristiani}, {Moscardini}, {Matarrese}, {Lucchin}, {Fontana}  \& {Giallongo}}{{Arnouts} et~al.}{1999}]{arnouts99}
{Arnouts} S.,  {Cristiani} S.,  {Moscardini} L.,  {Matarrese} S.,  {Lucchin} F.,  {Fontana} A.,   {Giallongo} E.,  1999, \mn@doi [\mnras] {10.1046/j.1365-8711.1999.02978.x}, \href {https://ui.adsabs.harvard.edu/abs/1999MNRAS.310..540A} {310, 540}

\bibitem[\protect\citeauthoryear{{Astropy Collaboration} et~al.,}{{Astropy Collaboration} et~al.}{2013}]{astropy:2013}
{Astropy Collaboration} et~al., 2013, \mn@doi [\aap] {10.1051/0004-6361/201322068}, \href {http://adsabs.harvard.edu/abs/2013A%26A...558A..33A} {558, A33}

\bibitem[\protect\citeauthoryear{{Astropy Collaboration} et~al.,}{{Astropy Collaboration} et~al.}{2018}]{astropy:2018}
{Astropy Collaboration} et~al., 2018, \mn@doi [\aj] {10.3847/1538-3881/aabc4f}, \href {https://ui.adsabs.harvard.edu/abs/2018AJ....156..123A} {156, 123}

\bibitem[\protect\citeauthoryear{{Astropy Collaboration} et~al.,}{{Astropy Collaboration} et~al.}{2022}]{astropy:2022}
{Astropy Collaboration} et~al., 2022, \mn@doi [\apj] {10.3847/1538-4357/ac7c74}, \href {https://ui.adsabs.harvard.edu/abs/2022ApJ...935..167A} {935, 167}

\bibitem[\protect\citeauthoryear{{Baker} et~al.,}{{Baker} et~al.}{2023}]{Baker23}
{Baker} W.~M.,  et~al., 2023, \mn@doi [arXiv e-prints] {10.48550/arXiv.2306.02472}, \href {https://ui.adsabs.harvard.edu/abs/2023arXiv230602472B} {p. arXiv:2306.02472}

\bibitem[\protect\citeauthoryear{{Bartlett} \& {Desmond}}{{Bartlett} \& {Desmond}}{2023}]{roxy}
{Bartlett} D.~J.,  {Desmond} H.,  2023, \mn@doi [The Open Journal of Astrophysics] {10.21105/astro.2309.00948}, \href {https://ui.adsabs.harvard.edu/abs/2023OJAp....6E..42B} {6, 42}

\bibitem[\protect\citeauthoryear{{Bertin}}{{Bertin}}{2011}]{psfex}
{Bertin} E.,  2011, in {Evans} I.~N.,  {Accomazzi} A.,  {Mink} D.~J.,   {Rots} A.~H.,  eds,  Astronomical Society of the Pacific Conference Series Vol. 442, Astronomical Data Analysis Software and Systems XX. p.~435

\bibitem[\protect\citeauthoryear{{Bertin} \& {Arnouts}}{{Bertin} \& {Arnouts}}{1996}]{sextractor}
{Bertin} E.,  {Arnouts} S.,  1996, \mn@doi [A\&AS] {10.1051/aas:1996164}, \href {https://ui.adsabs.harvard.edu/abs/1996A&AS..117..393B} {117, 393}

\bibitem[\protect\citeauthoryear{{Bian} et~al.,}{{Bian} et~al.}{2013}]{bian13}
{Bian} F.,  et~al., 2013, \mn@doi [\apj] {10.1088/0004-637X/774/1/28}, \href {https://ui.adsabs.harvard.edu/abs/2013ApJ...774...28B} {774, 28}

\bibitem[\protect\citeauthoryear{Bingham et~al.,}{Bingham et~al.}{2019}]{numpyro2}
Bingham E.,  et~al., 2019, J. Mach. Learn. Res., 20, 28:1

\bibitem[\protect\citeauthoryear{{Bouwens} et~al.,}{{Bouwens} et~al.}{2021}]{bouwens21}
{Bouwens} R.~J.,  et~al., 2021, \mn@doi [AJ] {10.3847/1538-3881/abf83e}, \href {https://ui.adsabs.harvard.edu/abs/2021AJ....162...47B} {162, 47}

\bibitem[\protect\citeauthoryear{{Bouwens}, {Illingworth}, {van Dokkum}, {Oesch}, {Stefanon}  \& {Ribeiro}}{{Bouwens} et~al.}{2022}]{Bouwens22}
{Bouwens} R.~J.,  {Illingworth} G.~D.,  {van Dokkum} P.~G.,  {Oesch} P.~A.,  {Stefanon} M.,   {Ribeiro} B.,  2022, \mn@doi [\apj] {10.3847/1538-4357/ac4791}, \href {https://ui.adsabs.harvard.edu/abs/2022ApJ...927...81B} {927, 81}

\bibitem[\protect\citeauthoryear{Bowler et~al.,}{Bowler et~al.}{2014}]{Bowler14}
Bowler R. A.~A.,  et~al., 2014, \mn@doi [\mnras] {10.1093/mnras/stu449}, 440, 2810

\bibitem[\protect\citeauthoryear{{Bowler}, {Dunlop}, {McLure}  \& {McLeod}}{{Bowler} et~al.}{2017}]{Bowler17}
{Bowler} R.~A.~A.,  {Dunlop} J.~S.,  {McLure} R.~J.,   {McLeod} D.~J.,  2017, \mn@doi [\mnras] {10.1093/mnras/stw3296}, \href {https://ui.adsabs.harvard.edu/abs/2017MNRAS.466.3612B} {466, 3612}

\bibitem[\protect\citeauthoryear{Bowler, Jarvis, Dunlop, McLure, McLeod, Adams, Milvang-Jensen  \& McCracken}{Bowler et~al.}{2020}]{Bowler20}
Bowler R. A.~A.,  Jarvis M.~J.,  Dunlop J.~S.,  McLure R.~J.,  McLeod D.~J.,  Adams N.~J.,  Milvang-Jensen B.,   McCracken H.~J.,  2020, \mn@doi [MNRAS] {10.1093/mnras/staa313}, 493, 2059

\bibitem[\protect\citeauthoryear{Bradley}{Bradley}{2023}]{photutils}
Bradley L.,  2023, astropy/photutils: 1.8.0, \mn@doi{10.5281/zenodo.7946442}, \url {https://doi.org/10.5281/zenodo.7946442}

\bibitem[\protect\citeauthoryear{{Bullock}, {Dekel}, {Kolatt}, {Kravtsov}, {Klypin}, {Porciani}  \& {Primack}}{{Bullock} et~al.}{2001}]{Bullock01}
{Bullock} J.~S.,  {Dekel} A.,  {Kolatt} T.~S.,  {Kravtsov} A.~V.,  {Klypin} A.~A.,  {Porciani} C.,   {Primack} J.~R.,  2001, \mn@doi [\apj] {10.1086/321477}, \href {https://ui.adsabs.harvard.edu/abs/2001ApJ...555..240B} {555, 240}

\bibitem[\protect\citeauthoryear{{Bushouse} et~al.,}{{Bushouse} et~al.}{2023}]{Bushouse23}
{Bushouse} H.,  et~al., 2023, {JWST Calibration Pipeline}, \mn@doi{10.5281/zenodo.7577320}

\bibitem[\protect\citeauthoryear{{Calzetti}, {Armus}, {Bohlin}, {Kinney}, {Koornneef}  \& {Storchi-Bergmann}}{{Calzetti} et~al.}{2000}]{calzetti}
{Calzetti} D.,  {Armus} L.,  {Bohlin} R.~C.,  {Kinney} A.~L.,  {Koornneef} J.,   {Storchi-Bergmann} T.,  2000, \mn@doi [ApJ] {10.1086/308692}, \href {https://ui.adsabs.harvard.edu/abs/2000ApJ...533..682C} {533, 682}

\bibitem[\protect\citeauthoryear{{Carnall}, {McLure}, {Dunlop}  \& {Dav{\'e}}}{{Carnall} et~al.}{2018}]{BAGPIPES}
{Carnall} A.~C.,  {McLure} R.~J.,  {Dunlop} J.~S.,   {Dav{\'e}} R.,  2018, \mn@doi [\mnras] {10.1093/mnras/sty2169}, \href {https://ui.adsabs.harvard.edu/abs/2018MNRAS.480.4379C} {480, 4379}

\bibitem[\protect\citeauthoryear{{Carnall} et~al.,}{{Carnall} et~al.}{2023a}]{Carnall23}
{Carnall} A.~C.,  et~al., 2023a, \mn@doi [\mnras] {10.1093/mnras/stad369}, \href {https://ui.adsabs.harvard.edu/abs/2023MNRAS.520.3974C} {520, 3974}

\bibitem[\protect\citeauthoryear{{Carnall} et~al.,}{{Carnall} et~al.}{2023b}]{Carnall23Nature}
{Carnall} A.~C.,  et~al., 2023b, \mn@doi [\nat] {10.1038/s41586-023-06158-6}, \href {https://ui.adsabs.harvard.edu/abs/2023Natur.619..716C} {619, 716}

\bibitem[\protect\citeauthoryear{{Chabrier}}{{Chabrier}}{2003}]{Chabrier03}
{Chabrier} G.,  2003, \mn@doi [\pasp] {10.1086/376392}, \href {https://ui.adsabs.harvard.edu/abs/2003PASP..115..763C} {115, 763}

\bibitem[\protect\citeauthoryear{{Conselice}}{{Conselice}}{2014}]{Conselice14}
{Conselice} C.~J.,  2014, \mn@doi [\araa] {10.1146/annurev-astro-081913-040037}, \href {https://ui.adsabs.harvard.edu/abs/2014ARA&A..52..291C} {52, 291}

\bibitem[\protect\citeauthoryear{{Conselice}, {Bershady}  \& {Jangren}}{{Conselice} et~al.}{2000}]{Conselice2000}
{Conselice} C.~J.,  {Bershady} M.~A.,   {Jangren} A.,  2000, \mn@doi [\apj] {10.1086/308300}, \href {https://ui.adsabs.harvard.edu/abs/2000ApJ...529..886C} {529, 886}

\bibitem[\protect\citeauthoryear{{Costantin} et~al.,}{{Costantin} et~al.}{2023}]{Constantin23}
{Costantin} L.,  et~al., 2023, \mn@doi [\apj] {10.3847/1538-4357/acb926}, \href {https://ui.adsabs.harvard.edu/abs/2023ApJ...946...71C} {946, 71}

\bibitem[\protect\citeauthoryear{{Cowie}, {Songaila}, {Hu}  \& {Cohen}}{{Cowie} et~al.}{1996}]{Cowie96}
{Cowie} L.~L.,  {Songaila} A.,  {Hu} E.~M.,   {Cohen} J.~G.,  1996, \mn@doi [\aj] {10.1086/118058}, \href {https://ui.adsabs.harvard.edu/abs/1996AJ....112..839C} {112, 839}

\bibitem[\protect\citeauthoryear{{Cuillandre} et~al.,}{{Cuillandre} et~al.}{2012}]{CFHTLS}
{Cuillandre} J.-C.~J.,  et~al., 2012, in {Peck} A.~B.,  {Seaman} R.~L.,   {Comeron} F.,  eds,  Society of Photo-Optical Instrumentation Engineers (SPIE) Conference Series Vol. 8448, Observatory Operations: Strategies, Processes, and Systems IV. p. 84480M, \mn@doi{10.1117/12.925584}

\bibitem[\protect\citeauthoryear{{Curtis-Lake} et~al.,}{{Curtis-Lake} et~al.}{2016}]{curtis_lake16}
{Curtis-Lake} E.,  et~al., 2016, \mn@doi [\mnras] {10.1093/mnras/stv3017}, \href {https://ui.adsabs.harvard.edu/abs/2016MNRAS.457..440C} {457, 440}

\bibitem[\protect\citeauthoryear{{Danovich}, {Dekel}, {Hahn}, {Ceverino}  \& {Primack}}{{Danovich} et~al.}{2015}]{Danovich15}
{Danovich} M.,  {Dekel} A.,  {Hahn} O.,  {Ceverino} D.,   {Primack} J.,  2015, \mn@doi [\mnras] {10.1093/mnras/stv270}, \href {https://ui.adsabs.harvard.edu/abs/2015MNRAS.449.2087D} {449, 2087}

\bibitem[\protect\citeauthoryear{{Davis} et~al.,}{{Davis} et~al.}{2023}]{Davis23}
{Davis} K.,  et~al., 2023, \mn@doi [arXiv e-prints] {10.48550/arXiv.2312.07799}, \href {https://ui.adsabs.harvard.edu/abs/2023arXiv231207799D} {p. arXiv:2312.07799}

\bibitem[\protect\citeauthoryear{{Dekel}, {Sari}  \& {Ceverino}}{{Dekel} et~al.}{2009}]{Dekel09}
{Dekel} A.,  {Sari} R.,   {Ceverino} D.,  2009, \mn@doi [\apj] {10.1088/0004-637X/703/1/785}, \href {https://ui.adsabs.harvard.edu/abs/2009ApJ...703..785D} {703, 785}

\bibitem[\protect\citeauthoryear{{Ding}, {Silverman}  \& {Onoue}}{{Ding} et~al.}{2022}]{Ding22}
{Ding} X.,  {Silverman} J.~D.,   {Onoue} M.,  2022, \mn@doi [ApJL] {10.3847/2041-8213/ac9c02}, \href {https://ui.adsabs.harvard.edu/abs/2022ApJ...939L..28D} {939, L28}

\bibitem[\protect\citeauthoryear{{Donnan} et~al.,}{{Donnan} et~al.}{2023}]{donnan22}
{Donnan} C.~T.,  et~al., 2023, \mn@doi [\mnras] {10.1093/mnras/stac3472}, \href {https://ui.adsabs.harvard.edu/abs/2023MNRAS.518.6011D} {518, 6011}

\bibitem[\protect\citeauthoryear{{Dunlop} et~al.,}{{Dunlop} et~al.}{2021}]{PRIMER}
{Dunlop} J.~S.,  et~al., 2021, {PRIMER: Public Release IMaging for Extragalactic Research}, JWST Proposal. Cycle 1, ID. \#1837

\bibitem[\protect\citeauthoryear{{Durkalec} et~al.,}{{Durkalec} et~al.}{2015}]{Durkalec15}
{Durkalec} A.,  et~al., 2015, \mn@doi [\aap] {10.1051/0004-6361/201425343}, \href {https://ui.adsabs.harvard.edu/abs/2015A&A...583A.128D} {583, A128}

\bibitem[\protect\citeauthoryear{{Euclid Collaboration} et~al.,}{{Euclid Collaboration} et~al.}{2022}]{Euclid22}
{Euclid Collaboration} et~al., 2022, \mn@doi [\aap] {10.1051/0004-6361/202141938}, \href {https://ui.adsabs.harvard.edu/abs/2022A&A...662A.112E} {662, A112}

\bibitem[\protect\citeauthoryear{{Fall} \& {Efstathiou}}{{Fall} \& {Efstathiou}}{1980}]{Fall80}
{Fall} S.~M.,  {Efstathiou} G.,  1980, \mn@doi [\mnras] {10.1093/mnras/193.2.189}, \href {https://ui.adsabs.harvard.edu/abs/1980MNRAS.193..189F} {193, 189}

\bibitem[\protect\citeauthoryear{{Ferguson} et~al.,}{{Ferguson} et~al.}{2004}]{Ferguson04}
{Ferguson} H.~C.,  et~al., 2004, \mn@doi [\apjl] {10.1086/378578}, \href {https://ui.adsabs.harvard.edu/abs/2004ApJ...600L.107F} {600, L107}

\bibitem[\protect\citeauthoryear{{Ferreira} et~al.,}{{Ferreira} et~al.}{2023}]{Ferreira23}
{Ferreira} L.,  et~al., 2023, \mn@doi [\apj] {10.3847/1538-4357/acec76}, \href {https://ui.adsabs.harvard.edu/abs/2023ApJ...955...94F} {955, 94}

\bibitem[\protect\citeauthoryear{{Finkelstein} et~al.,}{{Finkelstein} et~al.}{2017}]{CEERS}
{Finkelstein} S.~L.,  et~al., 2017, {The Cosmic Evolution Early Release Science (CEERS) Survey}, JWST Proposal ID 1345. Cycle 0 Early Release Science

\bibitem[\protect\citeauthoryear{{Gaia Collaboration} et~al.,}{{Gaia Collaboration} et~al.}{2023}]{GaiaDR3}
{Gaia Collaboration} et~al., 2023, \mn@doi [\aap] {10.1051/0004-6361/202243940}, \href {https://ui.adsabs.harvard.edu/abs/2023A&A...674A...1G} {674, A1}

\bibitem[\protect\citeauthoryear{{Garc{\'\i}a-Argum{\'a}nez} et~al.,}{{Garc{\'\i}a-Argum{\'a}nez} et~al.}{2023}]{Garcia-Argumanez23}
{Garc{\'\i}a-Argum{\'a}nez} {\'A}.,  et~al., 2023, \mn@doi [\apj] {10.3847/1538-4357/aca8ff}, \href {https://ui.adsabs.harvard.edu/abs/2023ApJ...944....3G} {944, 3}

\bibitem[\protect\citeauthoryear{{Grazian} et~al.,}{{Grazian} et~al.}{2012}]{Grazian12}
{Grazian} A.,  et~al., 2012, \mn@doi [\aap] {10.1051/0004-6361/201219669}, \href {https://ui.adsabs.harvard.edu/abs/2012A&A...547A..51G} {547, A51}

\bibitem[\protect\citeauthoryear{{Grogin} et~al.,}{{Grogin} et~al.}{2011}]{Candels1}
{Grogin} N.~A.,  et~al., 2011, \mn@doi [\apjs] {10.1088/0067-0049/197/2/35}, \href {https://ui.adsabs.harvard.edu/abs/2011ApJS..197...35G} {197, 35}

\bibitem[\protect\citeauthoryear{{Harikane} et~al.,}{{Harikane} et~al.}{2022}]{harikane22}
{Harikane} Y.,  et~al., 2022, \mn@doi [ApJS] {10.3847/1538-4365/ac3dfc}, \href {https://ui.adsabs.harvard.edu/abs/2022ApJS..259...20H} {259, 20}

\bibitem[\protect\citeauthoryear{{Hatfield}, {Bowler}, {Jarvis}  \& {Hale}}{{Hatfield} et~al.}{2018}]{Hatfield18}
{Hatfield} P.~W.,  {Bowler} R.~A.~A.,  {Jarvis} M.~J.,   {Hale} C.~L.,  2018, \mn@doi [\mnras] {10.1093/mnras/sty856}, \href {https://ui.adsabs.harvard.edu/abs/2018MNRAS.477.3760H} {477, 3760}

\bibitem[\protect\citeauthoryear{{Hathi}, {Malhotra}  \& {Rhoads}}{{Hathi} et~al.}{2008}]{Hathi08}
{Hathi} N.~P.,  {Malhotra} S.,   {Rhoads} J.~E.,  2008, \mn@doi [\apj] {10.1086/524836}, \href {https://ui.adsabs.harvard.edu/abs/2008ApJ...673..686H} {673, 686}

\bibitem[\protect\citeauthoryear{{Holwerda}, {Bouwens}, {Oesch}, {Smit}, {Illingworth}  \& {Labbe}}{{Holwerda} et~al.}{2015}]{Holwerda15}
{Holwerda} B.~W.,  {Bouwens} R.,  {Oesch} P.,  {Smit} R.,  {Illingworth} G.,   {Labbe} I.,  2015, \mn@doi [\apj] {10.1088/0004-637X/808/1/6}, \href {https://ui.adsabs.harvard.edu/abs/2015ApJ...808....6H} {808, 6}

\bibitem[\protect\citeauthoryear{{Huang}, {Ferguson}, {Ravindranath}  \& {Su}}{{Huang} et~al.}{2013}]{huang13}
{Huang} K.-H.,  {Ferguson} H.~C.,  {Ravindranath} S.,   {Su} J.,  2013, \mn@doi [\apj] {10.1088/0004-637X/765/1/68}, \href {https://ui.adsabs.harvard.edu/abs/2013ApJ...765...68H} {765, 68}

\bibitem[\protect\citeauthoryear{{Hubble}}{{Hubble}}{1926}]{Hubble1926}
{Hubble} E.~P.,  1926, \mn@doi [\apj] {10.1086/143018}, \href {https://ui.adsabs.harvard.edu/abs/1926ApJ....64..321H} {64, 321}

\bibitem[\protect\citeauthoryear{{Huertas-Company} et~al.,}{{Huertas-Company} et~al.}{2023}]{Huertas_company23}
{Huertas-Company} M.,  et~al., 2023, \mn@doi [arXiv e-prints] {10.48550/arXiv.2305.02478}, \href {https://ui.adsabs.harvard.edu/abs/2023arXiv230502478H} {p. arXiv:2305.02478}

\bibitem[\protect\citeauthoryear{{Ilbert} et~al.,}{{Ilbert} et~al.}{2006}]{ilbert06}
{Ilbert} O.,  et~al., 2006, \mn@doi [\aap] {10.1051/0004-6361:20065138}, \href {https://ui.adsabs.harvard.edu/abs/2006A&A...457..841I} {457, 841}

\bibitem[\protect\citeauthoryear{{Ito} et~al.,}{{Ito} et~al.}{2023}]{Ito23}
{Ito} K.,  et~al., 2023, \mn@doi [arXiv e-prints] {10.48550/arXiv.2307.06994}, \href {https://ui.adsabs.harvard.edu/abs/2023arXiv230706994I} {p. arXiv:2307.06994}

\bibitem[\protect\citeauthoryear{{Jacobs} et~al.,}{{Jacobs} et~al.}{2023}]{Jacobs23}
{Jacobs} C.,  et~al., 2023, \mn@doi [\apjl] {10.3847/2041-8213/accd6d}, \href {https://ui.adsabs.harvard.edu/abs/2023ApJ...948L..13J} {948, L13}

\bibitem[\protect\citeauthoryear{{Jarvis} et~al.,}{{Jarvis} et~al.}{2013}]{VIDEO}
{Jarvis} M.~J.,  et~al., 2013, \mn@doi [\mnras] {10.1093/mnras/sts118}, \href {https://ui.adsabs.harvard.edu/abs/2013MNRAS.428.1281J} {428, 1281}

\bibitem[\protect\citeauthoryear{{Ji} et~al.,}{{Ji} et~al.}{2024}]{Ji24}
{Ji} Z.,  et~al., 2024, \mn@doi [arXiv e-prints] {10.48550/arXiv.2401.00934}, \href {https://ui.adsabs.harvard.edu/abs/2024arXiv240100934J} {p. arXiv:2401.00934}

\bibitem[\protect\citeauthoryear{{Juod{\v{z}}balis} et~al.,}{{Juod{\v{z}}balis} et~al.}{2023}]{Juodzbalis23}
{Juod{\v{z}}balis} I.,  et~al., 2023, \mn@doi [MNRAS] {10.1093/mnras/stad2396}, \href {https://ui.adsabs.harvard.edu/abs/2023MNRAS.525.1353J} {525, 1353}

\bibitem[\protect\citeauthoryear{{Kannan}, {Garaldi}, {Smith}, {Pakmor}, {Springel}, {Vogelsberger}  \& {Hernquist}}{{Kannan} et~al.}{2022a}]{Kannan22}
{Kannan} R.,  {Garaldi} E.,  {Smith} A.,  {Pakmor} R.,  {Springel} V.,  {Vogelsberger} M.,   {Hernquist} L.,  2022a, \mn@doi [\mnras] {10.1093/mnras/stab3710}, \href {https://ui.adsabs.harvard.edu/abs/2022MNRAS.511.4005K} {511, 4005}

\bibitem[\protect\citeauthoryear{{Kannan}, {Garaldi}, {Smith}, {Pakmor}, {Springel}, {Vogelsberger}  \& {Hernquist}}{{Kannan} et~al.}{2022b}]{THESAN22}
{Kannan} R.,  {Garaldi} E.,  {Smith} A.,  {Pakmor} R.,  {Springel} V.,  {Vogelsberger} M.,   {Hernquist} L.,  2022b, \mn@doi [\mnras] {10.1093/mnras/stab3710}, \href {https://ui.adsabs.harvard.edu/abs/2022MNRAS.511.4005K} {511, 4005}

\bibitem[\protect\citeauthoryear{{Kartaltepe} et~al.,}{{Kartaltepe} et~al.}{2023}]{kartaltepe23}
{Kartaltepe} J.~S.,  et~al., 2023, \mn@doi [\apjl] {10.3847/2041-8213/acad01}, \href {https://ui.adsabs.harvard.edu/abs/2023ApJ...946L..15K} {946, L15}

\bibitem[\protect\citeauthoryear{{Katz} et~al.,}{{Katz} et~al.}{2023}]{katz23}
{Katz} H.,  et~al., 2023, \mn@doi [The Open Journal of Astrophysics] {10.21105/astro.2309.03269}, \href {https://ui.adsabs.harvard.edu/abs/2023OJAp....6E..44K} {6, 44}

\bibitem[\protect\citeauthoryear{{Kawamata}, {Ishigaki}, {Shimasaku}, {Oguri}, {Ouchi}  \& {Tanigawa}}{{Kawamata} et~al.}{2018}]{kawamata18}
{Kawamata} R.,  {Ishigaki} M.,  {Shimasaku} K.,  {Oguri} M.,  {Ouchi} M.,   {Tanigawa} S.,  2018, \mn@doi [\apj] {10.3847/1538-4357/aaa6cf}, \href {https://ui.adsabs.harvard.edu/abs/2018ApJ...855....4K} {855, 4}

\bibitem[\protect\citeauthoryear{{Kawinwanichakij} et~al.,}{{Kawinwanichakij} et~al.}{2021}]{Kawinwanichakij21}
{Kawinwanichakij} L.,  et~al., 2021, \mn@doi [\apj] {10.3847/1538-4357/ac1f21}, \href {https://ui.adsabs.harvard.edu/abs/2021ApJ...921...38K} {921, 38}

\bibitem[\protect\citeauthoryear{{Kocevski} et~al.,}{{Kocevski} et~al.}{2023}]{Kocevski23}
{Kocevski} D.~D.,  et~al., 2023, \mn@doi [\apjl] {10.3847/2041-8213/ace5a0}, \href {https://ui.adsabs.harvard.edu/abs/2023ApJ...954L...4K} {954, L4}

\bibitem[\protect\citeauthoryear{{Koekemoer} et~al.,}{{Koekemoer} et~al.}{2011}]{Candels2}
{Koekemoer} A.~M.,  et~al., 2011, \mn@doi [\apjs] {10.1088/0067-0049/197/2/36}, \href {https://ui.adsabs.harvard.edu/abs/2011ApJS..197...36K} {197, 36}

\bibitem[\protect\citeauthoryear{{Koposov} et~al.,}{{Koposov} et~al.}{2023}]{koposov23}
{Koposov} S.,  et~al., 2023, {joshspeagle/dynesty: v2.1.2}, Zenodo, \mn@doi{10.5281/zenodo.7995596}

\bibitem[\protect\citeauthoryear{{Kron}}{{Kron}}{1980}]{Kron80}
{Kron} R.~G.,  1980, \mn@doi [\apjs] {10.1086/190669}, \href {https://ui.adsabs.harvard.edu/abs/1980ApJS...43..305K} {43, 305}

\bibitem[\protect\citeauthoryear{{LaChance}, {Croft}, {Ni}, {Chen}, {Di Matteo}  \& {Bird}}{{LaChance} et~al.}{2024}]{LaChance24}
{LaChance} P.,  {Croft} R.,  {Ni} Y.,  {Chen} N.,  {Di Matteo} T.,   {Bird} S.,  2024, \mn@doi [arXiv e-prints] {10.48550/arXiv.2401.16608}, \href {https://ui.adsabs.harvard.edu/abs/2024arXiv240116608L} {p. arXiv:2401.16608}

\bibitem[\protect\citeauthoryear{{Labbe} et~al.,}{{Labbe} et~al.}{2023}]{Labbe23}
{Labbe} I.,  et~al., 2023, \mn@doi [arXiv e-prints] {10.48550/arXiv.2306.07320}, \href {https://ui.adsabs.harvard.edu/abs/2023arXiv230607320L} {p. arXiv:2306.07320}

\bibitem[\protect\citeauthoryear{{Lange} et~al.,}{{Lange} et~al.}{2015}]{Lange15}
{Lange} R.,  et~al., 2015, \mn@doi [\mnras] {10.1093/mnras/stu2467}, \href {https://ui.adsabs.harvard.edu/abs/2015MNRAS.447.2603L} {447, 2603}

\bibitem[\protect\citeauthoryear{{Law}, {Steidel}, {Shapley}, {Nagy}, {Reddy}  \& {Erb}}{{Law} et~al.}{2012}]{law12}
{Law} D.~R.,  {Steidel} C.~C.,  {Shapley} A.~E.,  {Nagy} S.~R.,  {Reddy} N.~A.,   {Erb} D.~K.,  2012, \mn@doi [\apj] {10.1088/0004-637X/745/1/85}, \href {https://ui.adsabs.harvard.edu/abs/2012ApJ...745...85L} {745, 85}

\bibitem[\protect\citeauthoryear{{Lawrence} et~al.,}{{Lawrence} et~al.}{2007}]{uds}
{Lawrence} A.,  et~al., 2007, \mn@doi [MNRAS] {10.1111/j.1365-2966.2007.12040.x}, \href {https://ui.adsabs.harvard.edu/abs/2007MNRAS.379.1599L} {379, 1599}

\bibitem[\protect\citeauthoryear{{Leethochawalit} et~al.,}{{Leethochawalit} et~al.}{2023}]{Leethochawalit23}
{Leethochawalit} N.,  et~al., 2023, \mn@doi [\apjl] {10.3847/2041-8213/ac959b}, \href {https://ui.adsabs.harvard.edu/abs/2023ApJ...942L..26L} {942, L26}

\bibitem[\protect\citeauthoryear{{Legrand} et~al.,}{{Legrand} et~al.}{2019}]{Legrand19}
{Legrand} L.,  et~al., 2019, \mn@doi [\mnras] {10.1093/mnras/stz1198}, \href {https://ui.adsabs.harvard.edu/abs/2019MNRAS.486.5468L} {486, 5468}

\bibitem[\protect\citeauthoryear{{Long} et~al.,}{{Long} et~al.}{2023}]{Long23}
{Long} A.~S.,  et~al., 2023, \mn@doi [arXiv e-prints] {10.48550/arXiv.2305.04662}, \href {https://ui.adsabs.harvard.edu/abs/2023arXiv230504662L} {p. arXiv:2305.04662}

\bibitem[\protect\citeauthoryear{{Lovell}, {Vijayan}, {Thomas}, {Wilkins}, {Barnes}, {Irodotou}  \& {Roper}}{{Lovell} et~al.}{2021}]{FLARES}
{Lovell} C.~C.,  {Vijayan} A.~P.,  {Thomas} P.~A.,  {Wilkins} S.~M.,  {Barnes} D.~J.,  {Irodotou} D.,   {Roper} W.,  2021, \mn@doi [\mnras] {10.1093/mnras/staa3360}, \href {https://ui.adsabs.harvard.edu/abs/2021MNRAS.500.2127L} {500, 2127}

\bibitem[\protect\citeauthoryear{{Marshall}, {Wilkins}, {Di Matteo}, {Roper}, {Vijayan}, {Ni}, {Feng}  \& {Croft}}{{Marshall} et~al.}{2022}]{Marshall22}
{Marshall} M.~A.,  {Wilkins} S.,  {Di Matteo} T.,  {Roper} W.~J.,  {Vijayan} A.~P.,  {Ni} Y.,  {Feng} Y.,   {Croft} R. A.~C.,  2022, \mn@doi [\mnras] {10.1093/mnras/stac380}, \href {https://ui.adsabs.harvard.edu/abs/2022MNRAS.511.5475M} {511, 5475}

\bibitem[\protect\citeauthoryear{{Matthee} et~al.,}{{Matthee} et~al.}{2023}]{matthee23}
{Matthee} J.,  et~al., 2023, \mn@doi [arXiv e-prints] {10.48550/arXiv.2306.05448}, \href {https://ui.adsabs.harvard.edu/abs/2023arXiv230605448M} {p. arXiv:2306.05448}

\bibitem[\protect\citeauthoryear{{McCracken} et~al.,}{{McCracken} et~al.}{2012}]{UltraVISTA}
{McCracken} H.~J.,  et~al., 2012, \mn@doi [\aap] {10.1051/0004-6361/201219507}, \href {https://ui.adsabs.harvard.edu/abs/2012A&A...544A.156M} {544, A156}

\bibitem[\protect\citeauthoryear{{McLeod}, {McLure}, {Dunlop}, {Cullen}, {Carnall}  \& {Duncan}}{{McLeod} et~al.}{2021}]{McLeod21}
{McLeod} D.~J.,  {McLure} R.~J.,  {Dunlop} J.~S.,  {Cullen} F.,  {Carnall} A.~C.,   {Duncan} K.,  2021, \mn@doi [\mnras] {10.1093/mnras/stab731}, \href {https://ui.adsabs.harvard.edu/abs/2021MNRAS.503.4413M} {503, 4413}

\bibitem[\protect\citeauthoryear{{Miyazaki} et~al.,}{{Miyazaki} et~al.}{2002}]{miyazaki02}
{Miyazaki} S.,  et~al., 2002, \mn@doi [\pasj] {10.1093/pasj/54.6.833}, \href {https://ui.adsabs.harvard.edu/abs/2002PASJ...54..833M} {54, 833}

\bibitem[\protect\citeauthoryear{{Mo}, {Mao}  \& {White}}{{Mo} et~al.}{1998}]{Mo98}
{Mo} H.~J.,  {Mao} S.,   {White} S. D.~M.,  1998, \mn@doi [\mnras] {10.1046/j.1365-8711.1998.01227.x}, \href {https://ui.adsabs.harvard.edu/abs/1998MNRAS.295..319M} {295, 319}

\bibitem[\protect\citeauthoryear{{Morishita} et~al.,}{{Morishita} et~al.}{2023}]{morishita23}
{Morishita} T.,  et~al., 2023, \mn@doi [arXiv e-prints] {10.48550/arXiv.2308.05018}, \href {https://ui.adsabs.harvard.edu/abs/2023arXiv230805018M} {p. arXiv:2308.05018}

\bibitem[\protect\citeauthoryear{{Mosleh}, {Williams}, {Franx}  \& {Kriek}}{{Mosleh} et~al.}{2011}]{mosleh10}
{Mosleh} M.,  {Williams} R.~J.,  {Franx} M.,   {Kriek} M.,  2011, \mn@doi [\apj] {10.1088/0004-637X/727/1/5}, \href {https://ui.adsabs.harvard.edu/abs/2011ApJ...727....5M} {727, 5}

\bibitem[\protect\citeauthoryear{{Nelson} et~al.,}{{Nelson} et~al.}{2016}]{Nelson16}
{Nelson} E.~J.,  et~al., 2016, \mn@doi [\apj] {10.3847/0004-637X/828/1/27}, \href {https://ui.adsabs.harvard.edu/abs/2016ApJ...828...27N} {828, 27}

\bibitem[\protect\citeauthoryear{Nightingale, Hayes  \& Griffiths}{Nightingale et~al.}{2021}]{pyautofit}
Nightingale J.~W.,  Hayes R.~G.,   Griffiths M.,  2021, \mn@doi [J. Open Source Softw.] {10.21105/joss.02550}, 6, 2550

\bibitem[\protect\citeauthoryear{Nightingale et~al.,}{Nightingale et~al.}{2023}]{pyautogalaxy}
Nightingale J.~W.,  et~al., 2023, \mn@doi [J. Open Source Softw.] {10.21105/joss.04475}, 8, 4475

\bibitem[\protect\citeauthoryear{{Oesch} et~al.,}{{Oesch} et~al.}{2010}]{Oesch10}
{Oesch} P.~A.,  et~al., 2010, \mn@doi [\apjl] {10.1088/2041-8205/709/1/L21}, \href {https://ui.adsabs.harvard.edu/abs/2010ApJ...709L..21O} {709, L21}

\bibitem[\protect\citeauthoryear{{Oke} \& {Gunn}}{{Oke} \& {Gunn}}{1983}]{oke83}
{Oke} J.~B.,  {Gunn} J.~E.,  1983, \mn@doi [\apj] {10.1086/160817}, \href {https://ui.adsabs.harvard.edu/abs/1983ApJ...266..713O} {266, 713}

\bibitem[\protect\citeauthoryear{{Ono} et~al.,}{{Ono} et~al.}{2018}]{ono18}
{Ono} Y.,  et~al., 2018, \mn@doi [PASJ] {10.1093/pasj/psx103}, \href {https://ui.adsabs.harvard.edu/abs/2018PASJ...70S..10O} {70, S10}

\bibitem[\protect\citeauthoryear{{Ono} et~al.,}{{Ono} et~al.}{2023}]{ono23_psf}
{Ono} Y.,  et~al., 2023, \mn@doi [ApJ] {10.3847/1538-4357/acd44a}, \href {https://ui.adsabs.harvard.edu/abs/2023ApJ...951...72O} {951, 72}

\bibitem[\protect\citeauthoryear{{Ormerod} et~al.,}{{Ormerod} et~al.}{2023}]{Ormerod23}
{Ormerod} K.,  et~al., 2023, \mn@doi [arXiv e-prints] {10.48550/arXiv.2309.04377}, \href {https://ui.adsabs.harvard.edu/abs/2023arXiv230904377O} {p. arXiv:2309.04377}

\bibitem[\protect\citeauthoryear{{Pandya} et~al.,}{{Pandya} et~al.}{2024}]{Pandya24}
{Pandya} V.,  et~al., 2024, \mn@doi [\apj] {10.3847/1538-4357/ad1a13}, \href {https://ui.adsabs.harvard.edu/abs/2024ApJ...963...54P} {963, 54}

\bibitem[\protect\citeauthoryear{{Peebles}}{{Peebles}}{1969}]{Peebles69}
{Peebles} P.~J.~E.,  1969, \mn@doi [\apj] {10.1086/149876}, \href {https://ui.adsabs.harvard.edu/abs/1969ApJ...155..393P} {155, 393}

\bibitem[\protect\citeauthoryear{{Peng} et~al.,}{{Peng} et~al.}{2010}]{Peng10}
{Peng} Y.-j.,  et~al., 2010, \mn@doi [\apj] {10.1088/0004-637X/721/1/193}, \href {https://ui.adsabs.harvard.edu/abs/2010ApJ...721..193P} {721, 193}

\bibitem[\protect\citeauthoryear{{Perrin}, {Sivaramakrishnan}, {Lajoie}, {Elliott}, {Pueyo}, {Ravindranath}  \& {Albert}}{{Perrin} et~al.}{2014}]{WebbPSF}
{Perrin} M.~D.,  {Sivaramakrishnan} A.,  {Lajoie} C.-P.,  {Elliott} E.,  {Pueyo} L.,  {Ravindranath} S.,   {Albert} L.,  2014, in {Oschmann} Jacobus~M. J.,  {Clampin} M.,  {Fazio} G.~G.,   {MacEwen} H.~A.,  eds,  Society of Photo-Optical Instrumentation Engineers (SPIE) Conference Series Vol. 9143, Space Telescopes and Instrumentation 2014: Optical, Infrared, and Millimeter Wave. p. 91433X, \mn@doi{10.1117/12.2056689}

\bibitem[\protect\citeauthoryear{{Petrosian}}{{Petrosian}}{1976}]{Petrosian76}
{Petrosian} V.,  1976, \mn@doi [\apjl] {10.1086/182301}, \href {https://ui.adsabs.harvard.edu/abs/1976ApJ...209L...1P} {210, L53}

\bibitem[\protect\citeauthoryear{Phan, Pradhan  \& Jankowiak}{Phan et~al.}{2019}]{numpyro1}
Phan D.,  Pradhan N.,   Jankowiak M.,  2019, arXiv preprint arXiv:1912.11554

\bibitem[\protect\citeauthoryear{{Pichon}, {Pogosyan}, {Kimm}, {Slyz}, {Devriendt}  \& {Dubois}}{{Pichon} et~al.}{2011}]{Pichon11}
{Pichon} C.,  {Pogosyan} D.,  {Kimm} T.,  {Slyz} A.,  {Devriendt} J.,   {Dubois} Y.,  2011, \mn@doi [\mnras] {10.1111/j.1365-2966.2011.19640.x}, \href {https://ui.adsabs.harvard.edu/abs/2011MNRAS.418.2493P} {418, 2493}

\bibitem[\protect\citeauthoryear{{Pillepich} et~al.,}{{Pillepich} et~al.}{2018}]{illustrisTNG}
{Pillepich} A.,  et~al., 2018, \mn@doi [\mnras] {10.1093/mnras/stx2656}, \href {https://ui.adsabs.harvard.edu/abs/2018MNRAS.473.4077P} {473, 4077}

\bibitem[\protect\citeauthoryear{{Pillepich} et~al.,}{{Pillepich} et~al.}{2019}]{TNG50}
{Pillepich} A.,  et~al., 2019, \mn@doi [\mnras] {10.1093/mnras/stz2338}, \href {https://ui.adsabs.harvard.edu/abs/2019MNRAS.490.3196P} {490, 3196}

\bibitem[\protect\citeauthoryear{{Roper}, {Lovell}, {Vijayan}, {Marshall}, {Irodotou}, {Kuusisto}, {Thomas}  \& {Wilkins}}{{Roper} et~al.}{2022}]{Roper22}
{Roper} W.~J.,  {Lovell} C.~C.,  {Vijayan} A.~P.,  {Marshall} M.~A.,  {Irodotou} D.,  {Kuusisto} J.~K.,  {Thomas} P.~A.,   {Wilkins} S.~M.,  2022, \mn@doi [\mnras] {10.1093/mnras/stac1368}, \href {https://ui.adsabs.harvard.edu/abs/2022MNRAS.514.1921R} {514, 1921}

\bibitem[\protect\citeauthoryear{{Roper} et~al.,}{{Roper} et~al.}{2023}]{Roper23}
{Roper} W.~J.,  et~al., 2023, \mn@doi [\mnras] {10.1093/mnras/stad2746}, \href {https://ui.adsabs.harvard.edu/abs/2023MNRAS.526.6128R} {526, 6128}

\bibitem[\protect\citeauthoryear{{Rujopakarn} et~al.,}{{Rujopakarn} et~al.}{2023}]{Rujopakarn23}
{Rujopakarn} W.,  et~al., 2023, \mn@doi [\apjl] {10.3847/2041-8213/accc82}, \href {https://ui.adsabs.harvard.edu/abs/2023ApJ...948L...8R} {948, L8}

\bibitem[\protect\citeauthoryear{{Scholtz} et~al.,}{{Scholtz} et~al.}{2023}]{Scholtz23}
{Scholtz} J.,  et~al., 2023, \mn@doi [arXiv e-prints] {10.48550/arXiv.2311.18731}, \href {https://ui.adsabs.harvard.edu/abs/2023arXiv231118731S} {p. arXiv:2311.18731}

\bibitem[\protect\citeauthoryear{{S{\'e}rsic}}{{S{\'e}rsic}}{1963}]{Sersic63}
{S{\'e}rsic} J.~L.,  1963, Boletin de la Asociacion Argentina de Astronomia La Plata Argentina, \href {https://ui.adsabs.harvard.edu/abs/1963BAAA....6...41S} {6, 41}

\bibitem[\protect\citeauthoryear{{Shen} et~al.,}{{Shen} et~al.}{2023}]{Shen23}
{Shen} L.,  et~al., 2023, \mn@doi [arXiv e-prints] {10.48550/arXiv.2310.13745}, \href {https://ui.adsabs.harvard.edu/abs/2023arXiv231013745S} {p. arXiv:2310.13745}

\bibitem[\protect\citeauthoryear{{Shen} et~al.,}{{Shen} et~al.}{2024}]{Shen24}
{Shen} X.,  et~al., 2024, \mn@doi [arXiv e-prints] {10.48550/arXiv.2402.08717}, \href {https://ui.adsabs.harvard.edu/abs/2024arXiv240208717S} {p. arXiv:2402.08717}

\bibitem[\protect\citeauthoryear{{Shibuya}, {Ouchi}  \& {Harikane}}{{Shibuya} et~al.}{2015}]{Shibuya15}
{Shibuya} T.,  {Ouchi} M.,   {Harikane} Y.,  2015, \mn@doi [\apjs] {10.1088/0067-0049/219/2/15}, \href {https://ui.adsabs.harvard.edu/abs/2015ApJS..219...15S} {219, 15}

\bibitem[\protect\citeauthoryear{{Shibuya}, {Ouchi}, {Kubo}  \& {Harikane}}{{Shibuya} et~al.}{2016}]{Shibuya16}
{Shibuya} T.,  {Ouchi} M.,  {Kubo} M.,   {Harikane} Y.,  2016, \mn@doi [\apj] {10.3847/0004-637X/821/2/72}, \href {https://ui.adsabs.harvard.edu/abs/2016ApJ...821...72S} {821, 72}

\bibitem[\protect\citeauthoryear{{Speagle}}{{Speagle}}{2020}]{Speagle20}
{Speagle} J.~S.,  2020, \mn@doi [\mnras] {10.1093/mnras/staa278}, \href {https://ui.adsabs.harvard.edu/abs/2020MNRAS.493.3132S} {493, 3132}

\bibitem[\protect\citeauthoryear{{Stanway} \& {Eldridge}}{{Stanway} \& {Eldridge}}{2018}]{BPASS}
{Stanway} E.~R.,  {Eldridge} J.~J.,  2018, \mn@doi [\mnras] {10.1093/mnras/sty1353}, \href {https://ui.adsabs.harvard.edu/abs/2018MNRAS.479...75S} {479, 75}

\bibitem[\protect\citeauthoryear{{Stefanon} et~al.,}{{Stefanon} et~al.}{2019}]{stefanon19}
{Stefanon} M.,  et~al., 2019, \mn@doi [ApJ] {10.3847/1538-4357/ab3792}, \href {https://ui.adsabs.harvard.edu/abs/2019ApJ...883...99S} {883, 99}

\bibitem[\protect\citeauthoryear{{Steidel}, {Adelberger}, {Giavalisco}, {Dickinson}  \& {Pettini}}{{Steidel} et~al.}{1999}]{Steidel99}
{Steidel} C.~C.,  {Adelberger} K.~L.,  {Giavalisco} M.,  {Dickinson} M.,   {Pettini} M.,  1999, \mn@doi [\apj] {10.1086/307363}, \href {https://ui.adsabs.harvard.edu/abs/1999ApJ...519....1S} {519, 1}

\bibitem[\protect\citeauthoryear{{Suess}, {Kriek}, {Price}  \& {Barro}}{{Suess} et~al.}{2019a}]{Suess19a}
{Suess} K.~A.,  {Kriek} M.,  {Price} S.~H.,   {Barro} G.,  2019a, \mn@doi [\apj] {10.3847/1538-4357/ab1bda}, \href {https://ui.adsabs.harvard.edu/abs/2019ApJ...877..103S} {877, 103}

\bibitem[\protect\citeauthoryear{{Suess}, {Kriek}, {Price}  \& {Barro}}{{Suess} et~al.}{2019b}]{Suess19b}
{Suess} K.~A.,  {Kriek} M.,  {Price} S.~H.,   {Barro} G.,  2019b, \mn@doi [\apjl] {10.3847/2041-8213/ab4db3}, \href {https://ui.adsabs.harvard.edu/abs/2019ApJ...885L..22S} {885, L22}

\bibitem[\protect\citeauthoryear{{Suess} et~al.,}{{Suess} et~al.}{2022}]{Suess22}
{Suess} K.~A.,  et~al., 2022, \mn@doi [\apjl] {10.3847/2041-8213/ac8e06}, \href {https://ui.adsabs.harvard.edu/abs/2022ApJ...937L..33S} {937, L33}

\bibitem[\protect\citeauthoryear{{Tacchella} et~al.,}{{Tacchella} et~al.}{2023}]{tacchella23}
{Tacchella} S.,  et~al., 2023, \mn@doi [ApJ] {10.3847/1538-4357/acdbc6}, \href {https://ui.adsabs.harvard.edu/abs/2023ApJ...952...74T} {952, 74}

\bibitem[\protect\citeauthoryear{{Trujillo} et~al.,}{{Trujillo} et~al.}{2006}]{Trujillo06}
{Trujillo} I.,  et~al., 2006, \mn@doi [\apj] {10.1086/506464}, \href {https://ui.adsabs.harvard.edu/abs/2006ApJ...650...18T} {650, 18}

\bibitem[\protect\citeauthoryear{{Varadaraj}, {Bowler}, {Jarvis}, {Adams}  \& {H{\"a}u{\ss}ler}}{{Varadaraj} et~al.}{2023}]{Varadaraj23}
{Varadaraj} R.~G.,  {Bowler} R.~A.~A.,  {Jarvis} M.~J.,  {Adams} N.~J.,   {H{\"a}u{\ss}ler} B.,  2023, \mn@doi [\mnras] {10.1093/mnras/stad2081}, \href {https://ui.adsabs.harvard.edu/abs/2023MNRAS.524.4586V} {524, 4586}

\bibitem[\protect\citeauthoryear{{Ventou} et~al.,}{{Ventou} et~al.}{2017}]{Ventou17}
{Ventou} E.,  et~al., 2017, \mn@doi [\aap] {10.1051/0004-6361/201731586}, \href {https://ui.adsabs.harvard.edu/abs/2017A&A...608A...9V} {608, A9}

\bibitem[\protect\citeauthoryear{{Vijayan}, {Lovell}, {Wilkins}, {Thomas}, {Barnes}, {Irodotou}, {Kuusisto}  \& {Roper}}{{Vijayan} et~al.}{2021}]{Vijayan21}
{Vijayan} A.~P.,  {Lovell} C.~C.,  {Wilkins} S.~M.,  {Thomas} P.~A.,  {Barnes} D.~J.,  {Irodotou} D.,  {Kuusisto} J.,   {Roper} W.~J.,  2021, \mn@doi [\mnras] {10.1093/mnras/staa3715}, \href {https://ui.adsabs.harvard.edu/abs/2021MNRAS.501.3289V} {501, 3289}

\bibitem[\protect\citeauthoryear{Virtanen et~al.,}{Virtanen et~al.}{2020}]{SciPy}
Virtanen P.,  et~al., 2020, \mn@doi [Nature Methods] {10.1038/s41592-019-0686-2}, \href {https://rdcu.be/b08Wh} {17, 261}

\bibitem[\protect\citeauthoryear{{Vogelsberger} et~al.,}{{Vogelsberger} et~al.}{2014}]{Vogelsberger14}
{Vogelsberger} M.,  et~al., 2014, \mn@doi [\nat] {10.1038/nature13316}, \href {https://ui.adsabs.harvard.edu/abs/2014Natur.509..177V} {509, 177}

\bibitem[\protect\citeauthoryear{{Ward} et~al.,}{{Ward} et~al.}{2023}]{Ward23}
{Ward} E.~M.,  et~al., 2023, \mn@doi [arXiv e-prints] {10.48550/arXiv.2311.02162}, \href {https://ui.adsabs.harvard.edu/abs/2023arXiv231102162W} {p. arXiv:2311.02162}

\bibitem[\protect\citeauthoryear{{Wechsler} \& {Tinker}}{{Wechsler} \& {Tinker}}{2018}]{wechsler18}
{Wechsler} R.~H.,  {Tinker} J.~L.,  2018, \mn@doi [\araa] {10.1146/annurev-astro-081817-051756}, \href {https://ui.adsabs.harvard.edu/abs/2018ARA&A..56..435W} {56, 435}

\bibitem[\protect\citeauthoryear{{Whitler}, {Stark}, {Endsley}, {Leja}, {Charlot}  \& {Chevallard}}{{Whitler} et~al.}{2023}]{whitler23}
{Whitler} L.,  {Stark} D.~P.,  {Endsley} R.,  {Leja} J.,  {Charlot} S.,   {Chevallard} J.,  2023, \mn@doi [MNRAS] {10.1093/mnras/stad004}, \href {https://ui.adsabs.harvard.edu/abs/2023MNRAS.519.5859W} {519, 5859}

\bibitem[\protect\citeauthoryear{{Whitney}, {Conselice}, {Bhatawdekar}  \& {Duncan}}{{Whitney} et~al.}{2019}]{whitney19}
{Whitney} A.,  {Conselice} C.~J.,  {Bhatawdekar} R.,   {Duncan} K.,  2019, \mn@doi [\apj] {10.3847/1538-4357/ab53d4}, \href {https://ui.adsabs.harvard.edu/abs/2019ApJ...887..113W} {887, 113}

\bibitem[\protect\citeauthoryear{{Wu}, {Dav{\'e}}, {Tacchella}  \& {Lotz}}{{Wu} et~al.}{2020}]{Wu20}
{Wu} X.,  {Dav{\'e}} R.,  {Tacchella} S.,   {Lotz} J.,  2020, \mn@doi [\mnras] {10.1093/mnras/staa1044}, \href {https://ui.adsabs.harvard.edu/abs/2020MNRAS.494.5636W} {494, 5636}

\bibitem[\protect\citeauthoryear{{Yang} et~al.,}{{Yang} et~al.}{2022}]{Yang22}
{Yang} L.,  et~al., 2022, \mn@doi [\mnras] {10.1093/mnras/stac1236}, \href {https://ui.adsabs.harvard.edu/abs/2022MNRAS.514.1148Y} {514, 1148}

\bibitem[\protect\citeauthoryear{{Zhuang} \& {Shen}}{{Zhuang} \& {Shen}}{2023}]{zhuang23}
{Zhuang} M.-Y.,  {Shen} Y.,  2023, \mn@doi [arXiv e-prints] {10.48550/arXiv.2304.13776}, \href {https://ui.adsabs.harvard.edu/abs/2023arXiv230413776Z} {p. arXiv:2304.13776}

\bibitem[\protect\citeauthoryear{{de Jong} \& {Lacey}}{{de Jong} \& {Lacey}}{2000}]{deJong00}
{de Jong} R.~S.,  {Lacey} C.,  2000, \mn@doi [\apj] {10.1086/317840}, \href {https://ui.adsabs.harvard.edu/abs/2000ApJ...545..781D} {545, 781}

\bibitem[\protect\citeauthoryear{{van der Wel} et~al.,}{{van der Wel} et~al.}{2014}]{vanderWel14}
{van der Wel} A.,  et~al., 2014, \mn@doi [\apj] {10.1088/0004-637X/788/1/28}, \href {https://ui.adsabs.harvard.edu/abs/2014ApJ...788...28V} {788, 28}

\makeatother
\end{thebibliography}




\appendix

\section{Comparison between S\'ersic and non-parametric sizes}
\label{sec:size measurement comparison}

\begin{figure}
    \centering
    \includegraphics[width=\columnwidth]{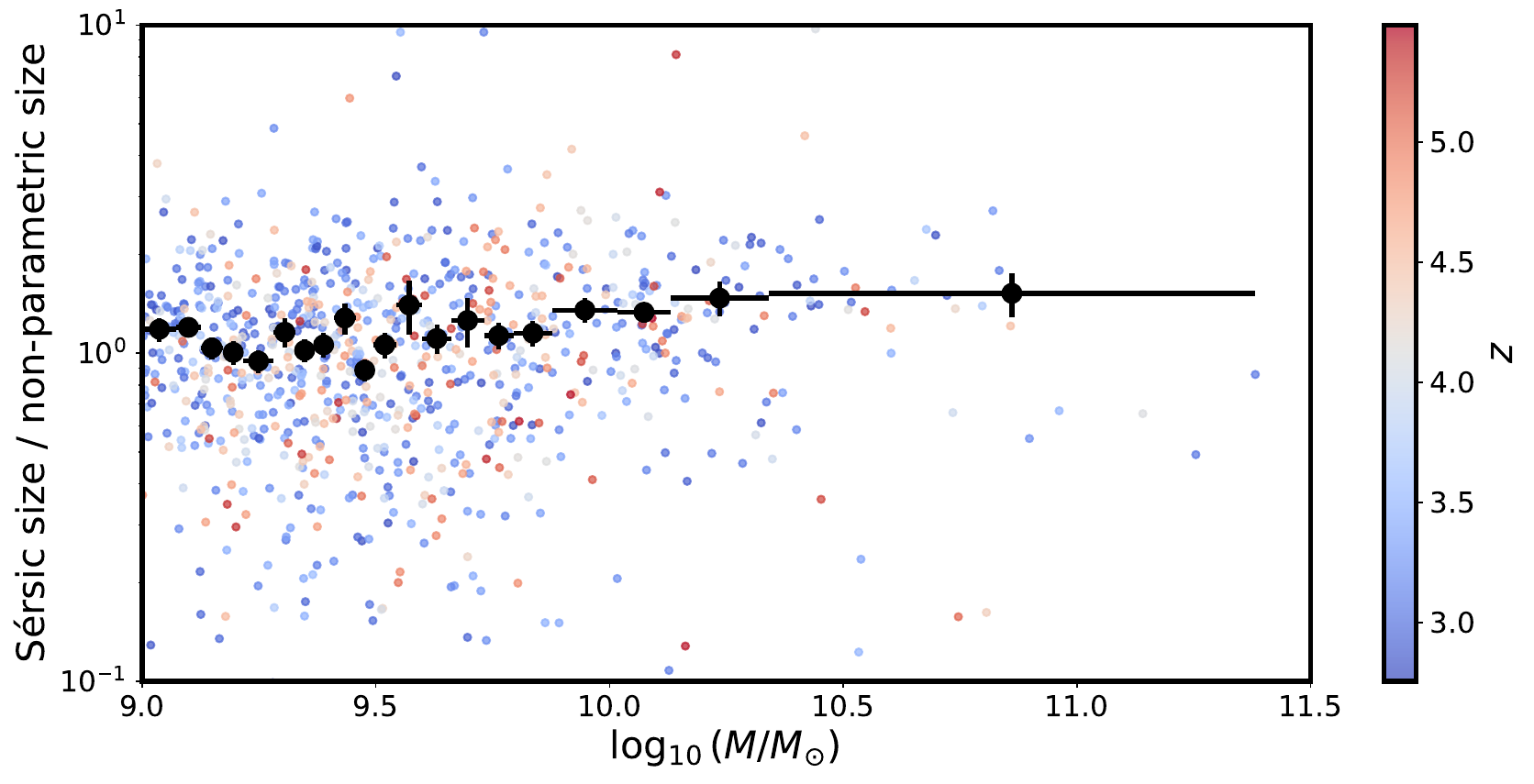}
    \includegraphics[width=\columnwidth]{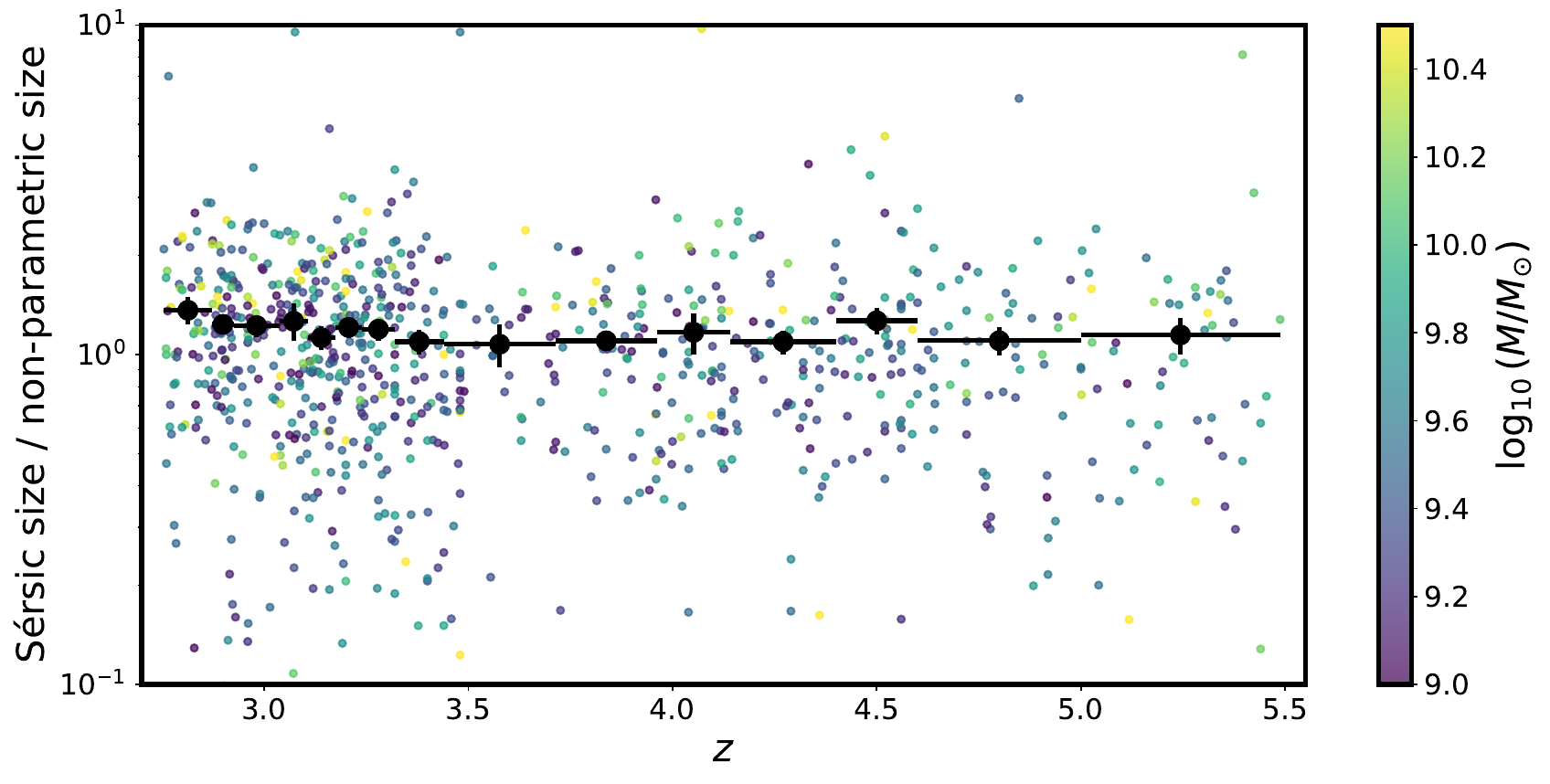}
    \caption{\textbf{Top:} the ratio of S\'ersic size to non-parametric size in F356W as a function of stellar mass for galaxies with a good S\'ersic fit. Individual galaxies are coloured by their photometric redshift. \textbf{Bottom:} now the ratio is plotted as a function of photometric redshift, with galaxies coloured by their stellar mass. In both panels, the black points show the results of binning the size ratio.}
    \label{fig:ratios}
\end{figure}

In this appendix we compare the S\'ersic sizes and non-parametric sizes measured from F356W.
In Fig. \ref{fig:ratios} we show the ratio of S\'ersic size to non-parametric size as a function of both stellar mass and redshift for objects with a good S\'ersic fit.
We bin the data such that there is an equal number of galaxies in each bin.
The errors are the standard error on the mean.
The mean ratios in the bins show little evolution across mass and redshift, remaining at a value just above one. 
As discussed in Section \ref{sec:size-luminosity}, this is expected as in some cases the S\'ersic fitting may tend to find larger sizes.
The amount of scatter appears to be consistent with simulations \citep[e.g.][]{Constantin23, LaChance24}.



\bsp	
\label{lastpage}
\end{document}